\documentclass[preprint,10pt]{elsarticle}

\usepackage{amssymb}

\usepackage{amsmath}
\usepackage{cancel}
\usepackage{epstopdf}
\usepackage{graphicx}
\usepackage{caption}
\usepackage{adjustbox}
\usepackage{lipsum}
\usepackage{subcaption}
\graphicspath{ }
\usepackage{float}
\journal{}

\makeatletter
\def\ps@pprintTitle{%
 \let\@oddhead\@empty
 \let\@evenhead\@empty
 \def\@oddfoot{}%
 \let\@evenfoot\@oddfoot}
\makeatother

\newcommand{\fr}{\displaystyle \frac}
\DeclareMathOperator*{\Max}{max}

\begin{document}

\begin{frontmatter}



\title{A Boltzmann scheme with physically relevant discrete velocities for Euler equations}


\author[iisc1]{N. Venkata Raghavendra}
\ead{venkata@aero.iisc.ernet.in}

\author[iisc2]{S. V. Raghurama Rao\corref{cor1}}
\ead{raghu@aero.iisc.ernet.in}
\cortext[cor1]{Corresponding author: Tel.: +91 80 2293 3031; fax: +91 80 2360 0134}

\address{Department of Aerospace Engineering, Indian Institute of Science, Bangalore}

\fntext[iisc1]{PhD research scholar}
\fntext[iisc2]{Associate Professor}
\begin{abstract} 
	Kinetic or Boltzmann schemes are interesting alternatives to the macroscopic numerical methods for solving the hyperbolic conservation laws of gas dynamics. They utilize the particle-based description instead of the wave propagation models.  While the continuous particle velocity based upwind schemes were developed in the earlier decades, the discrete velocity Boltzmann schemes introduced in the last decade are found to be simpler and are easier to handle.  In this work, we introduce a novel way of introducing discrete velocities which correspond to the physical wave speeds and formulate a discrete velocity Boltzmann scheme for solving Euler equations.      
\end{abstract}


\end{frontmatter}


\section{Introduction}  
      Kinetic or Boltzmann schemes, introduced during 1970s to 1990s, were interesting alternatives for the popular upwind schemes devised for solving the conservation laws of gas dynamics based on wave propagation methods.  The significant schemes in this category are due to Sanders \& Prendergast \cite{Prendergast1}, Pullin \cite{Pullin}, Reitz \cite{Reitz}, Deshpande \cite{Deshpande_AIAA}, Mandal \& Deshpande \cite{Mandal_Deshpande}, Kaniel \cite{Kaniel}, Perthame \cite{Perthame}, Prendergast \& Kun Xu \cite{Prendergast_KunXu}, Raghurama Rao and Deshpande \cite{SVRRao_SMD}.  The above schemes utilize the continuous molecular velocity for introducing upwinding, together with either Maxwellian distribution functions, Dirac delta functions or compactly supported distributions.  Discrete velocity Boltzmann schemes were introduced by Natalini \cite{Natalini}, Aregba-Driollet \& Natalini \cite{Driollet_Natalini}, Raghurama Rao \& Balakrishna \cite{SVRRao_Balakrishna}, Raghurama Rao \& Subba Rao \cite{SVRRao_SubbaRao}, Arun \emph{et al.} \cite{Arun_RaghuramaRao,Arun_RaghuramaRao_KRS,Arun_Maria}, among others.  The discrete velocity Boltzmann schemes present certain advantages compared to the continuous molecular velocity based upwind schemes by being simpler in design and analysis of numerical schemes.  In this work, we introduce discrete velocities by using a novel interpretation and a slight modification of the strategy used by Sanders \& Prendergast \cite{Prendergast1} and further introduce a discrete velocity Boltzmann scheme for solving Euler equations of gas dynamics.  

\section{Continuous and discrete velocity Boltzmann schemes}  

     Kinetic or Boltzmann schemes exploit the connection between the Boltzmann equation of kinetic theory of gases and the macroscopic conservation laws of gas dynamics, obtained as moments of the Boltzmann equation.  Since the Boltzmann equation contains a linear convection term, introducing upwinding is simpler in this framework.  The nonlinearity of the collision term leads to the nonlinear Euler equations after taking moments.  The Boltzmann equation, with B-G-K model \cite{BGK} for the collision term, is given in 1-D by 
\begin{equation}\label{bgk_eqn} 
\frac{\partial f}{\partial t} + v \frac{\partial f}{\partial x} = - \frac{1}{t_{R}} \left[ f - f^{eq} \right]   
\end{equation} 
     where $f$ is the molecular velocity distribution function, $v$ is the molecular velocity, $t_{R}$ is the relaxation time and $f^{eq}$ is the equilibrium distribution function typically being the Maxwellian.  The left hand side represents the convection term and the term on the right hand side is a model for the collision term.  Utilizing an operator splitting involving a convection step and a collision step and further simplifying the collision step by instantaneous relaxation to equilibrium ($t_{R} \rightarrow 0$),  we can rewrite the above equation as 
\begin{equation}
\frac{\partial f}{\partial t} + v \frac{\partial f}{\partial x} = 0, f = f^{eq} 
\end{equation} 
     The moment relations are defined by 
\begin{eqnarray} 
\textbf{U} = \left[ \begin{array}{c} \rho \\ \rho u \\ \rho E \end{array} \right] = 
\int_{0}^{\infty} \int_{- \infty}^{\infty} \Psi f \ dv \ d{I}   
\end{eqnarray} 
\begin{eqnarray} 
\textbf{G(U)} = \left[ \begin{array}{c} \rho u \\ p + \rho u^{2} \\ p u + \rho u E \end{array} \right] = 
\int_{0}^{\infty} \int_{- \infty}^{\infty} \Psi v f \ dv \ d{I}   
\end{eqnarray} 
where $\textbf{U}$ is the conserved variable vector and $\textbf{G}$ is the flux vector in the 1-D Euler equations, $\Psi$ is the moment function vector representing the collisional invariants and ${I}$ is the internal energy variable corresponding to non-translational degrees of freedom (introduced to take care of the poly-atomic nature of the gas).  The moment function vector consists of the mass, momentum and energy of the molecules that are conserved during collisions.  
\begin{equation}  
\Psi = \left[ \begin{array}{c} 
1 \\ v \\ {I} + \fr{1}{2} v^{2} 
\end{array} \right] 
\end{equation} 
The equilibrium distribution is typically the Maxwellian, given in 1-D by 
\begin{equation} \label{1d_maxwellian}
f^{eq} = \frac{\rho}{I_{0}} \left(\fr{\beta}{\pi}\right)^{\frac{1}{2}} e^{- \beta \left(v - u\right)^{2} } e^{-\frac{I}{I_{0}} }    
\end{equation} 
where 
\begin{align}
I_{0} &= \frac{(3-\gamma)RT}{2(\gamma-1)} \\
\beta &= \frac{1}{2RT}
\end{align}

The equilibrium distribution function $f^{eq}$ and the velocity $v$ in the Boltzmann equation are based on a molecular representation. The process of taking moments represents the macroscopic flow physics as a manifestation of the underlying molecular dynamics. This has two consequences for the numerical schemes based on the classical Boltzmann equation, herein referred to as \emph{continuous velocity Boltzmann schemes}:
\begin{enumerate}
\item They are based on discretizing the Boltzmann equation and then taking moments to obtain numerical methods for Euler equations, while the direct solution of the Boltzmann equation is not involved.  Typically upwinding based on molecular velocity involves splitting of the entire span of velocity space from $-\infty$ to $+\infty$ into two halves, leading to error functions and exponentials in the split fluxes, unless the Maxwellian is modified for introducing simplification.    
\item Since molecular dynamics is the premise for macroscopic physics, this is a bottom-up approach with the Boltzmann equation leading to the Euler equations. So, any concept derived directly from the Euler equations, like the Rankine-Hugoniot (R-H) conditions, cannot be easily introduced at the underlying molecular level in the design of continuous velocity Boltzmann schemes.  Thus, the accurate recognition of the shock waves, contact discontinuities and expansion waves is not easy in these  schemes.     
\end{enumerate}

Let us now consider an offshoot of Boltzmann schemes called \emph{discrete velocity Boltzmann schemes} or \emph{discrete kinetic schemes}.

Consider 1-D Euler equations governing compressible flows given by
\begin{equation}\label{euler_1d}
\frac{\partial{\mathbf{U}}}{\partial{t}}+\frac{\partial\mathbf{G(U)}}{\partial{x}} = 0
\end{equation}
with the initial condition
\begin{equation}\label{euler_1d_system_ic}
\textbf{U}(x,0) = \textbf{U}_0(x)
\end{equation}
Here $\textbf{U}$ is the vector of conserved variables and $\textbf{G(U)}$ is the flux vector, defined by
\begin{equation}\label{euler_1d_var_desc}
\textbf{U}\>=\left[{\begin{array}{cc}
                   U_1\\
                   U_2\\
                   U_3\\
                  \end{array}}\right]=\left[{\begin{array}{cc}
                   \rho\\
                   \rho u\\
                   \rho E\\
                  \end{array}}\right] \ \text{and} \ \mathbf{G(U)}=\left[{\begin{array}{cc}
                   G_1\\
                   G_2\\
                   G_3\\
                  \end{array}}\right]=\left[{\begin{array}{cc}
                   \rho u\\
                   p+\rho u^2\\
                   pu+\rho uE\\
                  \end{array}}\right]
\end{equation}
where $\rho$ is the density, $u$ is the velocity, $p$ is the pressure and $E$ is the total energy given by
\begin{equation}
E=\frac{p}{\rho(\gamma-1)}+\frac{u^2}{2}
\end{equation}             
with $\gamma$ being the ratio of specific heats.

The pressure, temperature and density are related by the equation of state:
\begin{equation}
p=\rho RT
\end{equation} 
where $R$ is the gas constant.  The speed of sound is then given by:
\begin{equation}
a=\sqrt{\frac{\gamma p}{\rho}}=\sqrt{\gamma RT}
\end{equation}

As discussed by Natalini and Aregba-Driollet (\hspace{1sp}\cite{Natalini},~\cite{Driollet_Natalini}), equation (\ref{euler_1d}) along with its initial condition (\ref{euler_1d_system_ic}) can be approximated by a sequence of semi-linear systems as:
\begin{equation}\label{euler_1d_semilinear}
\frac{\partial{\mathbf{f}}}{\partial{t}}+\mathbf{\Lambda} \frac{\partial{\mathbf{f}}}{\partial{x}}=-\frac{1}{\epsilon}[\mathbf{f}-\mathbf{f^\emph{eq}}]
\end{equation}
having initial condition $\textbf{f}(x,0)= \mathbf{f^\emph{eq}}(\textbf{U}_0(x))$.

Here $\textbf{f}$ represents the discrete distribution function, $\mathbf{f^\emph{eq}}$ is the corresponding local equilibrium distribution function, $\mathbf{\Lambda}$ is the diagonal matrix of $N$ discrete velocities ($\mathbf{\Lambda} = diag(\lambda_{q}), \ q=1, \cdots, N$) and $\epsilon$ is a relaxation parameter.

The equilibrium distribution function $\mathbf{f^\emph{eq}}$ and the matrix of discrete velocities $\mathbf{\Lambda}$ have to satisfy the following conditions of  the approximation:
\begin{enumerate}
\item  Consistency of the discrete kinetic approximation (\ref{euler_1d_semilinear}) with the system of Euler equations (\ref{euler_1d}) in the limit $\epsilon\rightarrow0$, leading to the moment relations:
\begin{equation}
\begin{aligned}\label{recover_euler_1d_system_thru_semilinear}
{U}_{l} & = \sum_{q=1}^{N} {f}_{q,l} \\ 
{G}_{l} & = \sum_{q=1}^{N} \lambda_{q} {f}_{q,l}    
\end{aligned}
\end{equation}
where $l \in [1,L] $. (Here $L=3$ in view of three conservation laws in 1-D Euler system.)

\item Stability of the approximation (\ref{euler_1d_semilinear}) through a non-negative diffusion in the model seen through the Chapman-Enskog type expansion \cite{Driollet_Natalini} or by satisfying the Bouchut's stability condition~\cite{Bouchut}.  The discrete kinetic system represents a vanishing viscosity model for the original set of hyperbolic conservation laws.  

\end{enumerate}

Equation (\ref{euler_1d_semilinear}) is referred to as the \emph{discrete velocity Boltzmann equation (DVBE)} and can be re-written using a splitting method~\cite{SVRRao_Balakrishna} as:

\emph{Collision Step:}
\begin{equation}\label{collision_step}
\frac{d{\mathbf{f}}}{dt} = -\frac{1}{\epsilon}[\textbf{f}-\mathbf{f^\emph{eq}}]
\end{equation}

and a \emph{Convection Step:}
\begin{equation}\label{convec_step}
\frac{\partial{\mathbf{f}}}{\partial{t}}+\mathbf{\Lambda}\frac{\partial\mathbf{f}}{\partial{x}} = 0
\end{equation}

In the limit $\epsilon\rightarrow0$, assuming \emph{instantaneous relaxation to equilibrium}, the DVBE represented by steps (\ref{collision_step}) and (\ref{convec_step}) becomes
\begin{equation}\label{eq_discrete_kinetic_form_euler_1d}
\textbf{f} = \mathbf{f^\emph{eq}}, \ \frac{\partial{\mathbf{f}}}{\partial{t}}+\mathbf{\Lambda}\frac{\partial\mathbf{f}}{\partial{x}} = 0
\end{equation}  
Numerical schemes based on the DVBE are called discrete kinetic schemes or discrete velocity Boltzmann schemes.  

Natalini and Aregba-Driollet (\hspace{1sp}\cite{Natalini},~\cite{Driollet_Natalini}) suggest that the equilibrium distribution function in the DVBE can be expressed as an algebraic combination of the conserved variable vector $\textbf{U}$ and the flux vector $\textbf{G}$. In 1-D, for example, we can have 
\begin{equation}
\mathbf{f^\emph{eq}} = \phi_1\mathbf{U} + \phi_2\mathbf{G(U)}
\end{equation}
where $\phi_1$, $\phi_2$ are some scalars.

 In contrast to continuous velocity Boltzmann schemes, discrete velocity Boltzmann schemes enjoy the following advantages:
\begin{enumerate}
\item They discretize the simpler discrete velocity Boltzmann equation. Subsequently, solutions for the Euler equations can be easily obtained using moments which are simple algebraic expressions, unlike the complicated integrals involved in the moments of the classic Boltzmann equation.  
\item The equilibrium distributions are simple algebraic functions of the macroscopic physical variables, unlike the Maxwellians which are Gaussians in classical kinetic theory.  
\item There is a two-way correspondence between the discrete velocity Boltzmann equation and the Euler equations.  In the top-down approach, starting from the Euler equations, a relaxation approximation can be introduced as in Jin and Xin \cite{Jin_Xin}.  In the limit of zero relaxation parameter, the diagonal form of such a relaxation system leads to discrete velocity Boltzmann system \cite{Natalini,Driollet_Natalini,SVRRao_Balakrishna} which further is useful in constructing discrete velocity Boltzmann schemes.  In the bottom-up approach, starting from a discrete velocity Boltzmann equation, introducing upwinding and then taking moments lead to an upwind scheme for macroscopic Euler equations.  This two-way correspondence can be useful in further analysis and design of better numerical methods.       
 
\end{enumerate}

\section{A novel discrete kinetic approximation with physically relevant discrete velocities for Euler equations}\label{sec:Euler_DKS}
The system of 1-D Euler equations (\ref{euler_1d}) with the definitions in equation (\ref{euler_1d_var_desc}) can be written using index-notation as three scalar conservation laws:
\begin{equation}\label{euler_1d_index}
\frac{\partial{U_l}}{\partial{t}}+\frac{\partial{G_l}}{\partial{x}} = 0, \ l = 1,2,3
\end{equation}
The discrete velocity Boltzmann equation (\ref{eq_discrete_kinetic_form_euler_1d}) for the 1-D Euler system then turns out to be:
\begin{equation}\label{eq_discrete_kinetic_form_euler_1d_index}
\textbf{f}_l = \mathbf{f^\emph{eq}_\emph{l}}(\mathbf{U}_0(x)), \ \frac{\partial{\mathbf{f}}_l}{\partial{t}}+\mathbf{\Lambda}\frac{\partial\mathbf{f}_l}{\partial{x}} = 0, \ l=1,2,3
\end{equation}

As shown by Aregba-Driollet and Natalini~\cite{Driollet_Natalini}, for $N$ discrete velocities, we can define the distribution function $\textbf{f}_l$, the corresponding local equilibrium distribution function $\mathbf{f^\emph{eq}_\emph{l}}$ and the diagonal matrix of discrete velocities $\mathbf{\Lambda}$ as
\begin{equation}\label{euler_1d_semilinear_desc}
\mathbf{f}_l=\left[{\begin{array}{c}
                   f_1\\
                   \vdots\\
                   f_N\\
                  \end{array}}\right]_l \>; \mathbf{f^\emph{eq}_\emph{l}}=\left[{\begin{array}{c}
                   f^{eq}_1\\
                   \vdots\\
                   f^{eq}_N\\
                  \end{array}}\right]_l \>; \mathbf{\Lambda}=\left[{\begin{array}{ccc}
                   \lambda_{1} & 0 & 0\\
                   0 & \ddots & 0\\
                   0 & 0 & \lambda_{N}\\
                  \end{array}}\right] \>, \ l=1,2,3
\end{equation}

As the discrete velocities are constant over a convection time step, using the definitions in (\ref{euler_1d_semilinear_desc}), we can alternatively write DVBE (\ref{eq_discrete_kinetic_form_euler_1d_index}) as:
\begin{equation}\label{discrete_kinetic_form_euler_1d_desc}
\textbf{f}_l = \mathbf{f^\emph{eq}_\emph{l}}, \ \frac{\partial{{f}}_{ql}}{\partial{t}}+\frac{\partial({\lambda_\emph{q}{f}_{ql}})}{\partial{x}} = 0, \ q=1,2,...,N; \ l=1,2,3
\end{equation}

We now need to make appropriate choices for $\mathbf{f^\emph{eq}_\emph{l}}$ and $\mathbf{\Lambda}$ so that the moment relations  (\ref{recover_euler_1d_system_thru_semilinear}) are satisfied for each of the Euler equations ($l=1,2,3$).  
\begin{align}
\label{recover_cons_euler_1d}
U_l =& \ \sum_{q=1}^N{f^\emph{eq}_\emph{ql}} \\
\label{recover_flux_euler_1d}
G_l =& \ \sum_{q=1}^N\lambda_\emph{q} f^\emph{eq}_\emph{ql}
\end{align}

We set $N=3$ in equation (\ref{euler_1d_semilinear_desc}) and set out to determine the discrete velocities. Note that according to the discrete velocity Boltzmann  approximation, $\lambda_{q}$ are constants to be determined such that stability of the approximation using Chapman-Enskog type expansion or Bouchut's condition is satisfied.  In this study, we take motivation from the work of Sanders and Prendergast~\cite{Prendergast1} and determine the discrete velocities by ensuring that the moment relations are satisfied but the discrete velocities mimic the eigenvalues of the flux Jacobian matrix for the original conservation laws.  It is to be noted however, that the equilibrium functions in our framework are not the same as the set of Dirac delta functions but will be determined differently based on the moment relations.  The resulting numerical method will still be in the framework of discrete velocity Boltzmann schemes but with different algebraic expressions for the equilibria.   

Sanders and Prendergast~\cite{Prendergast1} obtain the particle speeds as ($u-\sqrt{3RT}$, $u$, $u+\sqrt{3RT}$).  Note that the factor $\sqrt{3}$ in the particle speeds $u\pm\sqrt{3RT}$ is different from the  coefficient $\sqrt{\gamma}$ in the wave speeds for Euler equations as the sound speed is given by $\sqrt{\gamma RT}$.  This factor of $\sqrt{3}$ gets introduced through the fourth moment of the equilibrium distribution function used by Sanders and Prendergast, from one of the moments required to derive Navier-Stokes equations, together with the consistency conditions with Euler equations.  We note that strictly speaking this is not necessary as the fourth moment is relevant only in connection with obtaining the viscous stresses and heat flux vector for Navier-Stokes equations, based on Chapman-Enskog distribution function.  In the following, we present the modified derivation.  

\subsection{Derivation of discrete velocities}

Sanders and Prendergast~\cite{Prendergast1} devised their Beam scheme wherein they replaced the Maxwellian distribution function by a combination of Dirac delta functions called beams and set out to evaluate the beam weights and beam velocities such that Euler equations can still be recovered by taking moments of this distribution function. In the case of 1-D Euler equations, the distribution function $\bar{F}$ is a combination of three beams:
\begin{itemize}
\item A central beam with weight $\alpha$ and beam velocity $\tilde{u}$
\item Two side beams symmetrically located (in the space of molecular velocity $v$) with weights $\beta$ and beam velocities $\tilde{u}\pm\Delta{u}$
\end{itemize}
Mathematically
\begin{equation}\label{beam_dist_fn}
\bar{F} = \alpha\delta(v-\tilde{u})+\beta\delta(v-\tilde{u}+\Delta u)+\beta\delta(v-\tilde{u}-\Delta u)
\end{equation}

Clearly, in the formulation of the distribution function $\bar{F}$ in (\ref{beam_dist_fn}), there are four unknowns: $\alpha$, $\beta$, $\tilde{u}$ and $\Delta u$. To evaluate the unknowns, four equations are required for which Sanders and Prendergast~\cite{Prendergast1} choose the following moments:

\emph{First moment:}
\begin{equation}\label{beam_zero_moment}
	\rho = \int_{-\infty}^{\infty} \bar{F} dv 
\end{equation}
Substituting (\ref{beam_dist_fn}) in (\ref{beam_zero_moment}) and performing the integration, we get
\begin{equation}\label{beam_zero_moment_intg}
\rho = \alpha + 2 \beta
\end{equation}
\emph{Second moment:}
\begin{equation}\label{beam_first_moment}
\rho u = \int_{-\infty}^{\infty} v  \bar{F} dv 
\end{equation}
Substituting (\ref{beam_dist_fn}) in (\ref{beam_first_moment}) and performing the integration, we get
\begin{equation}\label{beam_first_moment_intg}
\rho u = (\alpha+2\beta)\tilde{u} \implies \tilde{u} = u \qquad \text{[using (\ref{beam_zero_moment_intg})]}
\end{equation}
\emph{Third moment:}
\begin{equation}\label{beam_second_moment}
p + \rho u^{2} = \int_{-\infty}^{\infty} v^{2} \bar{F} dv 
\end{equation}
Using (\ref{beam_dist_fn}) in (\ref{beam_second_moment}) and performing the integration, we get
\begin{equation}\label{beam_second_moment_intg}
p + \rho u^{2} = (\alpha+2\beta)\tilde{u}^2 + 2\beta(\Delta u)^2 \implies p = 2\beta(\Delta u)^2 \quad \text{[using (\ref{beam_zero_moment_intg}) and (\ref{beam_first_moment_intg})]}
\end{equation} 
The above three moments are obtained from the definitions of the conserved variable vector $U$ and the flux vector $G$, with two of the six relations being repetitive.  The fourth moment is as follows.
 
\emph{Fourth moment:}
\begin{equation} \label{beam_third_moment}
3 p u + \rho u^{3} = \int_{-\infty}^{\infty} v^{3} \bar{F} dv
\end{equation} 
Substituting (\ref{beam_dist_fn}) in (\ref{beam_third_moment}) and performing the integration, we get
\begin{equation}\label{beam_third_moment_intg}
3 p u + \rho u^{3} = (\alpha+2\beta)\tilde{u}^3 + 6\beta\tilde{u}(\Delta u)^2 \implies p = 2\beta(\Delta u)^2 \quad \text{[using (\ref{beam_zero_moment_intg}) and (\ref{beam_first_moment_intg})]}
\end{equation}
This is the same as (\ref{beam_second_moment_intg}). So, this moment is redundant. In Beam scheme~\cite{Prendergast1}, instead of the last integral (\ref{beam_third_moment}), the following moment is used to obtain the fourth equation:  

\emph{Fourth moment used in Beam scheme~\cite{Prendergast1}:}
\begin{equation} 
3 \frac{p^{2}}{\rho} = \int_{-\infty}^{\infty} \left( v - u \right)^4 \bar{F} dv 
\end{equation} 
However, the selection of this integral is arbitrary and this integral is not a part of the moments leading to Euler equations at all.  Instead, it is a part of the derivation of Navier-Stokes equations, based on Chapman-Enskog distribution function.  Here, we avoid using this  moment and propose an alternative way of deriving the discrete velocities (or the beam velocities).  

As of now, we have three equations (\ref{beam_zero_moment_intg}), (\ref{beam_first_moment_intg}) and (\ref{beam_second_moment_intg}) but four unknowns.  That means we are free to exercise one choice.  

Using the definition of the sound speed $a$, we have 
\begin{equation} \label{sound_sp_defn}
a^{2} = \frac{\gamma p}{\rho} \ \textrm{or} \ p = \frac{\rho a^{2}}{\gamma}    
\end{equation}
Comparing (\ref{sound_sp_defn}) and (\ref{beam_second_moment_intg}), we have
\begin{equation}
\frac{\rho a^{2}}{\gamma} = 2 \beta \left( \Delta u \right)^{2} 
\end{equation}
Using (\ref{beam_zero_moment_intg}), the above equation can be rewritten as
\begin{equation}
\left(\alpha + 2 \beta \right) \frac{a^{2}}{\gamma} = 2 \beta \left( \Delta u \right)^{2}  
\end{equation}
or 
\begin{equation}
\left(\Delta u\right)^{2} = a^{2} \frac{\alpha + 2\beta}{2 \beta \gamma}  
\end{equation}
Therefore 
\begin{equation}
\Delta u = \pm a \sqrt{\frac{\alpha + 2\beta}{2 \beta \gamma}} 
\end{equation}
Let us now exercise our choice and choose 
\begin{equation} \label{fourth_moment_choice}
\alpha + 2\beta = 2 \beta \gamma 
\end{equation}
so that 
\begin{equation}
\Delta u = \pm a = \pm \sqrt{\gamma RT}
\end{equation}
Therefore, the discrete velocities turn out to be $\lambda_1 = u-a$ and $\lambda_2 = u$ and $\lambda_3 = u + a$ where $a = \sqrt{\gamma R T}$.  
We thus obtain the discrete velocities which are physically more relevant in line with the wave speeds of the Euler system.  The discrete velocities in our formulation are similar to the particle speeds used by Tang and Xu~\cite{Tang_Xu} while our derivation and framework are distinctly different.  

We therefore have
\begin{equation}\label{euler_1d_semilinear_phy_RS}
\mathbf{f}_l=\left[{\begin{array}{c}
                   f_1\\
                   f_2\\
                   f_3\\
                  \end{array}}\right]_l \>; \mathbf{f^\emph{eq}_\emph{l}}=\left[{\begin{array}{c}
                   f^{eq}_1\\
                   f^{eq}_2\\
                   f^{eq}_3\\
                  \end{array}}\right]_l \>; \mathbf{\Lambda}=\left[{\begin{array}{ccc}
                   u-a & 0 & 0\\
                   0 & u & 0\\
                   0 & 0 & u+a\\
                  \end{array}}\right] \>, \ l=1,2,3
\end{equation}

\begin{figure}[H]
\includegraphics[width=5cm]{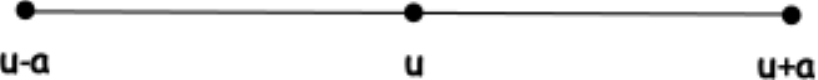}
\centering
\caption{Discrete velocities in 1-D velocity space}
\label{fig:1d_disc_velocities}
\end{figure}

The choice of discrete velocities in the velocity space is depicted in figure \ref{fig:1d_disc_velocities}.\\

We now set out to evaluate the equilibrium distribution functions $\textbf{f}^\emph{eq}_\emph{l}$ using the moment conditions (\ref{recover_cons_euler_1d}) and (\ref{recover_flux_euler_1d}). This will complete our discrete velocity Boltzmann system (\ref{discrete_kinetic_form_euler_1d_desc}), which can further be used for developing a numerical method for solving Euler equations.  

Applying consistency conditions with the continuity equation ($l=1$), we get
\begin{align}
f^{\emph{eq}}_{11}+f^{\emph{eq}}_{21}+f^{\emph{eq}}_{31}&=\rho \label{eval_feq_euler_1d_mass:1}\\
(u-a)f^{\emph{eq}}_{11}+uf^{\emph{eq}}_{21}+(u+a)f^{\emph{eq}}_{31}&=\rho u \label{eval_feq_euler_1d_mass:2}
\end{align}

Next applying consistency conditions with the momentum equation ($l=2$), we get
\begin{align}
f^{\emph{eq}}_{12}+f^{\emph{eq}}_{22}+f^{\emph{eq}}_{32}&=\rho u \label{eval_feq_euler_1d_mom:1}\\
(u-a)f^{\emph{eq}}_{12}+uf^{\emph{eq}}_{22}+(u+a)f^{\emph{eq}}_{32}&=(p+\rho u^2) \label{eval_feq_euler_1d_mom:2}
\end{align}

Observing equivalence between equations (\ref{eval_feq_euler_1d_mass:2}) and (\ref{eval_feq_euler_1d_mom:1}) for any general case, we set
\begin{align}
f^{\emph{eq}}_{12}&=(u-a)f^{\emph{eq}}_{11} \label{eval_feq_euler_1d_eqn_1}\\
f^{\emph{eq}}_{22}&=uf^{\emph{eq}}_{21} \label{eval_feq_euler_1d_eqn_2}\\
f^{\emph{eq}}_{32}&=(u+a)f^{\emph{eq}}_{31} \label{eval_feq_euler_1d_eqn_3}
\end{align}
Substituting the above three relations in (\ref{eval_feq_euler_1d_mom:2}), we get
\begin{align}
(u-a)^2f^{\emph{eq}}_{11}+u^2f^{\emph{eq}}_{21}+(u+a)^2f^{\emph{eq}}_{31}&=(p+\rho u^2) \label{eval_feq_euler_1d_eqn_4} \\
\implies u^2\cancel{(f^{\emph{eq}}_{11}+f^{\emph{eq}}_{21}+f^{\emph{eq}}_{31})}+a^2(f^{\emph{eq}}_{11}+f^{\emph{eq}}_{31})+2au(f^{\emph{eq}}_{31}-f^{\emph{eq}}_{11})&=(p+\cancel{\rho u^2}) \label{eval_feq_euler_1d_eqn_5}
\end{align}
\qquad \qquad \qquad \qquad \qquad \qquad \qquad \qquad \qquad \qquad \qquad \qquad (using equation (\ref{eval_feq_euler_1d_mass:1}))
\begin{equation}
\implies a^2(f^{\emph{eq}}_{11}+f^{\emph{eq}}_{31})+2au(f^{\emph{eq}}_{31}-f^{\emph{eq}}_{11})=p \label{eval_feq_euler_1d_eqn_6}
\end{equation}
Since the right hand side (RHS) of equation (\ref{eval_feq_euler_1d_eqn_6}) namely $p$ is purely a thermodynamic variable and not a function of $u$, the coefficient of the term containing $u$ on the left hand side (LHS) should be zero.
\begin{align}
\implies& 2au(f^{\emph{eq}}_{31}-f^{\emph{eq}}_{11})=0 \label{eval_feq_euler_1d_eqn_7} \\
\implies& f^{\emph{eq}}_{11}=f^{\emph{eq}}_{31} \label{eval_feq_euler_1d_eqn_8}
\end{align}
Substituting the above result (\ref{eval_feq_euler_1d_eqn_8}) in (\ref{eval_feq_euler_1d_eqn_6}), we get
\begin{equation}
f^{\emph{eq}}_{11}=f^{\emph{eq}}_{31}=\frac{p}{2a^2}=\frac{p}{2\gamma RT}=\frac{\rho}{2\gamma} \label{eval_feq_euler_1d_eqn_9}
\end{equation}
From (\ref{eval_feq_euler_1d_mass:1}), we then have
\begin{equation}
f^{\emph{eq}}_{21}=\rho\bigg(1-\frac{1}{\gamma}\bigg) \label{eval_feq_euler_1d_eqn_10}
\end{equation}
Using equations (\ref{eval_feq_euler_1d_eqn_9}) and (\ref{eval_feq_euler_1d_eqn_10}) in (\ref{eval_feq_euler_1d_eqn_1})-(\ref{eval_feq_euler_1d_eqn_3}), we get
\begin{align}
f^{\emph{eq}}_{12}&= \frac{\rho(u-a)}{2\gamma} \label{eval_feq_euler_1d_eqn_11} \\
f^{\emph{eq}}_{22}&= \rho u\bigg(1-\frac{1}{\gamma}\bigg) \label{eval_feq_euler_1d_eqn_12} \\
f^{\emph{eq}}_{32}&= \frac{\rho(u+a)}{2\gamma} \label{eval_feq_euler_1d_eqn_13} 
\end{align}

Next applying consistency conditions (\ref{recover_cons_euler_1d}) and (\ref{recover_flux_euler_1d}) with the energy equation ($l=3$), we get
\begin{align}
f^{\emph{eq}}_{13}+f^{\emph{eq}}_{23}+f^{\emph{eq}}_{33}&=\rho E=\frac{p}{(\gamma-1)}+\frac{\rho u^2}{2} \label{eval_feq_euler_1d_energy:1}\\
(u-a)f^{\emph{eq}}_{13}+uf^{\emph{eq}}_{23}+(u+a)f^{\emph{eq}}_{33}&=pu+\rho uE = pu \fr{\gamma}{\left(\gamma - 1\right)} + \fr{1}{2} \rho u^{3}  \label{eval_feq_euler_1d_energy:2}
\end{align}
From the above two equations (\ref{eval_feq_euler_1d_energy:1}) and (\ref{eval_feq_euler_1d_energy:2}), we get
\begin{align}
a(f^{\emph{eq}}_{33}-f^{\emph{eq}}_{13})&=pu=\rho RTu = \frac{\rho a^2u}{\gamma} \label{eval_feq_euler_1d_eqn_14} \\
\implies (f^{\emph{eq}}_{33}-f^{\emph{eq}}_{13})&=\frac{\rho au}{\gamma} \label{eval_feq_euler_1d_eqn_15}
\end{align}
From equation (\ref{eval_feq_euler_1d_energy:1}) we observe that the equilibrium distribution functions $f^{\emph{eq}}_{13}, \ f^{\emph{eq}}_{23}$ and $f^{\emph{eq}}_{33}$ should be linear combinations of internal and kinetic energies. We then introduce the following expressions conforming to equations (\ref{eval_feq_euler_1d_energy:1}) and (\ref{eval_feq_euler_1d_eqn_15}):
\begin{align}
f^{\emph{eq}}_{13}&=\frac{\rho (u-a)^2}{4\gamma} + c_1 I_0 \label{eval_feq_euler_1d_eqn_16}\\
f^{\emph{eq}}_{23}&=\frac{\rho u^2}{2}\bigg(1-\frac{1}{\gamma}\bigg) + c_2 I_0 \label{eval_feq_euler_1d_eqn_17}\\
f^{\emph{eq}}_{33}&=\frac{\rho (u+a)^2}{4\gamma} + c_3 I_0, \ c_3=c_1 \label{eval_feq_euler_1d_eqn_18}
\end{align}
where $I_0$ is the internal energy due to non-translational degrees of freedom and $c_1, \ c_2, \ c_3$ are constants to be determined.

We use the expression for $I_0$ from kinetic theory~\cite{Deshpande2} for 1-D given by
\begin{equation}
I_0=\frac{(3-\gamma)RT}{2(\gamma-1)} \label{I0_1D}
\end{equation}

Using expressions (\ref{eval_feq_euler_1d_eqn_16})-(\ref{eval_feq_euler_1d_eqn_18}) along with (\ref{I0_1D}) in (\ref{eval_feq_euler_1d_energy:2}), we get
\begin{align}
c_1+c_2+c_3 = \rho \label{eval_feq_euler_1d_eqn_19}\\
(u-a)c_1+uc_2+(u+a)c_3 = \rho u\label{eval_feq_euler_1d_eqn_20}\\
c_1=c_3 \label{eval_feq_euler_1d_eqn_21}
\end{align}
The above equations (\ref{eval_feq_euler_1d_eqn_19})-(\ref{eval_feq_euler_1d_eqn_21}) are similar to the equations (\ref{eval_feq_euler_1d_mass:1}), (\ref{eval_feq_euler_1d_mass:2}) and (\ref{eval_feq_euler_1d_eqn_8}) for $f^{\emph{eq}}_{11}, \ f^{\emph{eq}}_{21}, \ f^{\emph{eq}}_{31}$. Accordingly, we choose $c_1, \ c_2, \ c_3$ as
\begin{align}
c_1=f^{\emph{eq}}_{11}&= \frac{\rho}{2\gamma} \label{eval_feq_euler_1d_eqn_22} \\
c_2=f^{\emph{eq}}_{21}&= \rho \bigg(1-\frac{1}{\gamma}\bigg) \label{eval_feq_euler_1d_eqn_23} \\
c_3=f^{\emph{eq}}_{31}&= \frac{\rho}{2\gamma} \label{eval_feq_euler_1d_eqn_24} 
\end{align}

We therefore have the equilibrium distribution functions for the discrete velocity Boltzmann equation (\ref{discrete_kinetic_form_euler_1d_desc}) for 1-D Euler equations:
\begin{equation}\label{eval_feq_euler_1d_eqn_25}
\mathbf{f}^\emph{eq}_1=\left[{\begin{array}{c}
                   \frac{\rho}{2\gamma}\\
                   \rho \big(\frac{\gamma-1}{\gamma}\big)\\
                   \frac{\rho}{2\gamma}\\
                  \end{array}}\right] \>; \mathbf{f}^\emph{eq}_2=\left[{\begin{array}{c}
                   \frac{\rho(u-a)}{2\gamma}\\
                   \rho u\big(\frac{\gamma-1}{\gamma}\big)\\
                   \frac{\rho(u+a)}{2\gamma}\\
                  \end{array}}\right] \>; \mathbf{f}^\emph{eq}_3=\left[{\begin{array}{c}
                   \frac{\rho (u-a)^2}{4\gamma} + \frac{\rho}{2\gamma} I_0\\
                   \frac{\rho u^2}{2}\big(\frac{\gamma-1}{\gamma}\big) + \rho \big(\frac{\gamma-1}{\gamma}\big) I_0\\
                   \frac{\rho (u+a)^2}{4\gamma} + \frac{\rho}{2\gamma} I_0\\
                  \end{array}}\right] \>
\end{equation}

As remarked earlier, we note that the above set of equilibrium distribution functions are similar to the vectors of mass, momentum and energy in the three-particle proposition by Tang and Xu~\cite{Tang_Xu} for Steger-Warming flux vector-splitting scheme \cite{Steger} for 1-D Euler equations. But it may be noted that the elements corresponding to internal energy are different.

\subsection{Stability condition for the discrete kinetic approximation}
The stability of the discrete kinetic approximation (\ref{euler_1d_semilinear}) is studied by doing a Chapman-Enskog analysis for the approximation as discussed by Natalini and Aregba-Driollet (\hspace{1sp}\cite{Natalini},~\cite{Driollet_Natalini}). They show that the approximation is a vanishing viscosity model to the original system of 1-D Euler equations (\ref{euler_1d}) with a viscosity matrix $\mathbf{\Gamma}$ given by:
\begin{equation}
\mathbf{\Gamma} =  \mathbf{P \Lambda^2 \frac{\partial{\mathbf{f^\emph{eq}}}}{\partial{\mathbf{U}}}} - \bigg(\frac{\partial{\mathbf{G}}}{\partial{{\mathbf{U}}}}\bigg)^2
\end{equation}
where $\mathbf{P} = [I_L \ I_L \ \dots \ I_L]$ and $I_L$ is a $3\times3$ identity matrix (in view of three conservation laws in 1-D Euler system)
For the approximation to be stable, matrix $\mathbf{\Gamma}$ needs to be positive-definite.

However, for the approximation (\ref{euler_1d_semilinear}), using expressions (\ref{euler_1d_semilinear_phy_RS}) and (\ref{eval_feq_euler_1d_eqn_25}) when we evaluate matrix $\mathbf{\Gamma}$, it turns out to be non-symmetric. So, it is not possible to determine the positive-definiteness.  Aregba-Driollet and Natalini~\cite{Driollet_Natalini} suggest that in the general case, we can check for the positive-definiteness of $(\mathbf{\Gamma}+\mathbf{\Gamma}^{T})$ which is a symmetric matrix. However, this criterion does not give an explicit condition for the stability of approximation (\ref{euler_1d_semilinear}). We therefore use another simpler but stronger stability condition for the approximation given by Bouchut~\cite{Bouchut}. For stability, it states that
\begin{equation}\label{bouchut_stability}
\Omega\bigg(\frac{\partial{\mathbf{f^\emph{eq}}}}{\partial{\mathbf{U}}}\bigg) \subset [0,+\infty[
\end{equation}
where $\Omega$ denotes the eigenspectrum.  Let us now apply the stability condition (\ref{bouchut_stability}) for approximation (\ref{euler_1d_semilinear}).  Before evaluating the Jacobian $\big(\frac{\partial{\mathbf{f^\emph{eq}}}}{\partial{\mathbf{U}}}\big)$, we first note the relation between $\mathbf{f^\emph{eq}}$ in (\ref{bouchut_stability}) and  the equilibrium distribution function vectors ${\mathbf{f}^\emph{eq}_l}$ given in (\ref{eval_feq_euler_1d_eqn_25}):

\begin{equation}\label{eval_feq_euler_1d_eqn_26}
\mathbf{f^\emph{eq}}=\left[{\begin{array}{c}
                   \left[{\begin{array}{c}
                   f^\emph{eq}_{11}\\
                   f^\emph{eq}_{12}\\
                   f^\emph{eq}_{13}\\
                  \end{array}}\right]\\ \\
                  
                   \left[{\begin{array}{c}
                   f^\emph{eq}_{21}\\
                   f^\emph{eq}_{22}\\
                   f^\emph{eq}_{23}\\
                  \end{array}}\right]\\ \\
                  
                   \left[{\begin{array}{c}
                   f^\emph{eq}_{31}\\
                   f^\emph{eq}_{32}\\
                   f^\emph{eq}_{33}\\
                  \end{array}}\right]\\
                  \end{array}}\right] \\ \\
\end{equation}

We now express $\mathbf{f^\emph{eq}}$ in terms of the conserved variables $\textbf{U}$ defined in (\ref{euler_1d_var_desc}):

\begin{align}\label{eval_feq_euler_1d_eqn_27}
\mathbf{f}^\emph{eq}&=\left[{\begin{array}{c}
					\left[{\begin{array}{c}
                   \frac{U_1}{2\gamma}\\
					\frac{U_2}{2\gamma}-\frac{\sqrt{\gamma(\gamma-1)}}{2\gamma}\bigg(U_1U_3-\frac{U_2^2}{2}\bigg)^\frac{1}{2}\\                   
                   \frac{U_2^2}{4\gamma U_1}-\frac{\sqrt{\gamma(\gamma-1)}}{2\gamma}\frac{U_2}{U_1}\bigg(U_1U_3-\frac{U_2^2}{2}\bigg)^\frac{1}{2}+ \frac{(\gamma^2-2\gamma+3)}{4\gamma}\bigg(U_3-\frac{U_2^2}{2U_1}\bigg)\\
                  \end{array}}\right] \> \\ \\
\left[{\begin{array}{c}
				   U_1 \bigg(1-\frac{1}{\gamma}\bigg)\\
                   U_2\bigg(1-\frac{1}{\gamma}\bigg)\\
                   \frac{(\gamma-1)U_2^2}{2\gamma U_1}+ \frac{(3-\gamma)(\gamma-1)}{2\gamma}\bigg(U_3-\frac{U_2^2}{2U_1}\bigg)\\
                  \end{array}}\right] \> \\ \\
\left[{\begin{array}{c}
                   \frac{U_1}{2\gamma}\\
					\frac{U_2}{2\gamma}+\frac{\sqrt{\gamma(\gamma-1)}}{2\gamma}\bigg(U_1U_3-\frac{U_2^2}{2}\bigg)^\frac{1}{2}\\                   
                   \frac{U_2^2}{4\gamma U_1}+\frac{\sqrt{\gamma(\gamma-1)}}{2\gamma}\frac{U_2}{U_1}\bigg(U_1U_3-\frac{U_2^2}{2}\bigg)^\frac{1}{2}+ \frac{(\gamma^2-2\gamma+3)}{4\gamma}\bigg(U_3-\frac{U_2^2}{2U_1}\bigg)\\
                  \end{array}}\right] \>
                  \end{array}}\right] \>                  
\end{align}
We see that $\mathbf{f^\emph{eq}}$ is a set of three column vectors. Each column vector has its Jacobian with respect to the vector of conserved variables $\textbf{U}$. Each of the three Jacobians has its set of eigenvalues. Bouchut's condition (\ref{bouchut_stability}) stipulates that all of the eigenvalues be non-negative. Let us examine this further.

We have used $\text{Mathematica}^{\textregistered}$9.0 software \cite{Mathematica} to obtain the eigenvalues of the Jacobians.

The eigenvalues corresponding to the first and third column vector in (\ref{eval_feq_euler_1d_eqn_27}) are the same and are given by:
\begin{equation}\label{eigenspectra_first_third_1d}
\Omega_1 = \Omega_3 = \bigg\lbrace\frac{1}{2\gamma}, \ -2+\gamma+\frac{5\pm\sqrt{\gamma^4-4\gamma^3+14\gamma^2-12\gamma+1}}{\gamma} \ \bigg\rbrace
\end{equation}
Bouchut's condition (\ref{bouchut_stability}) requires that the above eigenspectrum is non-negative. This gives the stability condition:
\begin{equation}\label{bouchut_condn_first_third_1d}
\gamma = 3
\end{equation}
The eigenvalues corresponding to the second column vector in (\ref{eval_feq_euler_1d_eqn_27}) are:
\begin{equation}\label{eigenspectra_second_1d}
\Omega_2 = \bigg\lbrace\frac{\gamma-1}{\gamma}, \ \frac{\gamma-1}{\gamma}, \ \frac{(3-\gamma)(\gamma-1)}{2\gamma} \bigg\rbrace
\end{equation}
Again, Bouchut's condition (\ref{bouchut_stability}) gives 
\begin{equation}\label{bouchut_condn_second_1d}
1 \leq \gamma \leq 3
\end{equation}
Condition (\ref{bouchut_condn_first_third_1d}) is contained in (\ref{bouchut_condn_second_1d}). So we refer to the latter for stability condition and choose $\gamma=1.4$ for our 1-D numerical computations.

\subsection{Extension to two dimensions}
In the case of 2-D, Euler equations are given by
\begin{equation}\label{euler_2d}
\frac{\partial{\mathbf{U}}}{\partial{t}}+\frac{\partial\mathbf{G_1(U)}}{\partial{x}}+\frac{\partial\mathbf{G_2(U)}}{\partial{y}} = 0
\end{equation}
with the initial condition
\begin{equation}\label{euler_2d_system_ic}
\textbf{U}(x,y,0) = \textbf{U}_0(x,y)
\end{equation}
The conserved variable vector and flux vectors are given by
\begin{equation}\label{euler_2d_var_desc}
\textbf{U}\>=\left[{\begin{array}{cc}
                   U_1\\
                   U_2\\
                   U_4\\
                   U_3\\
                  \end{array}}\right]=\left[{\begin{array}{cc}
                   \rho\\
                   \rho u_1\\
                   \rho u_2\\
                   \rho E\\
                  \end{array}}\right]; \ \mathbf{G}_1=\left[{\begin{array}{cc}
                   \rho u_1\\
                   p+\rho u_1^2\\
                   \rho u_1u_2\\
                   pu_1+\rho u_1E\\
                  \end{array}}\right]; \ \mathbf{G}_2=\left[{\begin{array}{cc}
                   \rho u_2\\
                   \rho u_1u_2\\
                   p+\rho u_2^2\\
                   pu_2+\rho u_2E\\
                  \end{array}}\right]
\end{equation}
where the total energy $E$ is now given by
\begin{equation}
E=\frac{p}{\rho(\gamma-1)}+\frac{u_1^2+u_2^2}{2}
\end{equation}             

The discrete kinetic approximation for the above system is given by:
\begin{equation}\label{euler_2d_semilinear}
\frac{\partial{\mathbf{f}}}{\partial{t}}+\mathbf{\Lambda_1} \frac{\partial{\mathbf{f}}}{\partial{x}}+\mathbf{\Lambda_2} \frac{\partial{\mathbf{f}}}{\partial{y}}=-\frac{1}{\epsilon}[\mathbf{f}-\mathbf{f^\emph{eq}}]
\end{equation}
having initial condition $\textbf{f}(x,y,0)= \mathbf{f^\emph{eq}}(\textbf{U}_0(x,y))$.

We note that in this case we have diagonal matrices for discrete velocities in $x$ and $y$ directions: $\mathbf{\Lambda_1}$ and $\mathbf{\Lambda_2}$.

The necessary conditions for the discrete approximation (\ref{euler_2d_semilinear}) to converge to the Euler equations (\ref{euler_2d}) in the limit $\epsilon\rightarrow0$ are:
\begin{equation}
\begin{aligned}\label{recover_euler_2d_system_thru_semilinear}
\mathbf{U} =& \ \mathbf{P}\mathbf{f^\emph{eq}} \\
\mathbf{G_1} =& \ \mathbf{P}\mathbf{\Lambda_1}\mathbf{f^\emph{eq}} \\
\mathbf{G_2} =& \ \mathbf{P}\mathbf{\Lambda_2}\mathbf{f^\emph{eq}}
\end{aligned}
\end{equation}
where $\mathbf{P} = [I_L \ I_L \ \dots \ I_L]$ and $I_L$ is a $4\times4$ identity matrix (in view of four conservation laws in 2-D Euler system).

Under the assumption of \emph{instantaneous relaxation}, the discrete velocity Boltzmann equations (DVBEs) for the four conservation laws of 2-D Euler system $(l = 1,2,3,4)$ can be written as:
\begin{equation}\label{eq_discrete_kinetic_form_euler_2d_index}
\textbf{f}_l = \mathbf{f^\emph{eq}_\emph{l}}, \ \frac{\partial{\mathbf{f}}_l}{\partial{t}}+\mathbf{\Lambda_1}\frac{\partial\mathbf{f}_l}{\partial{x}}+\mathbf{\Lambda_2}\frac{\partial\mathbf{f}_l}{\partial{y}} = 0
\end{equation}
where for $l=1,2,3,4$:
\begin{align}\label{euler_2d_semilinear_desc}
&\mathbf{f_\emph{l}}=[f_1 \ f_2 \dots \ f_N]^T_l \\
&\mathbf{f^\emph{eq}_\emph{l}}=[f^{eq}_1 \ f^{eq}_2 \ \dots \ f^{eq}_N]^T_l \\
\label{euler_2d_semilinear_desc_2}
&\mathbf{\Lambda_1}=\left[{\begin{array}{ccc}
                   \lambda^x_{1} & 0 & 0\\
                   0 & \ddots & 0\\
                   0 & 0 & \lambda^x_{N}\\
                  \end{array}}\right] \> ;
\mathbf{\Lambda_2}=\left[{\begin{array}{ccc}
                   \lambda^y_{1} & 0 & 0\\
                   0 & \ddots & 0\\
                   0 & 0 & \lambda^y_{N}\\
                  \end{array}}\right] \>
\end{align}
As the discrete velocities are constant over a convection time step, using the definitions in (\ref{euler_2d_semilinear_desc})-(\ref{euler_2d_semilinear_desc_2}), we can alternatively write DVBE (\ref{eq_discrete_kinetic_form_euler_2d_index}) as:
\begin{equation}\label{discrete_kinetic_form_euler_2d_desc}
\textbf{f}_l = \mathbf{f^\emph{eq}_\emph{l}}, \ \frac{\partial{{f}}_{ql}}{\partial{t}}+\frac{\partial({\lambda^x_\emph{q}{f}_{ql}})}{\partial{x}}+\frac{\partial({\lambda^y_\emph{q}{f}_{ql}})}{\partial{y}} = 0, \ q=1,2,...,N; \ l=1,2,3,4
\end{equation}
We now need to make appropriate choices for $\mathbf{f^\emph{eq}_\emph{l}}$, $\mathbf{\Lambda_1}$ and $\mathbf{\Lambda_2}$, so that the consistency conditions (\ref{recover_euler_2d_system_thru_semilinear}) are satisfied for each of the Euler equations ($l=1,2,3,4$), which gives:
\begin{align}
\label{recover_cons_euler_2d}
U_l =& \ \sum_{q=1}^N{f^\emph{eq}_\emph{ql}} \\
\label{recover_G1}
G_{1l} =& \ \sum_{q=1}^N\lambda^x_\emph{q} f^\emph{eq}_\emph{ql} \\
\label{recover_G2}
G_{2l} =& \ \sum_{q=1}^N\lambda^y_\emph{q} f^\emph{eq}_\emph{ql}
\end{align}
We set $N=5$ in expressions (\ref{euler_2d_semilinear_desc})-(\ref{euler_2d_semilinear_desc_2}) and extend to the 2-D case the procedure followed for obtaining the discrete velocities in 1-D. 

\subsection{Derivation of discrete velocities in 2-D}
The 2-D Maxwellian distribution function \cite{Deshpande2} is given by:
\begin{equation} \label{2d_maxwellian}
f^{eq} = \frac{\rho}{I_{0}} \left(\fr{\beta}{\pi}\right) e^{- \beta \left(v_1 - u_1\right)^{2} } e^{- \beta \left(v_2 - u_2\right)^{2} } e^{-\frac{I}{I_{0}} }    
\end{equation} 
where 
\begin{align}
I_{0} &= \frac{(2-\gamma)RT}{(\gamma-1)} \\
\beta &= \frac{1}{2RT}
\end{align}
Comparing with the 1-D Maxwellian (\ref{1d_maxwellian}), we can see that the 2-D Maxwellian (\ref{2d_maxwellian}) involves a product of distributions, of molecular velocities, in $x$ and $y$ directions. Using this concept, a 2-D distribution function $\bar{F}$ is constructed with a set of five Dirac delta functions (beams)  where:
\begin{itemize}
\item The central beam has weight $\alpha$ and beam velocity $(\tilde{u}_1,\tilde{u}_2)$
\item Four side beams symmetrically located in the space of molecular velocity $(v_1,v_2)$ have weights $\beta$ and beam velocities $({\tilde{u}_1}\pm\Delta{u}, {\tilde{u}_2}), ({\tilde{u}_1}, {\tilde{u}_2}\pm\Delta{u})$
\end{itemize}
Then $\bar{F}$ can be expressed as
\begin{align}\label{2d_beam_dist_fn}
\bar{F} =& \ \beta\delta(v_1-\tilde{u}_1-\Delta u)\delta(v_2-\tilde{u}_2)+\beta\delta(v_1-\tilde{u}_1+\Delta u)\delta(v_2-\tilde{u}_2) \nonumber \\
 & \ +\alpha\delta(v_1-\tilde{u}_1)\delta(v_2-\tilde{u}_2) \nonumber \\
 & \ +\beta\delta(v_1-\tilde{u}_1)\delta(v_2-\tilde{u}_2-\Delta u)+\beta\delta(v_1-\tilde{u}_1)\delta(v_2-\tilde{u}_2+\Delta u)
\end{align}

In the formulation of the distribution function $\bar{F}$ in (\ref{2d_beam_dist_fn}), there are five unknowns: $\alpha$, $\beta$, $\tilde{u}_1$, $\tilde{u}_2$ and $\Delta u$. To evaluate the unknowns, five equations are required for which the following moment relations are used.

\emph{First moment:}
\begin{equation}\label{2d_beam_zero_moment}
	\rho = \int_{-\infty}^{\infty} \int_{-\infty}^{\infty} \bar{F} dv_1 dv_2 
\end{equation}
Substituting (\ref{2d_beam_dist_fn}) in (\ref{2d_beam_zero_moment}) and performing the integration, we get
\begin{equation}\label{2d_beam_zero_moment_intg}
\rho = \alpha + 4 \beta
\end{equation}
\emph{Second set of moments:}
\begin{align}\label{2d_beam_first_moment_1}
\rho u_1 &= \int_{-\infty}^{\infty} \int_{-\infty}^{\infty} v_1  \bar{F} dv_1 dv_2 \\
\label{2d_beam_first_moment_2}
\rho u_2 &= \int_{-\infty}^{\infty} \int_{-\infty}^{\infty} v_2  \bar{F} dv_1 dv_2 
\end{align}
Substituting (\ref{2d_beam_dist_fn}) in (\ref{2d_beam_first_moment_1}) and (\ref{2d_beam_first_moment_2}) and performing the integrations, we get
\begin{align}\label{2d_beam_first_moment_intg_1}
\tilde{u}_1 = u_1 \\
\label{2d_beam_first_moment_intg_2}
\tilde{u}_2 = u_2
\end{align}
\emph{Third set of moments:}
\begin{align}\label{2d_beam_second_moment_1}
p + \rho u_1^{2} = \int_{-\infty}^{\infty} \int_{-\infty}^{\infty} v_1^{2} \bar{F} dv_1 dv_2 \\
\label{2d_beam_second_moment_2}
p + \rho u_2^{2} = \int_{-\infty}^{\infty} \int_{-\infty}^{\infty} v_2^{2} \bar{F} dv_1 dv_2 
\end{align}
Using (\ref{2d_beam_dist_fn}) in either (\ref{2d_beam_second_moment_1}) or (\ref{2d_beam_second_moment_2}) gives on integration,
\begin{equation}\label{2d_beam_second_moment_intg}
2\beta(\Delta u)^2 = p
\end{equation} 
As of now, we have four equations (\ref{2d_beam_zero_moment_intg}), (\ref{2d_beam_first_moment_intg_1}), (\ref{2d_beam_first_moment_intg_2}) and (\ref{2d_beam_second_moment_intg}) but five unknowns. The next set of moments related to Euler equations (from the energy equation) do not give any additional relations between the unknowns. We are therefore free to exercise one choice.  

\begin{figure}[H]
\includegraphics[width=8cm]{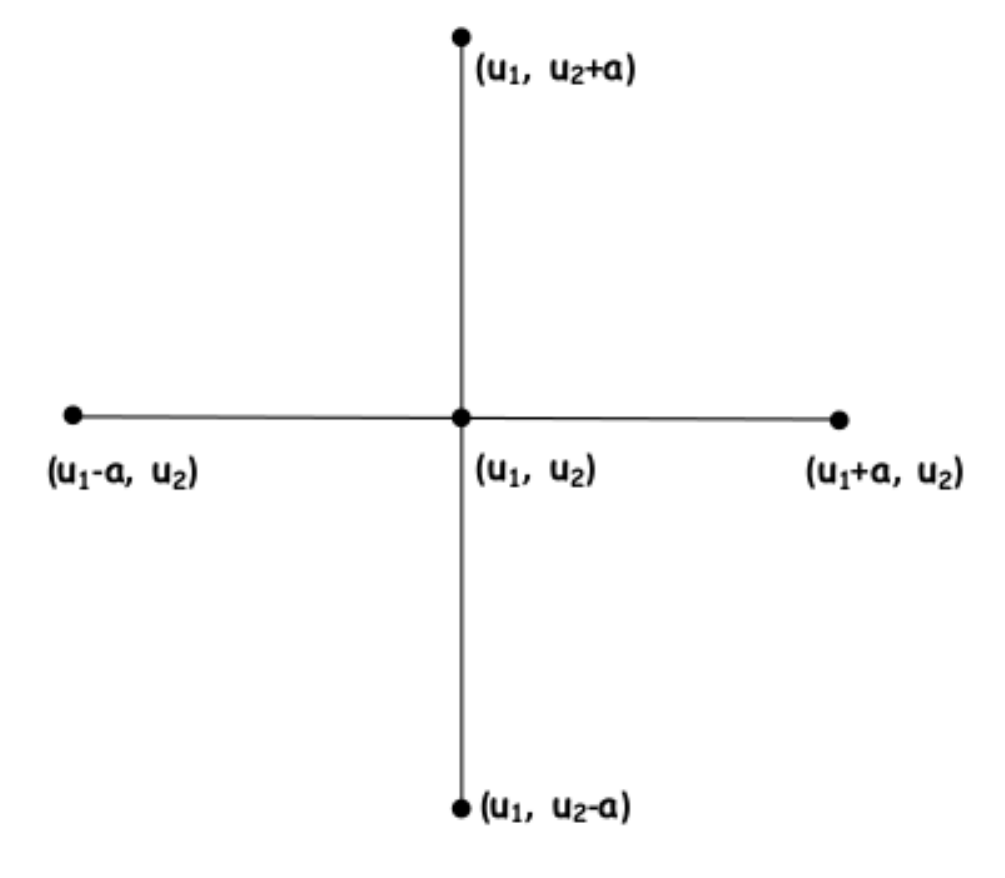}
\centering
\caption{Discrete velocities in 2-D velocity space}
\label{fig:2d_disc_velocities}
\end{figure}

Using the definition of the sound speed $a$, we have 
\begin{equation} \label{2d_sound_sp_defn}
a^{2} = \frac{\gamma p}{\rho} \ \textrm{or} \ p = \frac{\rho a^{2}}{\gamma}    
\end{equation}
Comparing (\ref{2d_sound_sp_defn}) and (\ref{2d_beam_second_moment_intg}), we have
\begin{equation}
\frac{\rho a^{2}}{\gamma} = 2 \beta \left( \Delta u \right)^{2} 
\end{equation}
Using (\ref{2d_beam_zero_moment_intg}), the above equation can be rewritten as
\begin{equation}
\left(\alpha + 4 \beta \right) \frac{a^{2}}{\gamma} = 2 \beta \left( \Delta u \right)^{2}  
\end{equation}
or 
\begin{equation}
\left(\Delta u\right)^{2} = a^{2} \frac{\alpha + 4\beta}{2 \beta \gamma}  
\end{equation}
Therefore 
\begin{equation}
\Delta u = \pm a \sqrt{\frac{\alpha + 4\beta}{2 \beta \gamma}} 
\end{equation}
Let us now exercise our choice and choose 
\begin{equation} \label{2d_fourth_moment_choice}
\alpha + 4\beta = 2 \beta \gamma 
\end{equation}
so that 
\begin{equation}
\Delta u = \pm a = \pm \sqrt{\fr{\gamma p}{\rho}} = \pm \sqrt{\gamma RT}
\end{equation}
Therefore, the five discrete velocities turn out to be:

$(u_1,u_2),(u_1\pm a,u_2),(u_1,u_2\pm a)$ where $a = \sqrt{\gamma R T}$. \\ 
We thus obtain the discrete velocities which mimic the eigenvalues of the two flux Jacobians for 2-D Euler equations. This choice of discrete velocities in 2-D velocity space is depicted in figure \ref{fig:2d_disc_velocities}.

We therefore have
\begin{equation}\label{euler_2d_semilinear_phy_RS}
 \mathbf{\Lambda_1}=\left[{\begin{array}{ccccc}
                   u_1-a & 0 & 0 & 0 & 0\\
                   0 & u_1 & 0 & 0 & 0\\
                   0 & 0 & u_1 & 0 & 0\\
                   0 & 0 & 0 & u_1 & 0\\
                   0 & 0 & 0 & 0 & u_1+a\\
                  \end{array}}\right] \>; \mathbf{\Lambda_2}=\left[{\begin{array}{ccccc}
                   u_2 & 0 & 0 & 0 & 0\\
                   0 & u_2-a & 0 & 0 & 0\\
                   0 & 0 & u_2 & 0 & 0\\
                   0 & 0 & 0 & u_2+a & 0\\
                   0 & 0 & 0 & 0 & u_2\\
                  \end{array}}\right] \>
\end{equation}
With the above choices, we can obtain the expressions for $\mathbf{f^\emph{eq}_\emph{l}}=[f^{eq}_1 f^{eq}_2\ f^{eq}_3\ f^{eq}_4\ f^{eq}_5]^T_l$ from consistency conditions (\ref{recover_cons_euler_2d}), (\ref{recover_G1}) and (\ref{recover_G2}). Here, the approach is similar to the one followed for 1-D. We thereby obtain:
\begin{equation}\label{eval_feq_euler_2d_eqn_1}
\mathbf{f}^\emph{eq}_1=\left[{\begin{array}{c}
                   \frac{\rho}{2\gamma}\\
                    \frac{\rho}{2\gamma}\\
                  \rho \big(1-\frac{2}{\gamma}\big)\\
                   \frac{\rho}{2\gamma}\\
                   \frac{\rho}{2\gamma}\\
                  \end{array}}\right] \>; \mathbf{f}^\emph{eq}_2=\left[{\begin{array}{c}
                   \frac{\rho(u_1-a)}{2\gamma}\\
                    \frac{\rho u_1}{2\gamma}\\
                  \rho u_1 \big(1-\frac{2}{\gamma}\big)\\
                   \frac{\rho u_1}{2\gamma}\\
                   \frac{\rho (u_1+a)}{2\gamma}\\
                  \end{array}}\right] \>
\end{equation}
\begin{equation}\label{eval_feq_euler_2d_eqn_2}
\mathbf{f}^\emph{eq}_3=\left[{\begin{array}{c}
                   \frac{\rho u_2}{2\gamma}\\
                    \frac{\rho (u_2-a)}{2\gamma}\\
                  \rho u_2 \big(1-\frac{2}{\gamma}\big)\\
                   \frac{\rho (u_2+a)}{2\gamma}\\
                   \frac{\rho u_2}{2\gamma}\\
                  \end{array}}\right] \>; \mathbf{f}^\emph{eq}_4=\left[{\begin{array}{c}
                   \frac{\rho [{(u_1-a)}^2+u_2^2]}{4\gamma}+\frac{\rho I_0}{2\gamma}\\
                   \frac{\rho [{u_1^2+(u_2-a)}^2]}{4\gamma}+\frac{\rho I_0}{2\gamma}\\
                                     \frac{\rho(u_1^2+u_2^2)}{2} \big(1-\frac{2}{\gamma}\big)+\rho I_0 \big(1-\frac{2}{\gamma}\big)\\
                   \frac{\rho [{u_1^2+(u_2+a)}^2]}{4\gamma}+\frac{\rho I_0}{2\gamma}\\
                   \frac{\rho [{(u_1+a)}^2+u_2^2]}{4\gamma}+\frac{\rho I_0}{2\gamma}\\

                                     \end{array}}\right] \>
\end{equation}

In the above expressions, $I_0$ for 2-D as obtained from kinetic theory~\cite{Deshpande2} is:
\begin{equation}
I_0=\frac{(4-2\gamma)RT}{2(\gamma-1)} \label{I0_2D}
\end{equation}

Having determined the expressions for $\mathbf{f^\emph{eq}_\emph{l}}$, $\mathbf{\Lambda_1}$ and $\mathbf{\Lambda_2}$, the discrete velocity Boltzmann equation (\ref{eq_discrete_kinetic_form_euler_2d_index}) can be solved numerically for the distribution functions $\mathbf{f_\emph{l}}=[f_1 \ f_2\ f_3\ f_4\ f_5]^T_l$. Once the values of $\mathbf{f_\emph{l}}$ are obtained, the updated values of conserved variables of Euler equations can be recovered from moment relations.  

In the next section, we shall formulate an upwind scheme for numerical solution of the discrete velocity Boltzmann equation (\ref{discrete_kinetic_form_euler_1d_desc}) in 1-D and (\ref{discrete_kinetic_form_euler_2d_desc}) in 2-D. But before we do this, we need to ascertain the stability of the discrete kinetic approximation (\ref{euler_2d_semilinear}) in 2-D.

To examine stability, we again apply Bouchut's condition (\ref{bouchut_stability}). From (\ref{eval_feq_euler_2d_eqn_1})-(\ref{eval_feq_euler_2d_eqn_2}), we express $\mathbf{f^\emph{eq}}$ in terms of components of conserved variable $\mathbf{U}$ as:
\begin{equation}\label{eval_feq_euler_2d_eqn_3}
\mathbf{f^\emph{eq}}=\left[{\begin{array}{c}
					\left[{\begin{array}{c}
                   \frac{U_1}{2\gamma}\\
					\frac{U_2}{2\gamma}-\frac{\sqrt{\gamma(\gamma-1)}}{2\gamma}\bigg(U_1U_4-\frac{U_2^2+U_3^2}{2}\bigg)^\frac{1}{2}\\                   
                   \frac{U_3}{2\gamma}\\
                   \frac{U_2^2+U_3^2}{4\gamma U_1}-\frac{\sqrt{\gamma(\gamma-1)}}{2\gamma}\frac{U_2}{U_1}\bigg(U_1U_4-\frac{U_2^2+U_3^2}{2}\bigg)^\frac{1}{2}+ \frac{(\gamma^2-3\gamma+4)}{4\gamma}\bigg(U_4-\frac{U_2^2+U_3^2}{2U_1}\bigg)\\
                  \end{array}}\right] \> \\ \\
					\left[{\begin{array}{c}
                   \frac{U_1}{2\gamma}\\
                   \frac{U_2}{2\gamma}\\
					\frac{U_3}{2\gamma}-\frac{\sqrt{\gamma(\gamma-1)}}{2\gamma}\bigg(U_1U_4-\frac{U_2^2+U_3^2}{2}\bigg)^\frac{1}{2}\\                   
                   \frac{U_2^2+U_3^2}{4\gamma U_1}-\frac{\sqrt{\gamma(\gamma-1)}}{2\gamma}\frac{U_3}{U_1}\bigg(U_1U_4-\frac{U_2^2+U_3^2}{2}\bigg)^\frac{1}{2}+ \frac{(\gamma^2-3\gamma+4)}{4\gamma}\bigg(U_4-\frac{U_2^2+U_3^2}{2U_1}\bigg)\\
                  \end{array}}\right] \> \\ \\
					\left[{\begin{array}{c}
				   U_1 \bigg(1-\frac{2}{\gamma}\bigg)\\
                   U_2\bigg(1-\frac{2}{\gamma}\bigg)\\
                   U_2\bigg(1-\frac{2}{\gamma}\bigg)\\
                   \frac{(\gamma-2)}{2\gamma}\frac{U_2^2+U_3^2}{U_1}+ \frac{(4-2\gamma)(\gamma-2)}{2\gamma}\bigg(U_4-\frac{U_2^2+U_3^2}{2U_1}\bigg)\\
                  \end{array}}\right] \> \\ \\
					\left[{\begin{array}{c}
                   \frac{U_1}{2\gamma}\\
                   \frac{U_2}{2\gamma}\\
					\frac{U_3}{2\gamma}+\frac{\sqrt{\gamma(\gamma-1)}}{2\gamma}\bigg(U_1U_4-\frac{U_2^2+U_3^2}{2}\bigg)^\frac{1}{2}\\                   
                   \frac{U_2^2+U_3^2}{4\gamma U_1}+\frac{\sqrt{\gamma(\gamma-1)}}{2\gamma}\frac{U_3}{U_1}\bigg(U_1U_4-\frac{U_2^2+U_3^2}{2}\bigg)^\frac{1}{2}+ \frac{(\gamma^2-3\gamma+4)}{4\gamma}\bigg(U_4-\frac{U_2^2+U_3^2}{2U_1}\bigg)\\
                  \end{array}}\right] \> \\ \\
					\left[{\begin{array}{c}
                   \frac{U_1}{2\gamma}\\
					\frac{U_2}{2\gamma}+\frac{\sqrt{\gamma(\gamma-1)}}{2\gamma}\bigg(U_1U_4-\frac{U_2^2+U_3^2}{2}\bigg)^\frac{1}{2}\\                   
                   \frac{U_3}{2\gamma}\\
                   \frac{U_2^2+U_3^2}{4\gamma U_1}+\frac{\sqrt{\gamma(\gamma-1)}}{2\gamma}\frac{U_2}{U_1}\bigg(U_1U_4-\frac{U_2^2+U_3^2}{2}\bigg)^\frac{1}{2}+ \frac{(\gamma^2-3\gamma+4)}{4\gamma}\bigg(U_4-\frac{U_2^2+U_3^2}{2U_1}\bigg)\\
                  \end{array}}\right] \>
                  \end{array}}\right] \>                  
\end{equation}
\clearpage
We see that $\mathbf{f^\emph{eq}}$ is a set of five column vectors. Each column vector has its Jacobian with the vector of conserved variables $\textbf{U}$. Each of the five Jacobians has its set of eigenvalues. Bouchut's condition (\ref{bouchut_stability}) stipulates that all of these eigenvalues be non-negative. Let us examine this further.

The eigenvalues corresponding to the third column vector in (\ref{eval_feq_euler_2d_eqn_3}) are:
\begin{equation}\label{eigenspectra_1_2_4_5_2d}
\Omega_3 = \bigg\lbrace\frac{\gamma-2}{\gamma}, \ \frac{\gamma-2}{\gamma},\ \frac{\gamma-2}{\gamma}, \ -\frac{(\gamma-2)^2}{2\gamma} \bigg\rbrace
\end{equation}
Here, Bouchut's condition (\ref{bouchut_stability}) gives 
\begin{equation}\label{bouchut_condn_one_2d}
\gamma=2
\end{equation}
The eigenvalues corresponding to the first, second, fourth and fifth column vectors in (\ref{eval_feq_euler_2d_eqn_3}) are the same and are given by:
\begin{equation}\label{eigenspectra_3_2d}
\Omega_1 = \Omega_2 = \Omega_4 = \Omega_5 = \bigg\lbrace\frac{1}{2\gamma}, \ \frac{1}{2\gamma}, \ \frac{6+\gamma(\gamma-3)}{8\gamma}\pm\frac{\sqrt{\gamma^4-6\gamma^3+21\gamma^2-20\gamma+4}}{8\gamma} \ \bigg\rbrace
\end{equation}
Bouchut's condition (\ref{bouchut_stability}) requires that the above eigenspectrum is non-negative. This gives the stability condition:
\begin{equation}\label{bouchut_condn_two_2d}
\gamma \leq 2
\end{equation}
Condition (\ref{bouchut_condn_one_2d}) is contained in (\ref{bouchut_condn_two_2d}). So we refer to the latter for the stability condition and choose $\gamma=1.4$ for our 2-D numerical computations.

\section{Upwind discrete velocity Boltzmann scheme for one-dimensional flows} 
The upwind schemes devised in this work are based on solution of the discrete velocity Boltzmann equation. We first formulate an upwind scheme for the 1-D equation (\ref{discrete_kinetic_form_euler_1d_desc}) leading to solution of 1-D Euler equations. In the next section, we discuss extension of the upwind scheme for two-dimensional flows.

\begin{figure}[H]
\includegraphics[width=5cm]{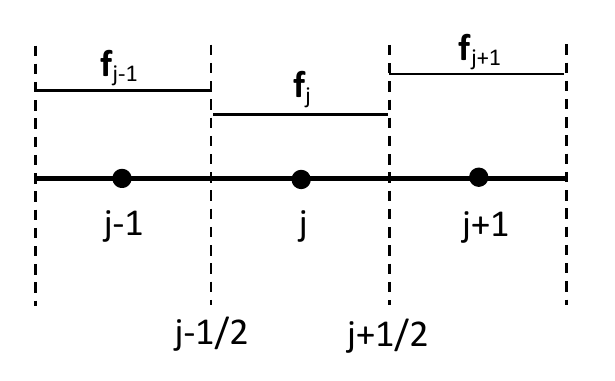}
\centering
\caption{Three-point finite volume stencil}
\label{fig:three_lt_stencil}
\end{figure}

Consider a 3-point stencil as shown in figure \ref{fig:three_lt_stencil} depicting piecewise constant approximation of distribution functions in each finite volume.

At the beginning of each time step $\Delta t$, we assume that the collision step in the discrete velocity Boltzmann equation in 1-D  (\ref{discrete_kinetic_form_euler_1d_desc}) results in instantaneous relaxation of the distribution functions $\textbf{f}^n_{l,j}$ to the equilibrium distribution functions $\mathbf{f}^\emph{eq,n}_{l,j}$ defined in (\ref{eval_feq_euler_1d_eqn_25}):
\begin{equation}\label{discrete_kinetic_form_euler_1d_coll}
\mathbf{f}^n_{l,j} = \mathbf{f}^\emph{eq}_\emph{l}{(\mathbf{U}(x_j,t^n))}, \ l=1,2,3
\end{equation}
Then, at the end of $\Delta t$ determined by CFL condition, the distribution functions are evolved to $\textbf{f}^{n+1}_{l,j}$ by integrating the convection steps over the finite-volume centered at node $j$ to obtain
\begin{equation}\label{fvm_f_1d}
{\bar{\mathbf{f}}^{n+1}}_{l,j} = {\bar{\mathbf{f}}^{n}}_{l,j} - \frac{\Delta t}{\Delta x}[{\mathbf{h}}^{n}_{l,j+\frac{1}{2}} - {\mathbf{h}}^{n}_{l,j-\frac{1}{2}}]
\end{equation}
where $\Delta x = x_{j+\frac{1}{2}}-x_{j-\frac{1}{2}}$ and the quantities with a bar denote cell-integral averages defined by 
\begin{equation}
{\bar{[\bullet]}} = \frac{1}{\Delta x}\int_{x_{j-\frac{1}{2}}}^{x_{j+\frac{1}{2}}}[\bullet] dx
\end{equation}
Using upwinding, we now formulate the interface fluxes ${\mathbf{h}}^{n}_{l,j\pm\frac{1}{2}}$ in equation (\ref{fvm_f_1d}) to solve for the updated distribution functions ${\bar{\mathbf{f}}^{n+1}}_{l,j}$.

Following the collision step (\ref{discrete_kinetic_form_euler_1d_coll}), the interface flux for the finite-volume discretization of convection step in equation (\ref{discrete_kinetic_form_euler_1d_desc}) can be written as:
\begin{equation}\label{h_upwind_flux}
 {h}^{n}_{ql,j+\frac{1}{2}} \equiv (\lambda_{q} f^{eq}_{ql})^n_{j+\frac{1}{2}}, \ l=1,2,3
\end{equation}
As the discrete velocities $\lambda_q$ in expression (\ref{h_upwind_flux}) are constant over a time step, we can write the above interface flux in split flux form as:
\begin{equation}\label{h_upwind_thru_flux_split}
 {h}^{n}_{ql,j+\frac{1}{2}} = (\lambda^{+}_{q} f^{eq}_{ql})^n_{j} + (\lambda^{-}_{q} f^{eq}_{ql})^n_{j+1}
\end{equation}
where the split wave speeds are defined as:
\begin{equation}\label{wave_sp_exp}
\lambda^{\pm}_q = \frac{\lambda_q\pm|\lambda_q|}{2}
\end{equation}

The interface flux in equation (\ref{h_upwind_thru_flux_split}) can be rewritten in vector notation as
\begin{equation}\label{h_vector_upwind_thru_flux_split}
{\mathbf{h}}^{n}_{l,j+\frac{1}{2}} = \mathbf{\Lambda}_j^{+}\mathbf{f}^{eq,n}_{l,j} + \mathbf{\Lambda}_{j+1}^{-}\mathbf{f}^{eq,n}_{l,j+1}, \ l=1,2,3
\end{equation}
where
\begin{equation}
 \mathbf{\Lambda}^{\pm}_j \equiv diag(\lambda^{\pm}_{q,j}) 
\end{equation}
Once we solve equation (\ref{fvm_f_1d}) for the distribution functions ${\bar{\mathbf{f}}^{n+1}}_{l,j}$ at the end of $\Delta t$, the updated values of the conserved variables of the Euler equations can be recovered using moment relations (\ref{recover_cons_euler_1d}).

We note that for the specific case of our new discrete kinetic system with physically relevant discrete velocities,
\begin{equation}
\mathbf{\Lambda}^{\pm}_j \equiv diag\left((u_j-a_j)^{\pm},(u_j)^{\pm},(u_j+a_j)^{\pm}\right)
\end{equation}

\subsection{Positivity analysis of the upwind scheme in 1-D}
The finite-volume update formula for 1-D Euler equations (\ref{euler_1d}) is given by:
\begin{equation}\label{fvm_f_1d_positivity}
{\mathbf{U}^{n+1}_{j}} = {\mathbf{U}^{n}_{j}} - \frac{\Delta t}{\Delta x}[{\mathbf{G}}^{n}_{j+\frac{1}{2}} - {\mathbf{G}}^{n}_{j-\frac{1}{2}}]
\end{equation}
The interface fluxes in (\ref{fvm_f_1d_positivity}) are prescribed based on upwinding as: 
\begin{align}\label{split_flux_positivity}
\begin{split}
\mathbf{G}_{j-\frac{1}{2}} = \mathbf{G}^{+}_{j-1} + \mathbf{G}^{-}_{j} \\
\mathbf{G}_{j+\frac{1}{2}} = \mathbf{G}^{+}_{j} + \mathbf{G}^{-}_{j+1}
\end{split}
\end{align}
Using moment relations (\ref{recover_flux_euler_1d}) and three discrete velocities ($N=3$) relevant to our new discrete kinetic system, the above interfaces fluxes can be expressed as:
\begin{align}\label{split_flux_eq_dist_positivity}
\begin{split}
G_{l,j-\frac{1}{2}} = \sum_{q=1}^{3}{\lambda_{q,j-1}^{+}}f_{ql,j-1} + \sum_{q=1}^{3}{\lambda_{q,j}^{-}}f_{ql,j} \\
G_{l,j+\frac{1}{2}} = \sum_{q=1}^{3}{\lambda_{q,j}^{+}}f_{ql,j} + \sum_{q=1}^{3}{\lambda_{q,j+1}^{-}}f_{ql,j+1}
\end{split}
\end{align}
Let us use $\sigma$ to denote $\frac{\Delta t}{\Delta x}$. Then we have the CFL condition:
\begin{equation}\label{cfd_condn_positivity}
\sigma \Max_{j \in Z}\{|\lambda_{1,j}|,|\lambda_{2,j}|,|\lambda_{3,j}|\} \leq 1
\end{equation}
Now, substituting the expressions from (\ref{split_flux_eq_dist_positivity}) in (\ref{fvm_f_1d_positivity}) for the continuity equation ($l=1$), we obtain:
\begin{equation}\label{rho_positivity_eq_1}
{\rho^{n+1}_{j}} = {{\rho}^{n}_{j}} - \sigma\sum_{q=1}^{3}\left[-{\lambda_{q,j-1}^{+}}f_{q1,j-1}^{n} + \left({\lambda_{q,j}^{+}}f_{q1,j}^{n}-{\lambda_{q,j}^{-}}f_{q1,j}^{n}\right)+{\lambda_{q,j+1}^{-}}f_{q1,j+1}^{n}\right]
\end{equation}
Also, from (\ref{recover_cons_euler_1d}) we have
\begin{equation}\label{rho_split_positivity}
{\rho}^{n}_{j} = \sum_{q=1}^{3}f_{q1,j}^n, \ {(\rho u)}^{n}_{j} = \sum_{q=1}^{3}f_{q2,j}^n, \ {(\rho E)}^{n}_{j} = \sum_{q=1}^{3}f_{q3,j} ^n
\end{equation}
Using the expression for density from (\ref{rho_split_positivity}) in (\ref{rho_positivity_eq_1}), we get
\begin{equation}\label{rho_positivity_eq_2}
{\rho^{n+1}_{j}} = \sum_{q=1}^{3}\left[\sigma{\lambda_{q,j-1}^{+}}f_{q1,j-1}^{n} + \left(1-\sigma|\lambda_{q,j}|\right)f_{q1,j}^{n}+\sigma{(-\lambda^{-}_{q,j+1})}f_{q1,j+1}^{n}\right]
\end{equation}
From the expressions for $\textbf{f}^{eq}_l$ in (\ref{eval_feq_euler_1d_eqn_25}), we can see that all the elements $f_{ql}^{n}$ corresponding to $l=1$ and $q=1,2,3$ are positive with the stability condition (\ref{bouchut_condn_second_1d}), CFL condition (\ref{cfd_condn_positivity}) and an initial positive density $\rho^n_j$. Consequently all the terms on the right hand side of (\ref{rho_positivity_eq_2}) are positive. Hence, ${\rho^{n+1}_{j}} \geq 0 \ \forall j$ which completes positivity proof for density.

We next examine positivity of internal energy. By definition, we have
\begin{align}\label{int_energy_positivity_1}
\rho e &= \rho E - \frac{\rho u^2}{2} \\\label{int_energy_positivity_1_2}
\implies 2 \rho^{n+1}_j (\rho e)^{n+1}_j &= 2\rho^{n+1}_j(\rho E)^{n+1}_j - \left[(\rho u)_j^{n+1}\right]^2
\end{align}
Now, substituting the expressions from (\ref{split_flux_eq_dist_positivity}) in (\ref{fvm_f_1d_positivity}) for the momentum equation ($l=2$) and using the expression for momentum from (\ref{rho_split_positivity}), we obtain:
\begin{equation}\label{int_energy_positivity_2}
{{(\rho u)}^{n+1}_{j}} = \sum_{q=1}^{3}\left[\sigma{\lambda_{q,j-1}^{+}}f_{q2,j-1}^{n} + \left(1-\sigma|\lambda_{q,j}|\right)f_{q2,j}^{n}+\sigma{(-\lambda^{-}_{q,j+1})}f_{q2,j+1}^{n}\right]
\end{equation}

Similarly, substituting the expressions from (\ref{split_flux_eq_dist_positivity}) in (\ref{fvm_f_1d_positivity}) for the energy equation ($l=3$) and using the expression for total energy from (\ref{rho_split_positivity}), we get:
\begin{equation}\label{int_energy_positivity_3}
{{(\rho E)}^{n+1}_{j}} = \sum_{q=1}^{3}\left[\sigma{\lambda_{q,j-1}^{+}}f_{q3,j-1}^{n} + \left(1-\sigma|\lambda_{q,j}|\right)f_{q3,j}^{n}+\sigma{(-\lambda^{-}_{q,j+1})}f_{q3,j+1}^{n}\right]
\end{equation}

We substitute the expressions for the equilibrium distribution functions corresponding to $l=2$ and $l=3$ from (\ref{eval_feq_euler_1d_eqn_25}) in (\ref{int_energy_positivity_2}) and (\ref{int_energy_positivity_3}) respectively. The resulting expressions for ${{(\rho u)}^{n+1}_{j}}$ and ${{(\rho E)}^{n+1}_{j}}$ are then plugged into (\ref{int_energy_positivity_1_2}). On performing the required algebra, we get
\begin{align}
2 \rho^{n+1}_j (\rho e)^{n+1}_j &= \frac{(3-\gamma)}{2}{\rho^{n+1}_j}\sum_{q=1}^3 \kappa_q\sum_{k=-1}^1{\alpha_{q,j+k}}{(\rho e)^{n}_{j+k}} \nonumber \\
&\ \ \ + \frac{1}{2}\sum_{m=1}^3\sum_{n=1}^3 \kappa_{m}\kappa_{n}B_{mn}\label{positive_energy}
\end{align}
where:
\begin{align}
& \ \ \ \kappa_1=\frac{1}{2\gamma}, \kappa_2=\frac{\gamma-1}{\gamma}, \kappa_3=\frac{1}{2\gamma} \nonumber \\
& \ \ \ \alpha_{q,j+k} \ \text{are non-negative coefficients under CFL condition, similar to } \nonumber \\
& \ \ \ \ \ \ \text{expressions defined by Tang and Xu \cite{Tang_Xu}} \nonumber \\
& \ \ \ B_{mn} \geq 0  \ \forall m, n \in \{1,2,3\} \ \text{are algebraic expressions similar to those} \nonumber \\
& \ \ \ \ \ \ \text{defined by Tang and Xu \cite{Tang_Xu}}
\end{align}
Then, all the terms on the right hand side of equation (\ref{positive_energy}) are non-negative. Hence the internal energy remains non-negative over time. This completes the positivity proof for the upwind scheme with our 1-D discrete kinetic system.

\subsection{Entropy fix for the upwind scheme}\label{sec:sonic_pt_entropy_fix}
Let us examine the case in equation (\ref{h_upwind_thru_flux_split}) where the wave speed under consideration is $\lambda = (u-a)$ and $\lambda_j < 0$ while $\lambda_{j+1} > 0$. This is the case of an expansive sonic point and is depicted in the figure below.
\begin{figure}[H]
\includegraphics[width=5cm]{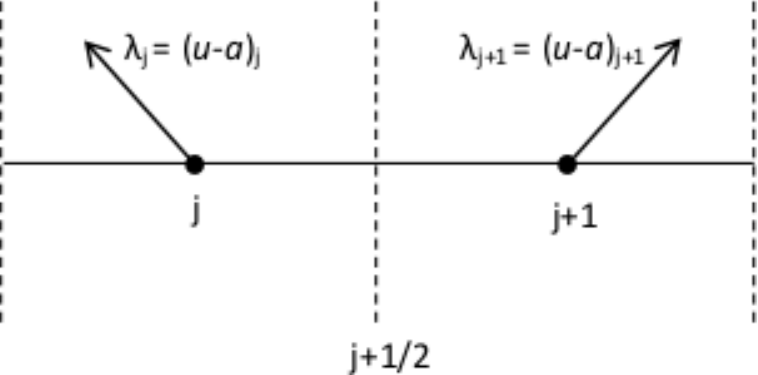}
\centering
\caption{Expansive sonic point in $x$-$t$ plane}
\label{fig:expansive_sonic_pt}
\end{figure}
In this case:
\begin{align}
&\lambda_j^{+} = 0 \\
&\lambda_{j+1}^{-} = 0
\end{align}
So the interface flux calculated becomes zero. This can result in formation of expansion shock while physically this represents an expansive region. A similar situation can result with the wave speed $\lambda = (u+a)$ when $u_j < -a_j$ and $u_{j+1} > -a_{j+1}$. In such situations, we use the entropy fix given by Steger and Warming (see Laney~\cite{Laney}):
\begin{equation}
\lambda_j^{\pm}=\frac{1}{2}\bigg(\lambda_j\pm\sqrt{\lambda_j^2+\delta^2}\bigg)
\end{equation}
$\delta$ is a user-defined positive number taken as 0.1 in this work. The above fix ensures that the split wave speeds are not zero.
\section{Upwind discrete kinetic scheme for two-dimensional flows}
In the preceding section, we devised an upwind scheme to solve the discrete velocity Boltzmann equation (\ref{discrete_kinetic_form_euler_1d_desc}) using a finite-volume framework in 1-D.   We now extend this scheme to 2-D.

In 2-D, the DVBEs (\ref{eq_discrete_kinetic_form_euler_2d_index}) take the form
\begin{equation}\label{eq_new_DVBE_euler_2d}
\mathbf{f}_l=\mathbf{f^\emph{eq}_\emph{l}}, \frac{\partial{\mathbf{f}_l}}{\partial{t}}+ \mathbf{\Lambda_1} \frac{\partial{\mathbf{f}_l}}{\partial{x}}+ \mathbf{\Lambda_2} \frac{\partial{\mathbf{f}_l}}{\partial{y}}=0, \ l = 1,2,3,4
\end{equation}
with $\mathbf{f^\emph{eq}_\emph{l}}$ defined in (\ref{eval_feq_euler_2d_eqn_1}) and (\ref{eval_feq_euler_2d_eqn_2}) and $\mathbf{\Lambda}_1$, $\mathbf{\Lambda}_2$ given by (\ref{euler_2d_semilinear_phy_RS}).

\begin{figure}[H]
\includegraphics[width=8cm]{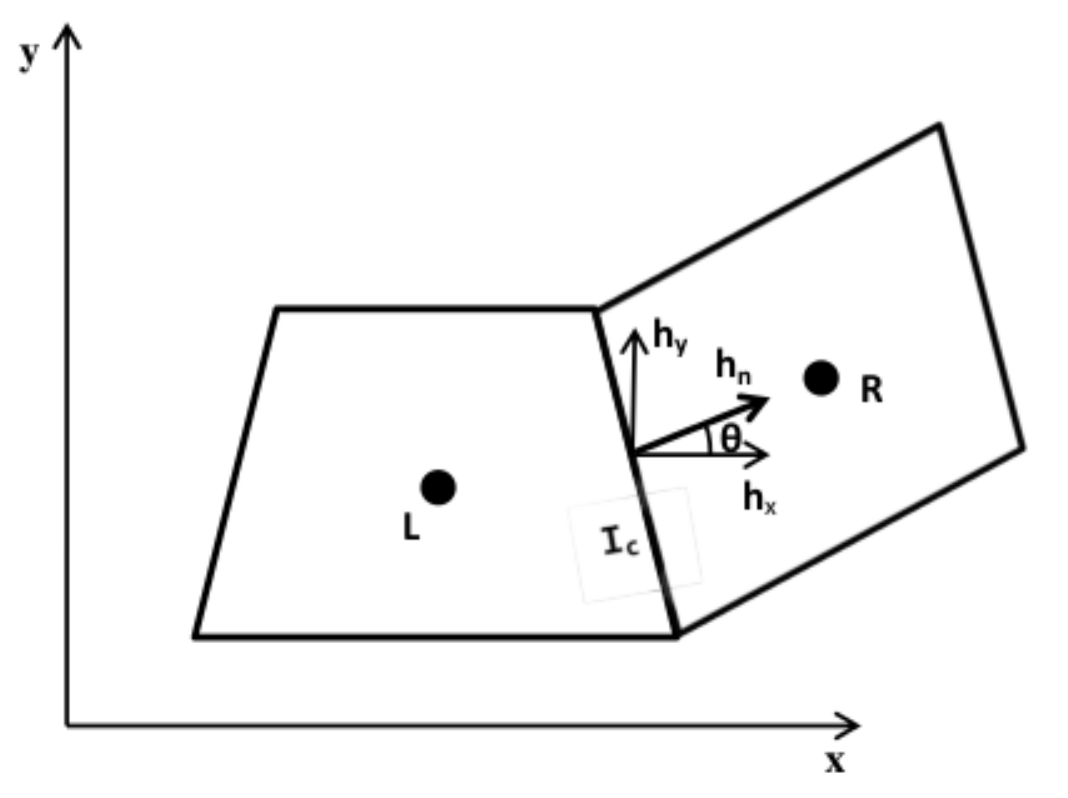} 
\centering
\caption{Finite volume framework in 2D}
\label{fig:rot_flux}
\end{figure}

As the discrete velocities are constant over the time step $\Delta t$, we can express the $x$ and $y$ component fluxes in (\ref{eq_new_DVBE_euler_2d}) as:
\begin{equation}\label{RS_flux_vector}
\mathbf{h}_x = \mathbf{\Lambda}_1\mathbf{f}, \mathbf{h}_y = \mathbf{\Lambda}_2 \mathbf{f}
\end{equation}  
The net interface flux $\mathbf{h}_n$ is normal to the cell face \emph{$I_c$} in a locally 1-D sense as shown in figure \ref{fig:rot_flux}.  
From the above figure, clearly the interface flux can be expressed in terms of the component fluxes as
\begin{equation}\label{resolve_norm_int_flux_fvm_2d}
\mathbf{h}_{nI_c} = (\mathbf{h}_xcos\theta + \mathbf{h}_ysin\theta)_{I_{c}}
\end{equation}
Using the same argument, the fluxes on the left state  \emph{L} and right state \emph{R} in the direction of the normal to the cell interface under consideration are
\begin{align}\label{resolve_norm_left_right_flux_fvm_2d}
\begin{split}
\mathbf{h}_{nL} = (\mathbf{h}_xcos\theta + \mathbf{h}_ysin\theta)_L \\
\mathbf{h}_{nR} = (\mathbf{h}_xcos\theta + \mathbf{h}_ysin\theta)_R
\end{split}
\end{align}
From equations (\ref{resolve_norm_int_flux_fvm_2d}) and (\ref{resolve_norm_left_right_flux_fvm_2d}), we have a locally 1-D definition of the interface flux and the fluxes on the left and right states, all aligned along the normal to the cell face under consideration. The interface flux can now be expressed using upwinding as:
\begin{equation}\label{int_flux_fvm_2d_upwind}
\mathbf{h}_{nI_c} = \mathbf{h}_{nL}^{+} + \mathbf{h}_{nR}^{-}
\end{equation}
Using equations (\ref{resolve_norm_left_right_flux_fvm_2d}) and (\ref{RS_flux_vector}), the above interface flux for the DVBEs corresponding to each of the four conservation laws $(l=1,2,3,4)$ of the Euler system can be expressed as
\begin{eqnarray}\label{h_upwind_thru_flux_split_2d_euler}
(\mathbf{h}_l)_{nI_c} = (\mathbf{\Lambda}_{1L}cos\theta)^{+}(\mathbf{f^\emph{eq}_\emph{l}})_{L} + (\mathbf{\Lambda}_{2L}sin\theta)^{+}(\mathbf{f^\emph{eq}_\emph{l}})_{L} \nonumber \\  +  (\mathbf{\Lambda}_{1R}cos\theta)^{-}(\mathbf{f^\emph{eq}_\emph{l}})_{R} + (\mathbf{\Lambda}_{2R}sin\theta)^{-}(\mathbf{f^\emph{eq}_\emph{l}})_{R}
\end{eqnarray}
Once the interface fluxes are evaluated at all the four cell faces $(I_c=1,2,3,4)$ of the finite volume, the DVBEs are numerically solved for the updated distribution function at the cell center $(i,j)$ after a discrete time step $\Delta t$ by using the 2-D finite-volume update formula:
\begin{equation}\label{fvm_euler_2d}
\mathbf{f}^{n+1}_{i,j} = \mathbf{f}^{eq,n}_{i,j} - \frac{\Delta t}{A_{i,j}}\sum_{I_c=1}^4\mathbf{h}_{nI_c}{\Delta s}_{I_{c}} 
\end{equation}
where: $A_{i,j}$ is the area of the cell centered at $(i,j)$ \\
\hspace*{1cm} ${\Delta s}_{I_{c}}$ is the length of the cell face \emph{$I_c$}

Subsequently the updated conserved variables of the respective conservation laws can be obtained using equation (\ref{recover_cons_euler_2d}).

\subsection{Positivity preservation by the upwind scheme in 2-D}\label{sec:positivity_pres}
Some of the elements of the 2-D equilibrium distribution functions in (\ref{eval_feq_euler_2d_eqn_1}) and (\ref{eval_feq_euler_2d_eqn_2}) can assume negative values when $\gamma < 2$, for e.g., $\rho (1-\frac{2}{\gamma})$. However, this is not a problem in the discrete kinetic framework. But it is important to ascertain the positivity of density, pressure and internal energy.

Unfortunately, a positivity proof in 2-D may not be straightforward. To ensure positivity preservation, we test our upwind scheme for the new discrete kinetic system on specifically designed test cases provided by Parent~\cite{B_Parent}. These test cases involve strong expansions which are zones of low pressure and density. Parent~\cite{B_Parent} states that these test cases provide an excellent test bed to assess the capability of numerical schemes at maintaining positivity-preservation in multidimensional flow fields.  

\section{Second-order accuracy} \label{sec:Second-order accuracy}
Till now, we assumed a piece-wise constant approximation of the conserved or primitive variables. As a result, all the schemes obtained are first-order accurate in space. To obtain second-order accuracy, piece-wise linear approximation of the variables is assumed as follows:
\begin{equation}\label{piecewise_linear_approx} 
U(x,t^n) = [U^n_j + \Big(\frac{\partial{U}}{\partial{x}}\Big)_j^n(x-x_j)]
\end{equation}
The values of the variables at the interfaces are then obtained by setting  $x = x_{j\pm\frac{1}{2}} = x_j \pm \frac{\Delta x}{2}$.\\
The non-oscillatory behaviour of the scheme depends on the appropriate choice of approximate derivatives, and we use the one-parameter family of minmod limiter~\cite{Kurganov} for this purpose given by
\begin{equation}\label{second_order_limiter}
\Big(\frac{\partial{U}}{\partial{x}}\Big)_j := minmod\Big(\zeta\frac{{U}_j - {U}_{j-1}}{\Delta x}, \frac{{U}_{j+1} - {U}_{j-1}}{2\Delta x}, \zeta\frac{{U}_{j+1} - {U}_j}{\Delta x}\Big), \ 1\leq \zeta \leq 2
\end{equation}
The numerical values of the cell-interface fluxes required for the finite-volume update formula are computed using the reconstructed values obtained from equation (\ref{piecewise_linear_approx}).

\section{Results and Discussion}
Results are presented for 1-D and 2-D test cases for inviscid compressible flows. To evaluate the performance of our new discrete kinetic scheme, we use, for comparison, a benchmark discrete kinetic scheme called as {\em Upwind Relaxation Scheme} (URS), which is an upwind scheme applied to \emph{an isotropic relaxation system} introduced by Raghurama Rao which was utilized by Jayaraj \cite{Jayaraj}, Arun \emph{et al.} \cite{Arun_RaghuramaRao} and Raghurama Rao {\em et al.} \cite{SVRRao_Rohan_Sourabh}.    

\subsection{1-D Shock tube problems}
\begin{figure}[htb!]
\centering
\begin{subfigure}{.5\textwidth}
\centering
\includegraphics[trim = 4mm 4mm 10mm 10mm,clip,width=1.05\linewidth]{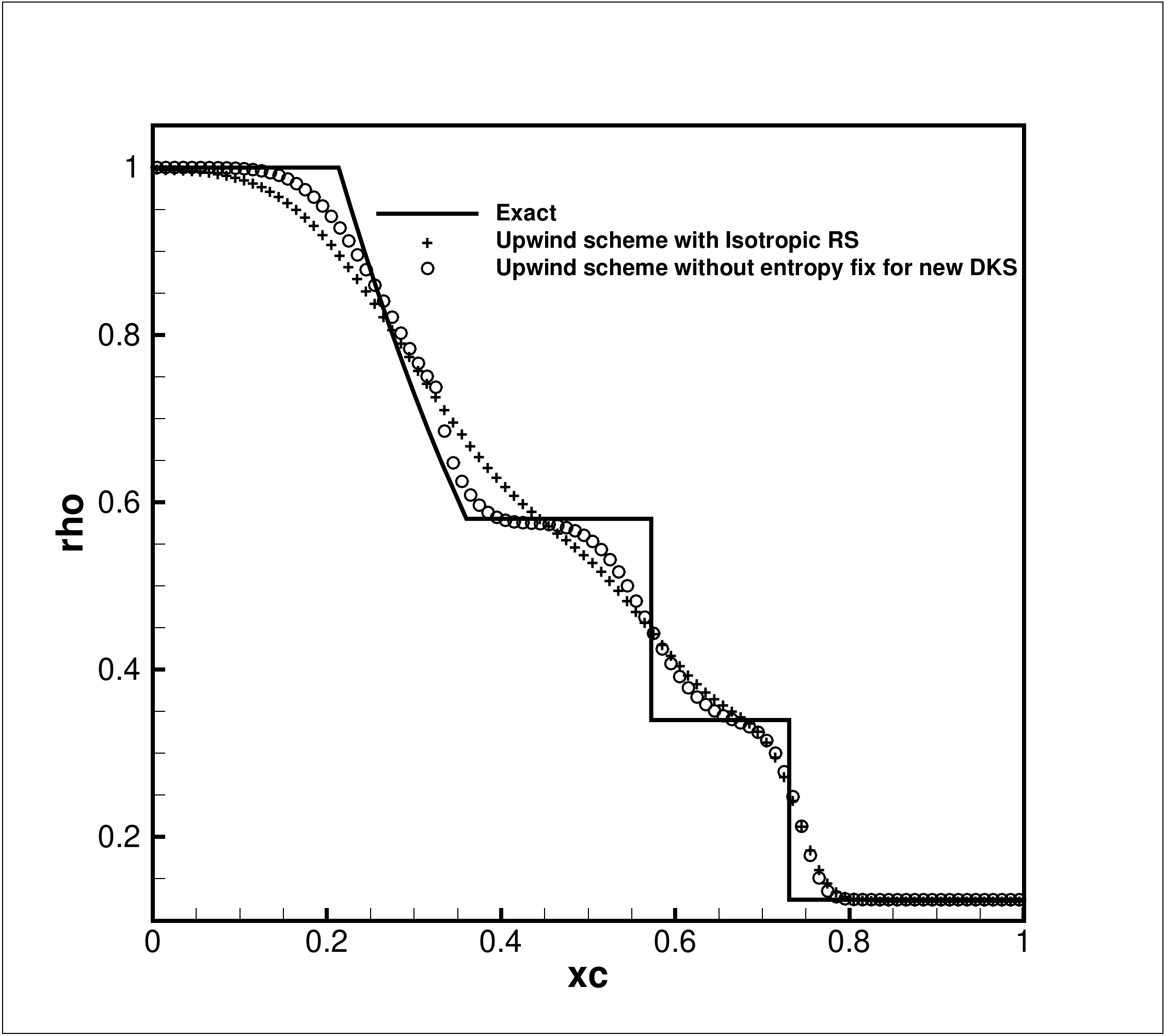}
\caption{Without entropy fix}
  \label{fig:Comparison_Toro_TestCase1}
\end{subfigure}%
\begin{subfigure}{.5\textwidth}
\centering
\includegraphics[trim = 4mm 4mm 10mm 10mm,clip,width=1.05\linewidth]{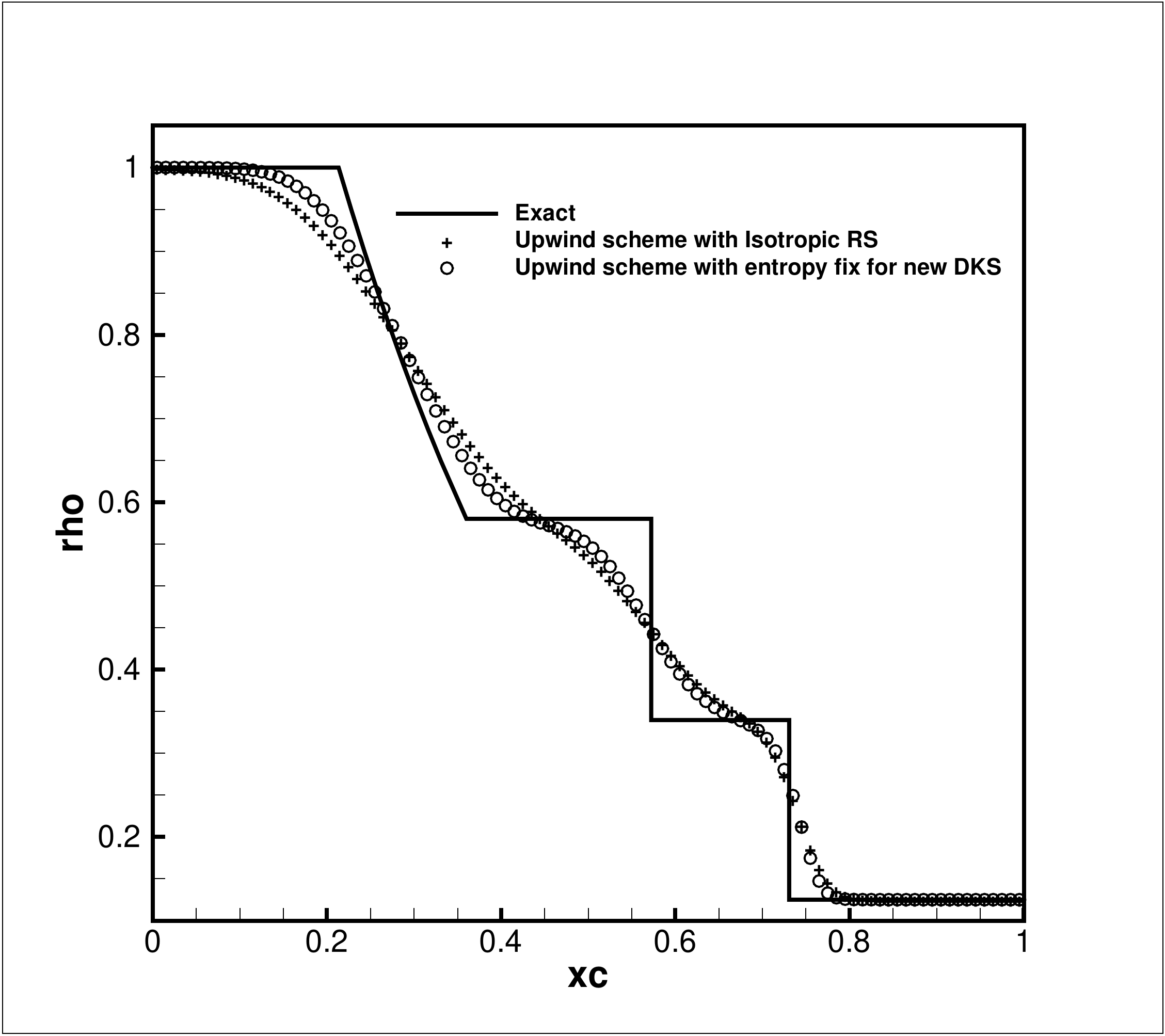}
\caption{With entropy fix (sec.\ref{sec:sonic_pt_entropy_fix})}
  \label{fig:Comparison_Toro_TestCase1_diff_accuracy}
\end{subfigure}%
\caption{Comparison of results for Shock tube problem with sonic point}
\label{fig:Sod_shock_tube_results}
\end{figure}

\begin{figure}[htb!]
\centering
\begin{subfigure}{.5\textwidth}
\centering
\includegraphics[trim = 4mm 4mm 10mm 10mm,clip,width=1.05\linewidth]{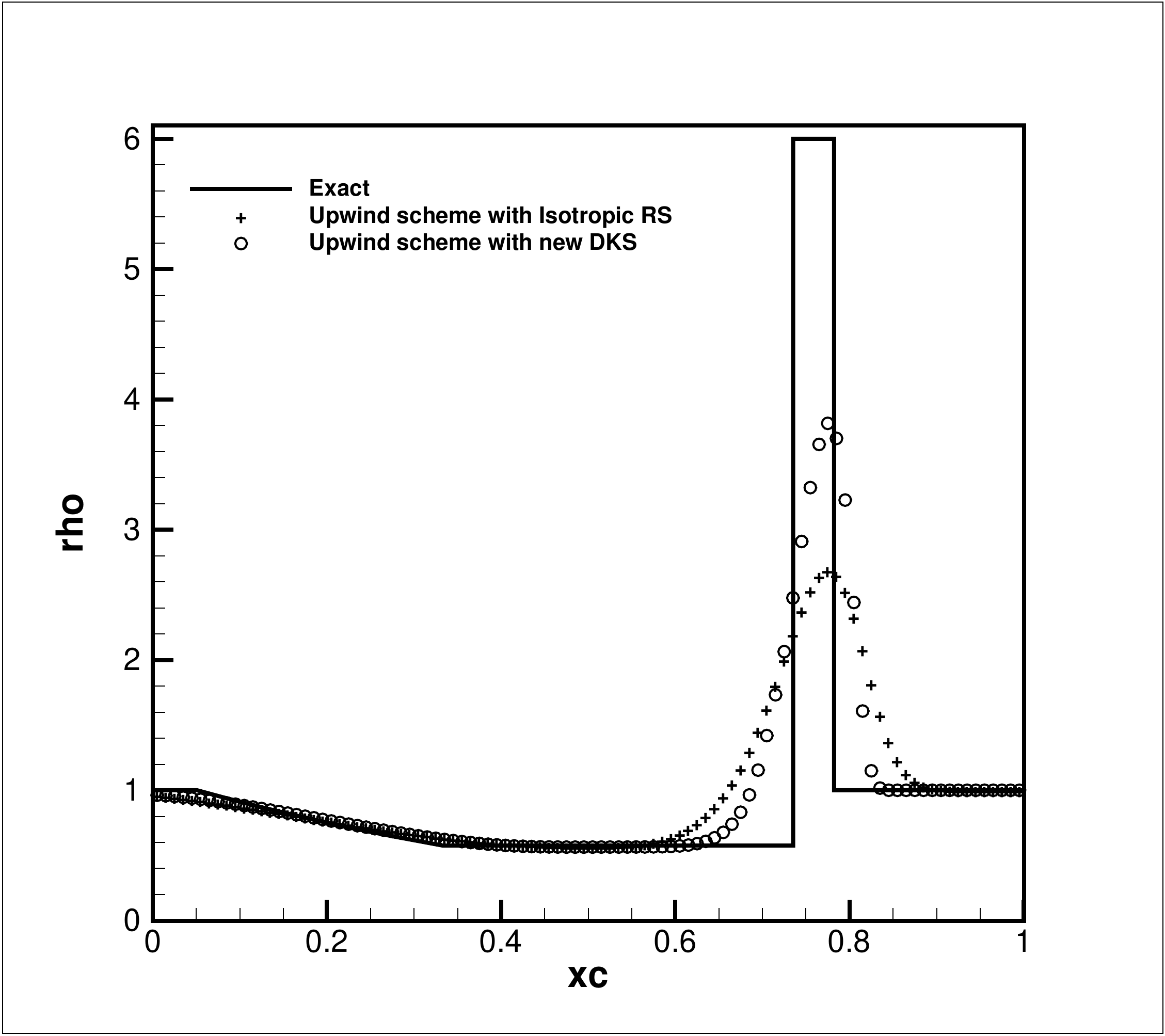}
  \caption{Strong shock test case}
  \label{fig:Comparison_Toro_TestCase3}
\end{subfigure}%
\begin{subfigure}{.5\textwidth}
\centering
\includegraphics[trim = 4mm 4mm 10mm 10mm,clip,width=1.07\linewidth]{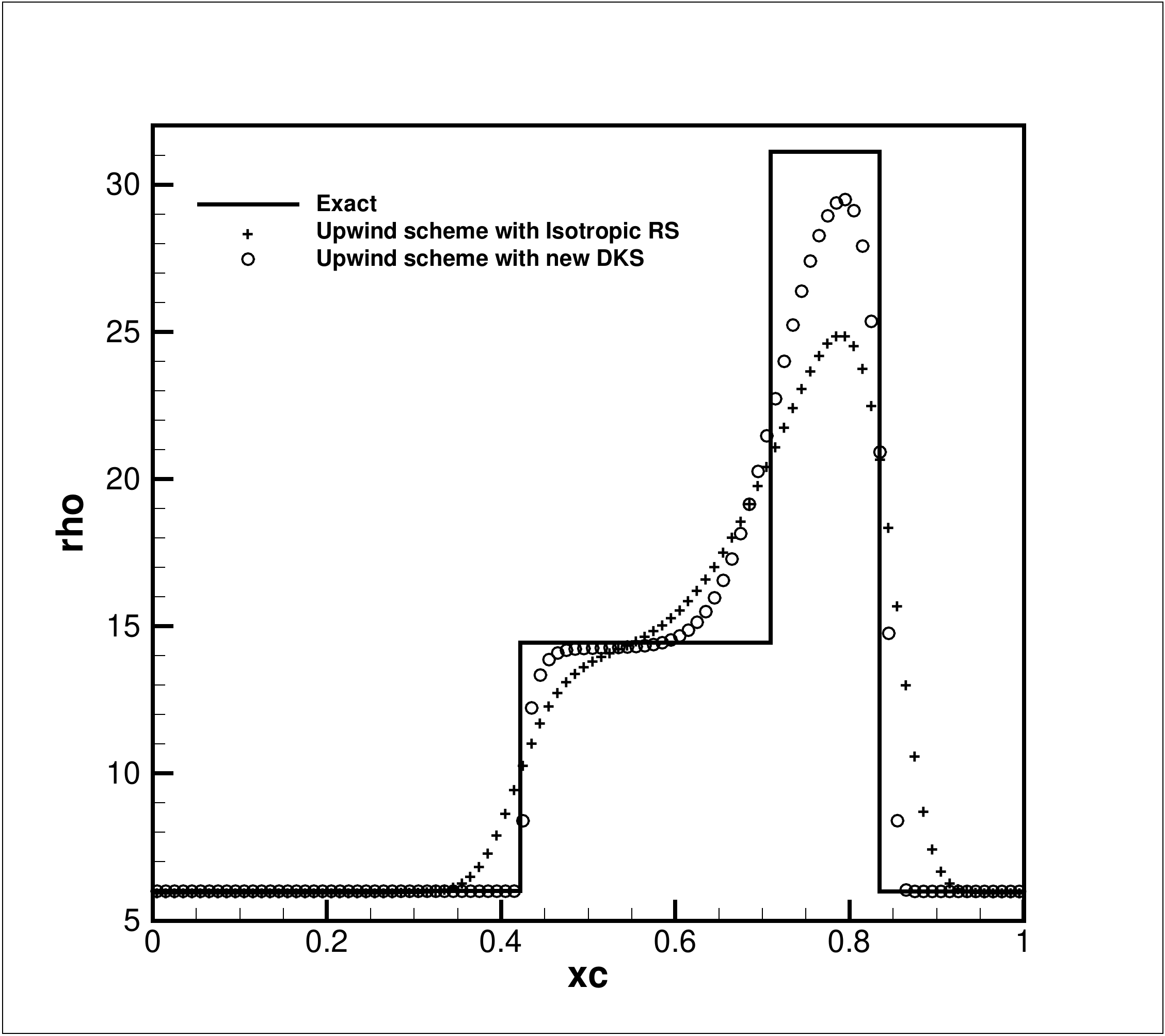}
  \caption{Strong discontinuities test case}
  \label{fig:Comparison_Toro_TestCase4}
\end{subfigure}
\label{fig:Sod_shock_tube_results_2}
\end{figure}

\begin{figure}
\centering
\begin{subfigure}{.5\textwidth}
\centering
\includegraphics[trim = 4mm 4mm 10mm 10mm,clip,width=1.07\linewidth]{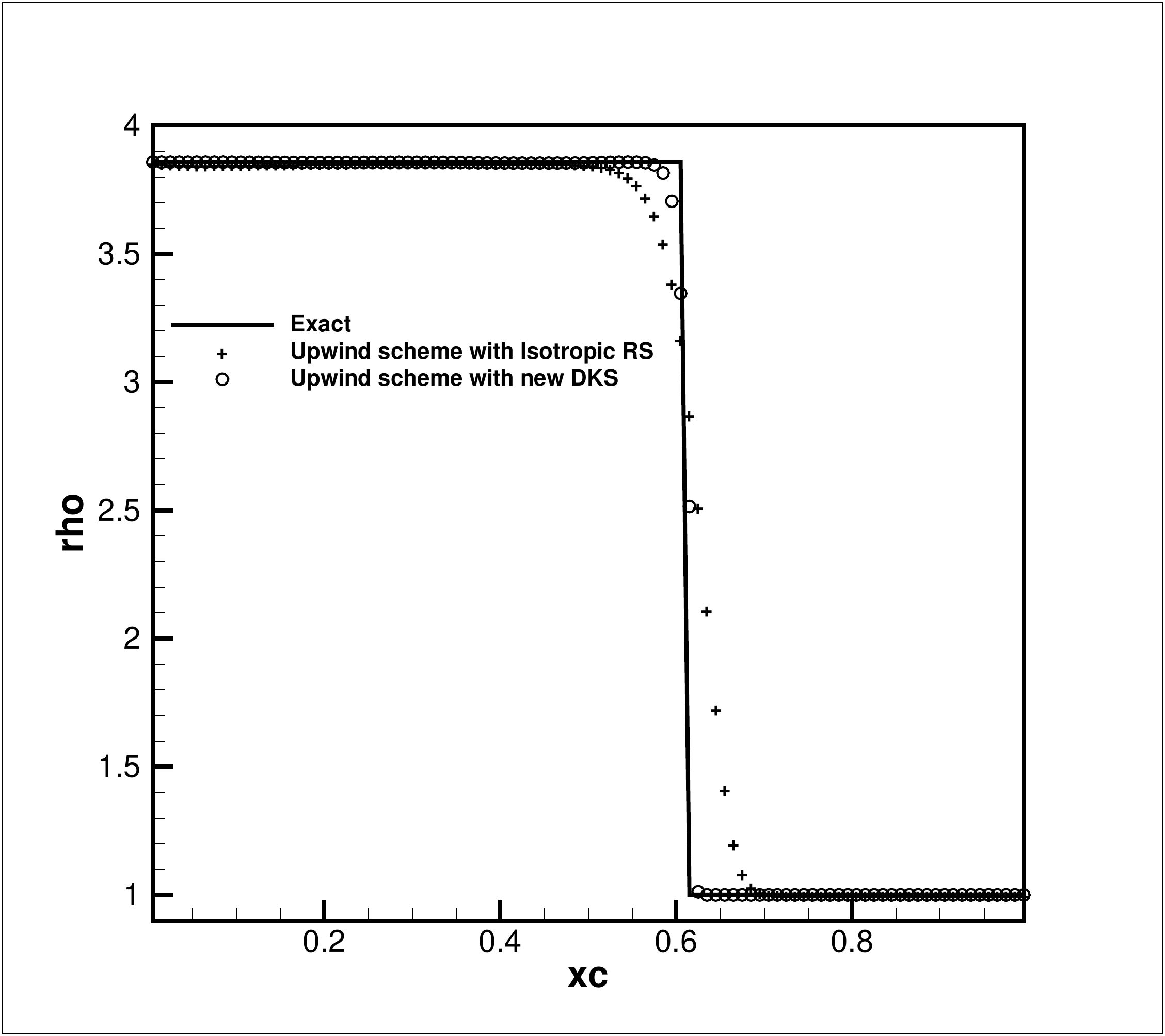}
  \caption{Slow moving shock at time=1}
  \label{fig:Comparison_Toro_TestCase5}
\end{subfigure}%
\begin{subfigure}{.5\textwidth}
\centering
\includegraphics[trim = 4mm 4mm 10mm 10mm,clip,width=1.07\linewidth]{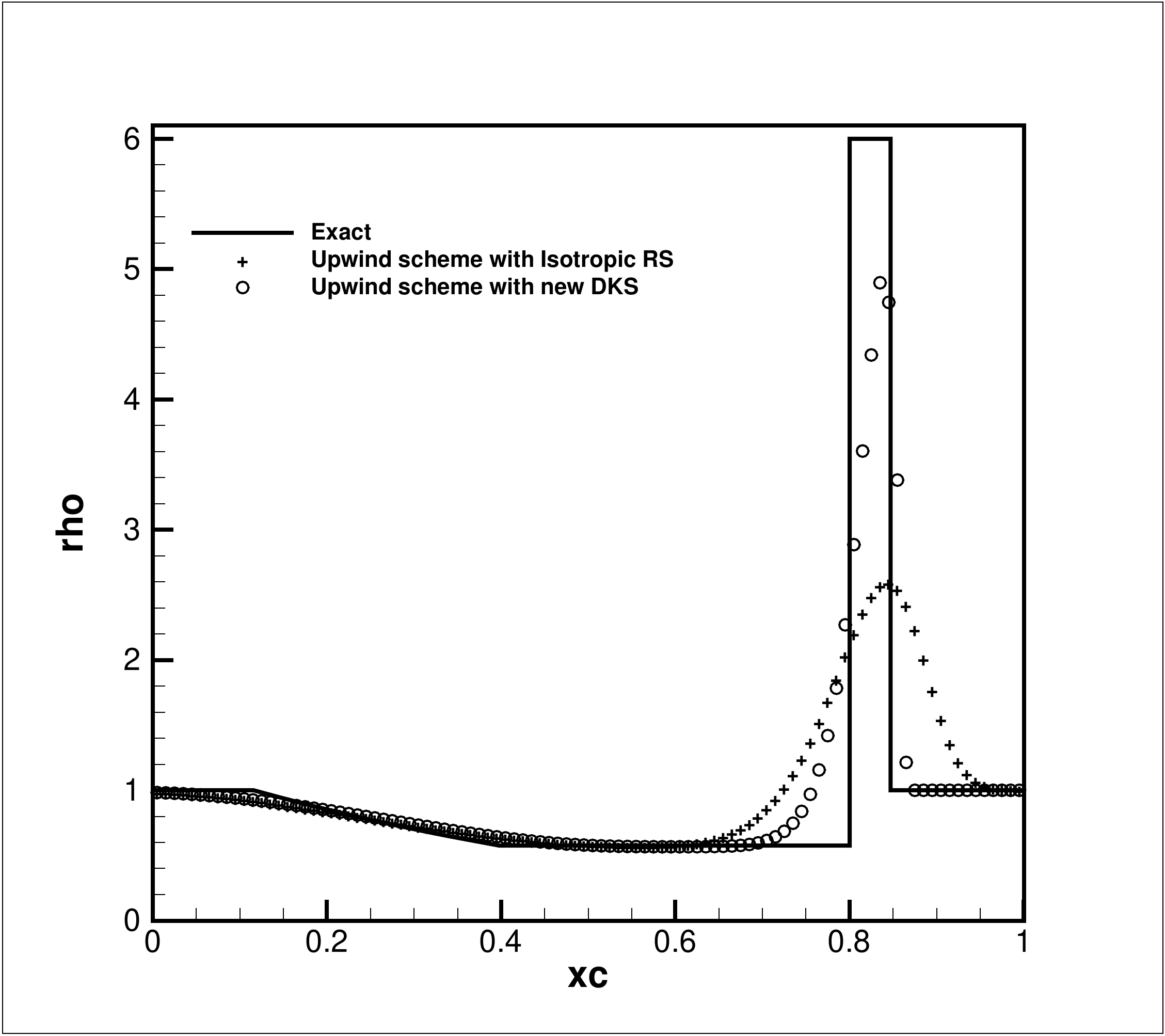}
  \caption{Slowly moving contact at time=0.012}
  \label{fig:Comparison_SlowMovingShock_TestCase}
\end{subfigure}
\caption{Slowly moving discontinuities}
\label{fig:Sod_shock_tube_results_3}
\end{figure}

\begin{figure}
\begin{subfigure}{.5\textwidth}
\centering
\includegraphics[trim = 4mm 4mm 10mm 10mm,clip,width=1.07\linewidth]{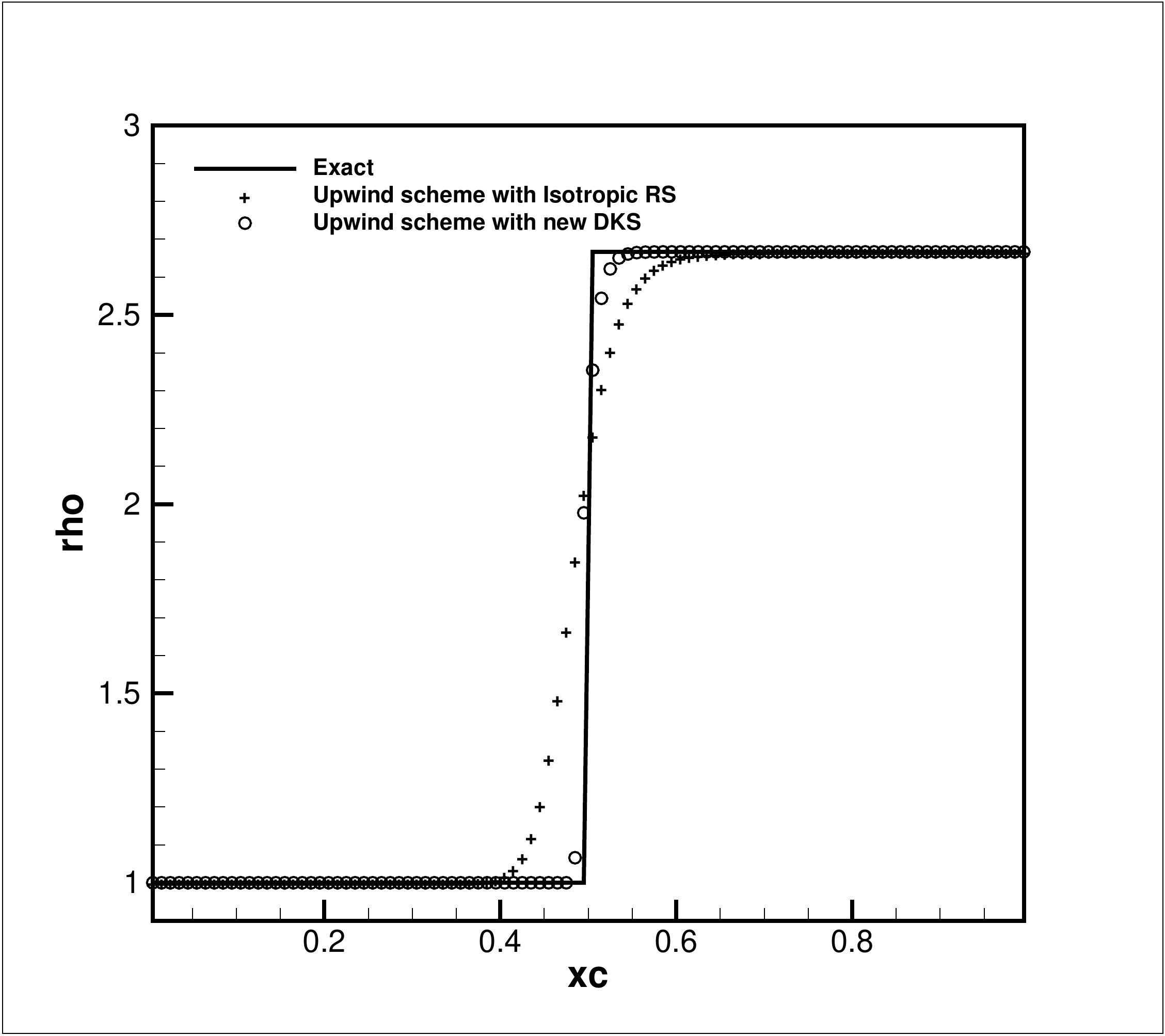}
  \caption{Steady shock test case}
  \label{fig:Comparison_SteadyShock_TestCase}
\end{subfigure}%
\begin{subfigure}{.5\textwidth}
\centering
\includegraphics[trim = 4mm 4mm 10mm 10mm,clip,width=1.07\linewidth]{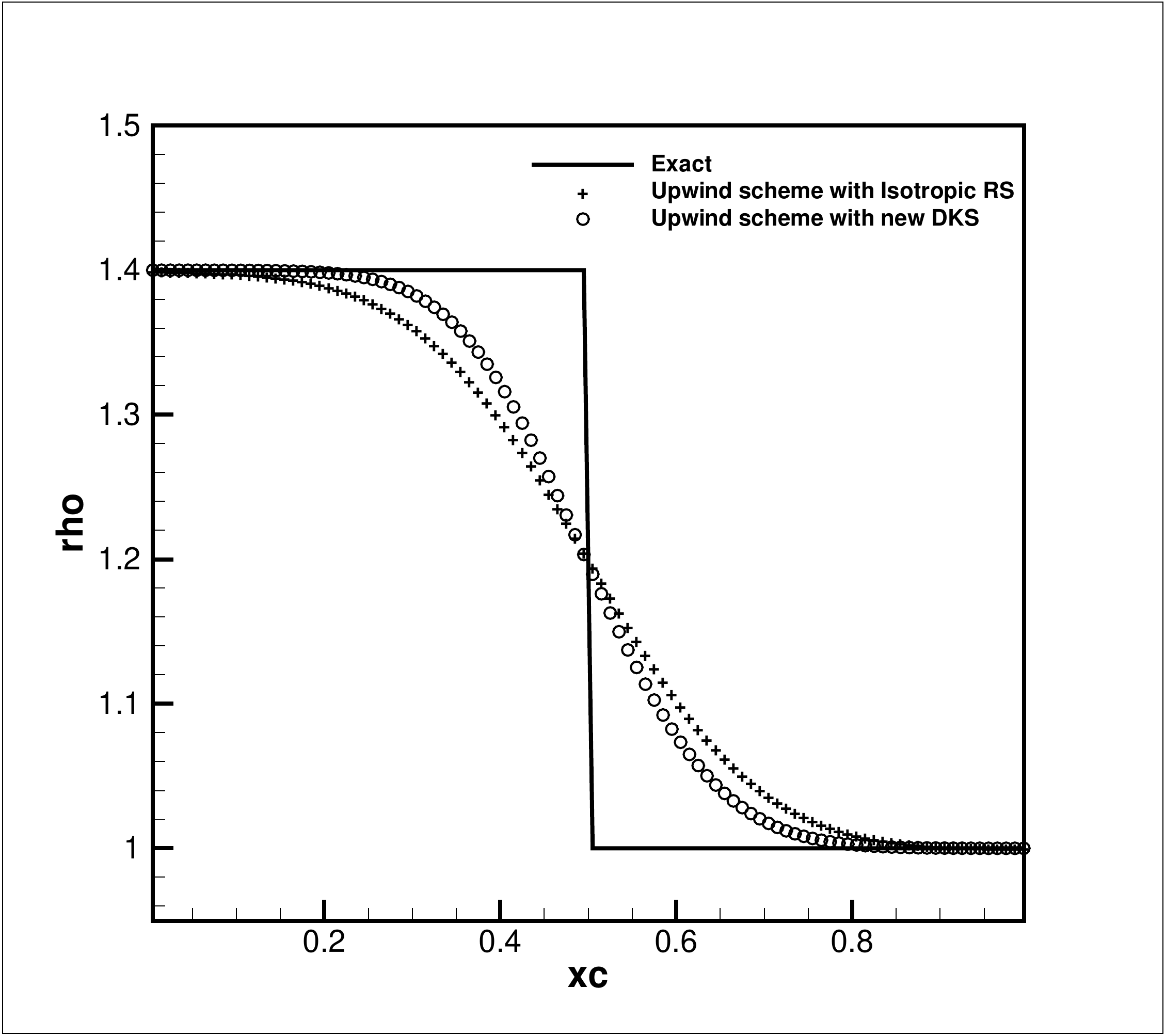}
  \caption{Steady contact test case}
  \label{fig:Comparison_SteadyContact_TestCase}
\end{subfigure}%
\caption{Steady discontinuities}
\label{fig: Comparison_Toro_TestCases}
\end{figure}

These problems include test cases with - sonic point, strong shock of Mach 198, strong discontinuities, slowly moving shock (discussed by Quirk \cite{Quirk}, Jin \emph{et al.} \cite{Jin_Liu}), slowly  moving contact discontinuity - as provided by Toro \cite{Toro}, steady shock test case \cite{Zhang_Shu} and steady contact discontinuity test case. From the results in figures \ref{fig:Sod_shock_tube_results} to \ref{fig: Comparison_Toro_TestCases}, it is seen that the new discrete kinetic scheme (DKS) works successfully for these tests and in fact outperforms URS in terms of accuracy.   
\clearpage
\subsection{2-D Euler test cases}
The present schemes are tested on various benchmark problems governed by 2-D Euler equations. The problems are chosen to assess the numerical schemes for their accuracy and robustness, for testing their capacities to avoid shock instabilities. 
\subsubsection{Regular shock reflection}
\begin{figure}[htb!]
\makebox[\textwidth][c]{%
\begin{subfigure}[b]{.45\textwidth}
\centering
\includegraphics[width=1.1\textwidth]{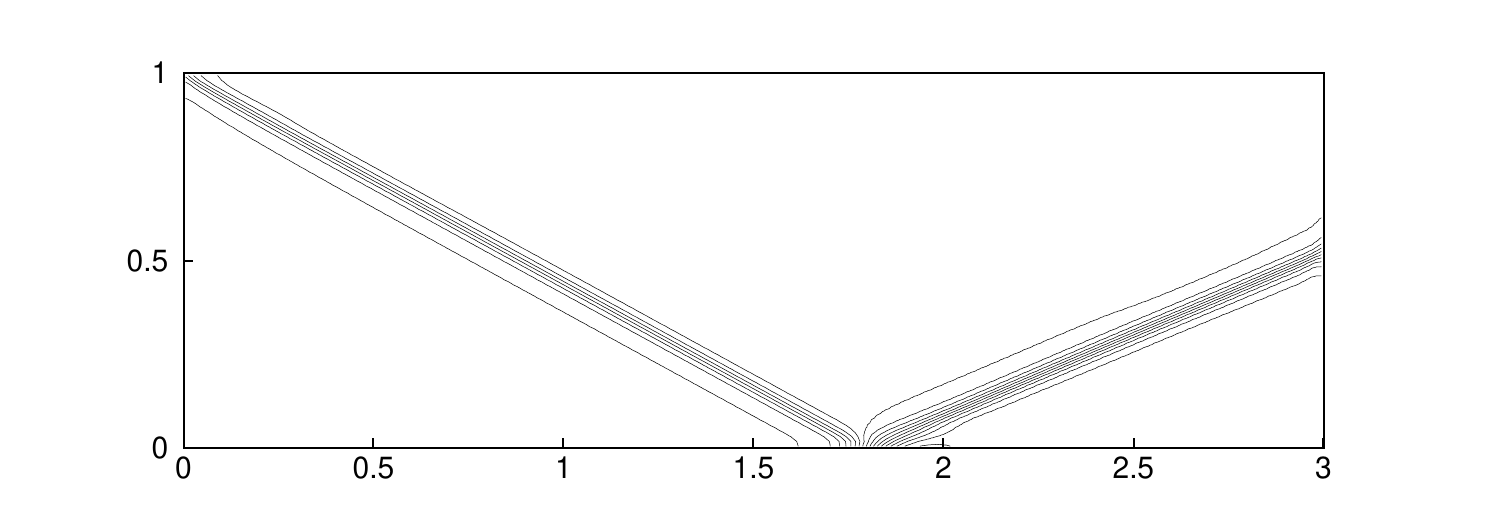}
\caption{URS}
\end{subfigure}%
\begin{subfigure}[b]{.45\textwidth}
\centering
\includegraphics[width=1.1\textwidth]{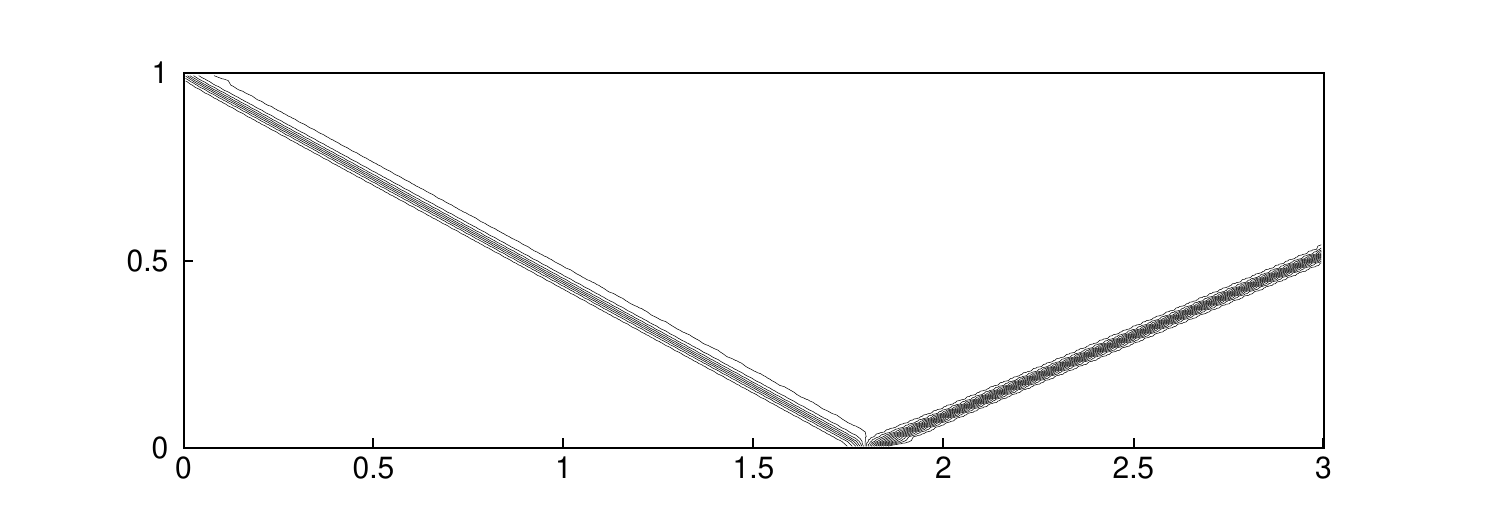}
\caption{Upwind scheme with new DKS}
\end{subfigure}
}
\label{fig:Regular_shock_reflection_SO_240_80}
\caption{Comparison of first-order (top row) and second-order (bottom row) results - Density contours (0.81:0.1:2.81) - for Regular shock reflection on a 240$\times$80 grid}
\label{fig:Regular_shock_reflection_240_80}
\end{figure}

This test case~\cite{Jin_Xin} involves capturing the flow features of an oblique shock incident upon a solid wall and getting reflected back. Figure \ref{fig:Regular_shock_reflection_240_80} shows comparison of results with first-order and second-order accuracy for the oblique shock reflection problem. Clearly, the upwind scheme with the new DKS captures the shocks more crisply than URS.

\begin{figure}[htb!]
\makebox[\textwidth][c]{%
\begin{subfigure}[b]{.45\textwidth}
\centering
\includegraphics[width=1.1\textwidth]{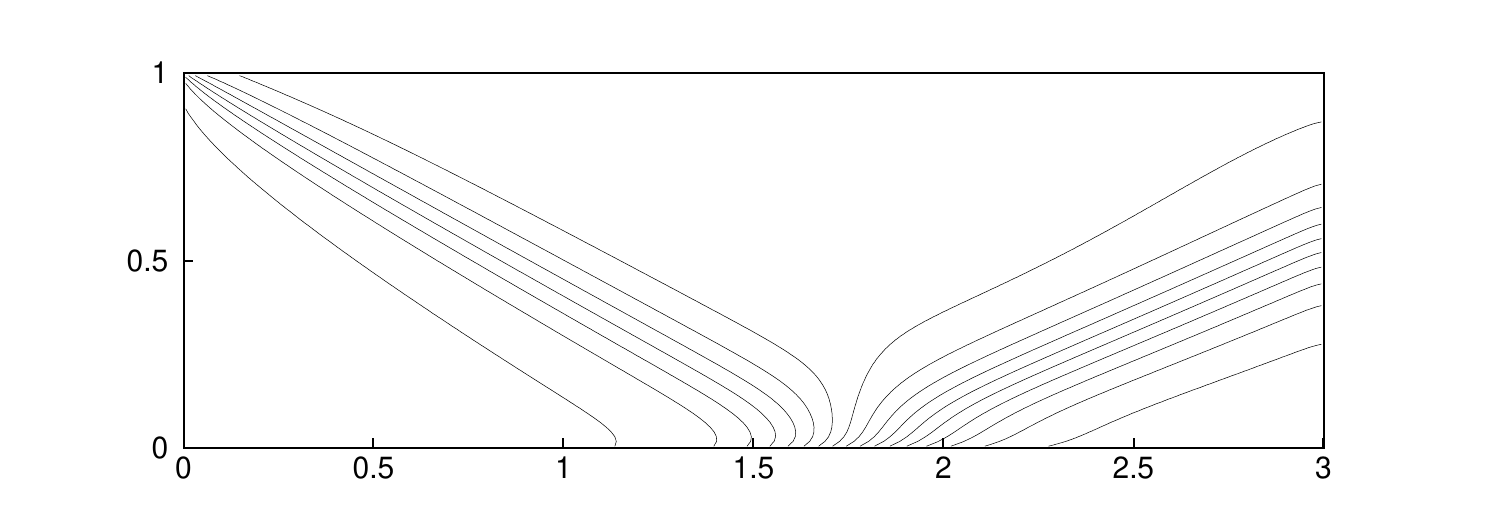}
\caption{URS}
\end{subfigure}%
\begin{subfigure}[b]{.45\textwidth}
\centering
\includegraphics[width=1.1\textwidth]{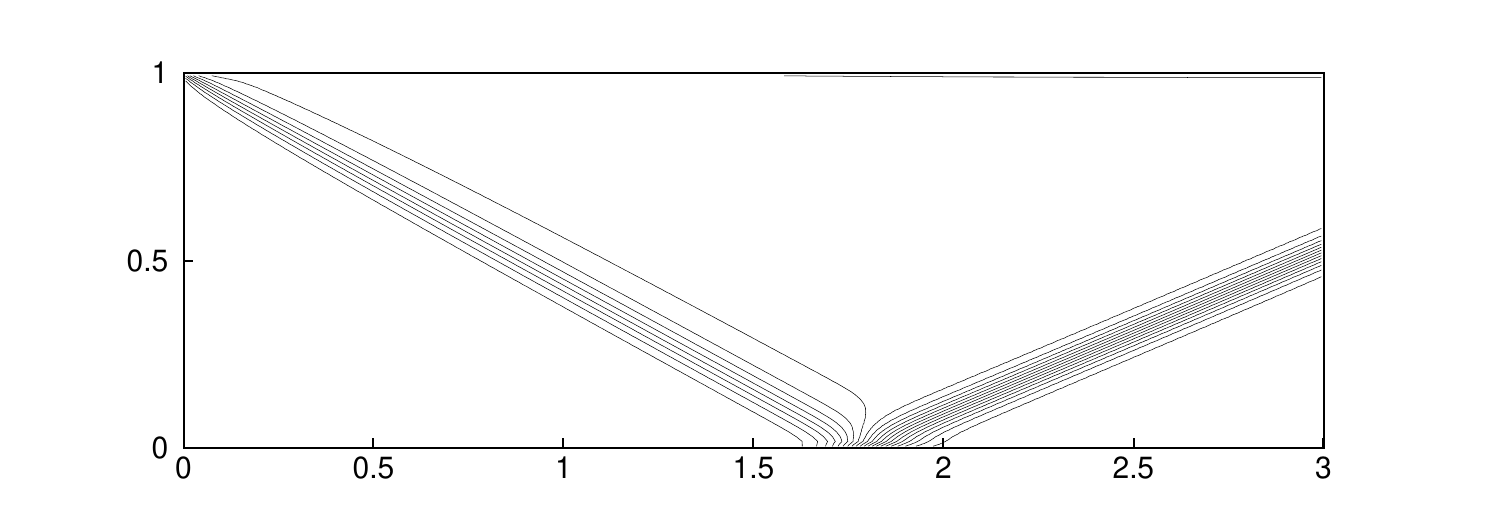}
\caption{Upwind scheme with new DKS}
\end{subfigure}
}
\label{fig:Regular_shock_reflection_FO_240_80}
\end{figure}

\subsubsection{Forward-facing step}
\begin{figure}[h!]
\makebox[\textwidth][c]{%
\begin{subfigure}[b]{.5\textwidth}
\centering
\includegraphics[width=1.1\textwidth]{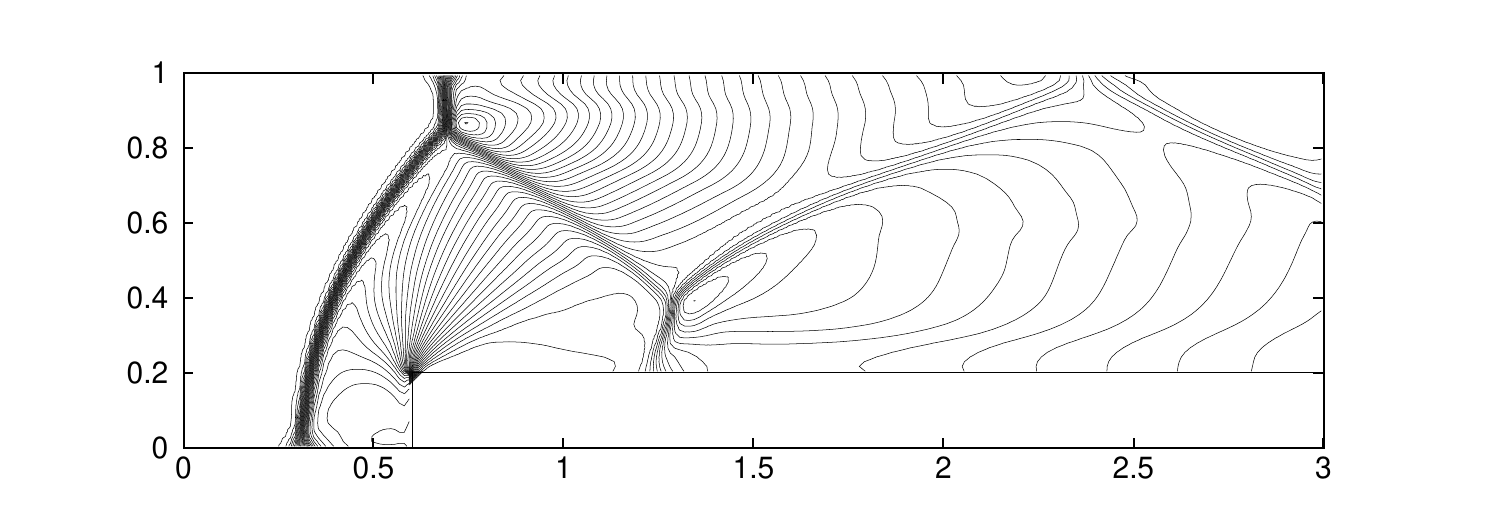}
\caption{URS}
\end{subfigure}%
\begin{subfigure}[b]{.5\textwidth}
\centering
\includegraphics[width=1.1\textwidth]{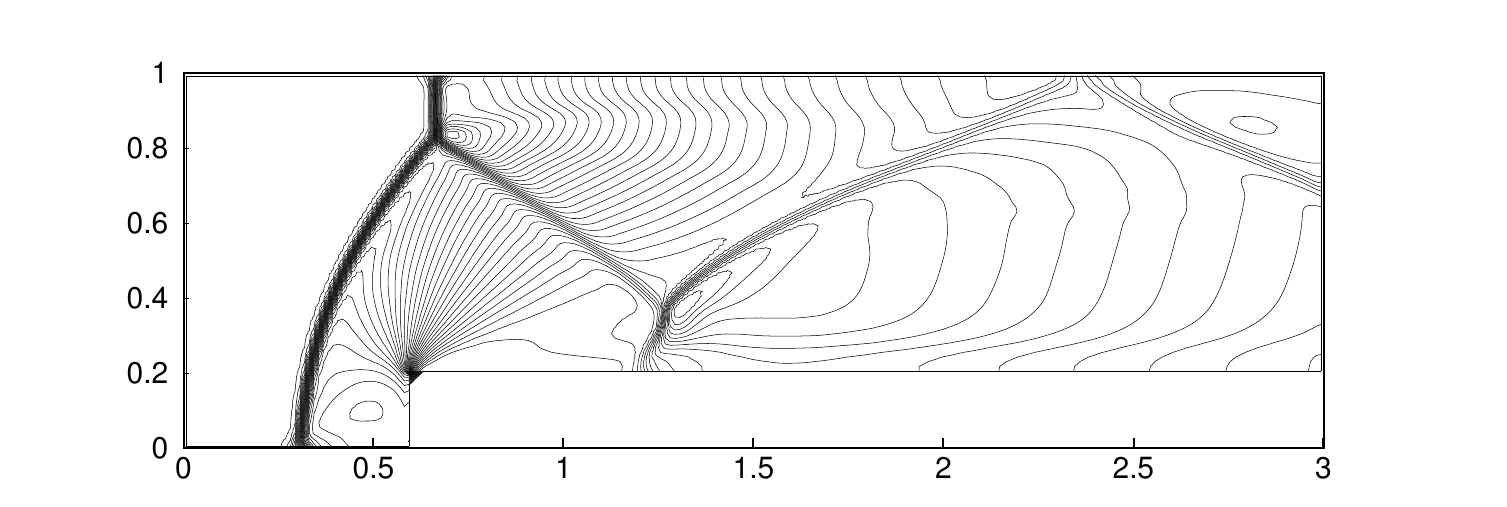}
\caption{Upwind scheme with new DKS}
\end{subfigure}
}
\caption{Comparison of second-order results - Density contours (1.0:0.15:6.4) at time=4 -~for Forward-facing step on a 240x$\times$80 grid}
\label{fig:Forward_facing_step_SO_240_80}
\end{figure}

In this unsteady test case~\cite{Woodward}, a Mach 3 flow enters a wind tunnel containing a forward-facing step. At time t=4.0, a lambda shock develops. A slip stream can also be seen beyond the triple point. Results are presented in figure \ref{fig:Forward_facing_step_SO_240_80}. Clearly, the upwind scheme with the new DKS captures the lambda shock, slip stream and reflected shocks more crisply than URS.

\subsubsection{Slip flow}
\begin{figure}[h!]
\centering
\begin{subfigure}{.5\textwidth}
\centering
\includegraphics[trim = 4mm 4mm 10mm 10mm,clip,width=0.95\linewidth]{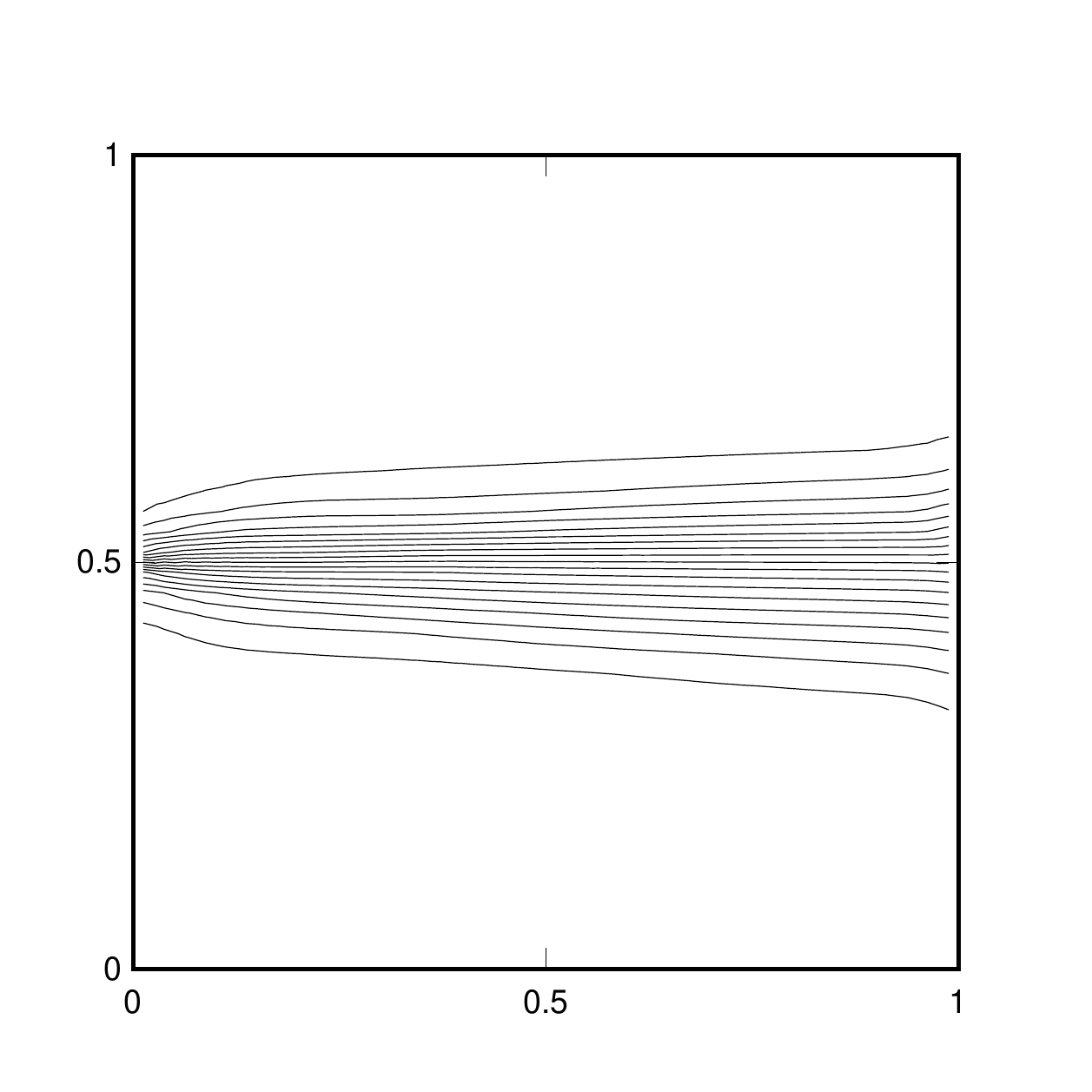}
  \caption{URS}
  \label{fig:URS_2nd_Order_2-D_Slip_Flow}
\end{subfigure}%
\begin{subfigure}{.5\textwidth}
\centering
\includegraphics[trim = 4mm 4mm 10mm 10mm,clip,width=0.95\linewidth]{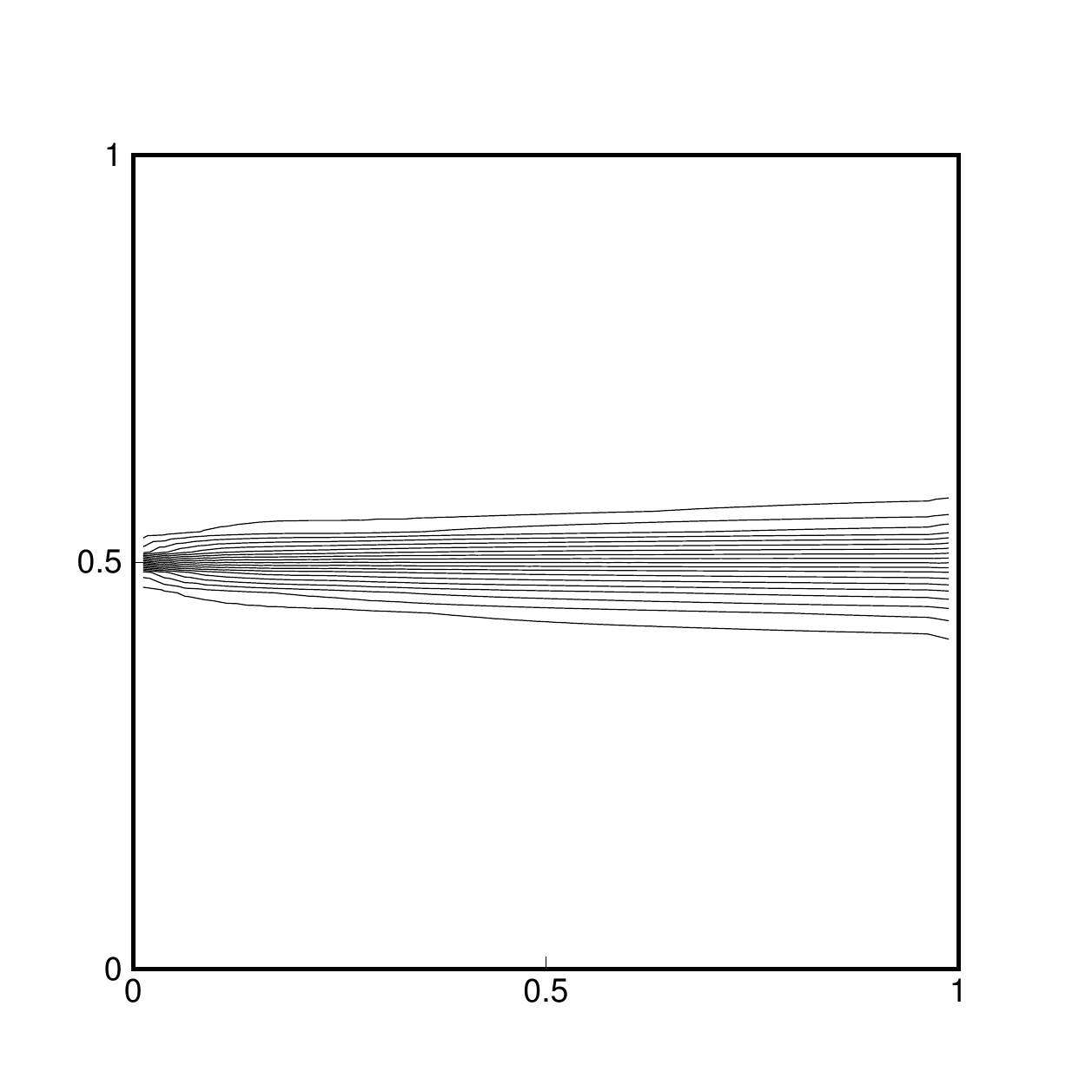}
  \caption{Upwind scheme with new DKS}
  \label{fig:Upwind_scheme_new_DKS_2-D_Slip_Flow}
\end{subfigure}
\caption{Comparison of results for horizontal slip flow, Mach contours on a 40$\times$40 grid}
\label{fig: Comparison_Slip_flow_TestCases}
\end{figure}

In this test case~\cite{Manna}, a Mach 3 flow slips over a Mach 2 flow. There is no jump in density and pressure across the interface. This problem tests the ability of the numerical scheme to capture grid aligned discontinuities. Results for this test case are presented in figure \ref{fig: Comparison_Slip_flow_TestCases}. The upwind scheme with the new DKS diffuses the grid-aligned slip stream but is far less diffusive compared to URS.

\subsubsection{Odd-even decoupling}
\begin{figure}[h!]
\centering
\begin{subfigure}{.5\textwidth}
\centering
\includegraphics[trim = 4mm 4mm 10mm 10mm,clip,width=1.07\linewidth]{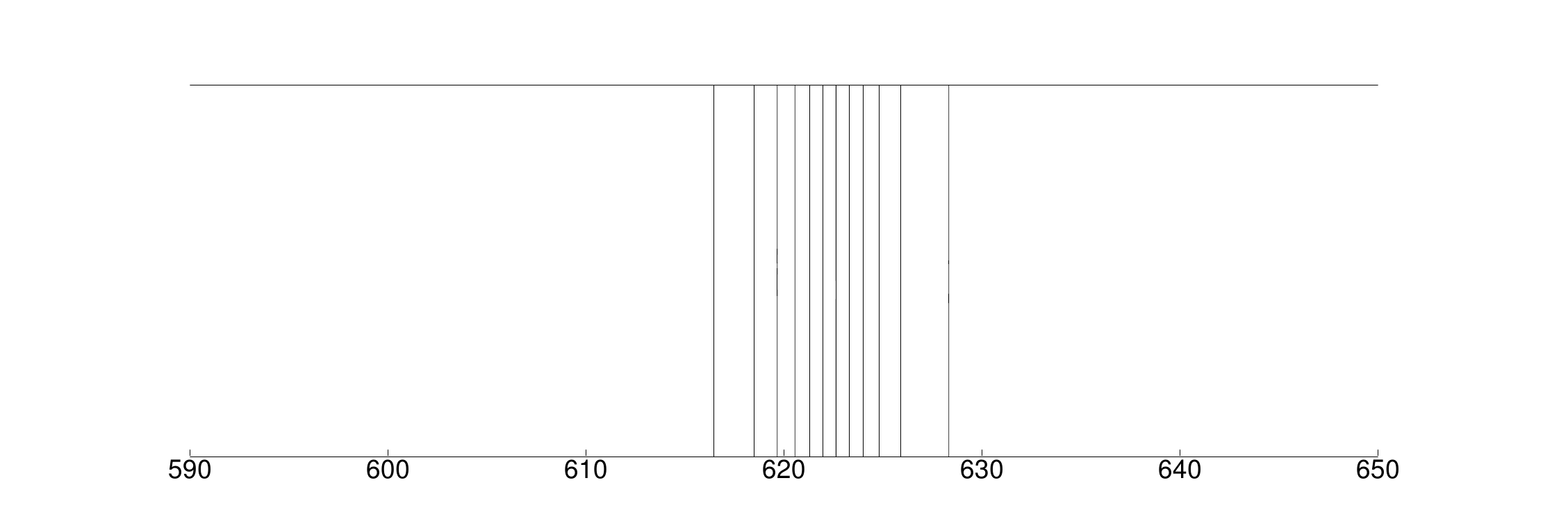}
  \caption{URS}
\end{subfigure}\\
\begin{subfigure}{.5\textwidth}
\centering
\includegraphics[trim = 4mm 4mm 10mm 10mm,clip,width=1.07\linewidth]{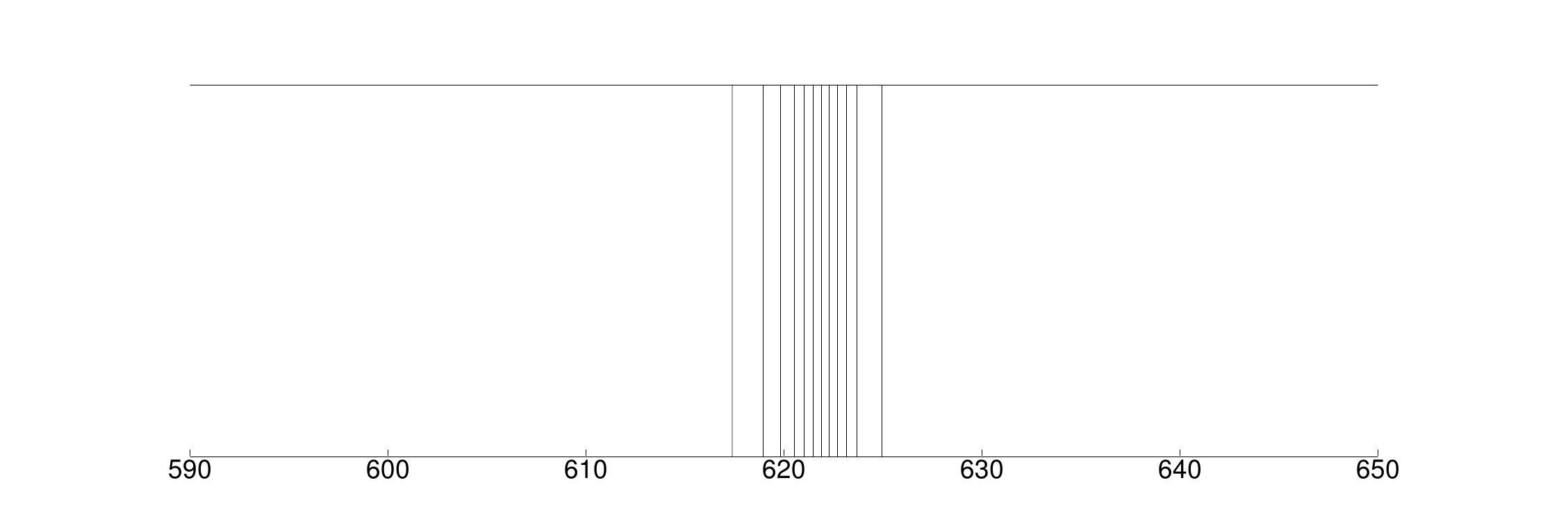}
  \caption{Upwind scheme with new DKS}
\end{subfigure}
\caption{Comparison of results for odd-even decoupling test case, density contours on a 20$\times$800 grid at time t=100}
\label{fig: Comparison_odd_even_decoupling}
\end{figure}

This is yet another test case~\cite{Quirk} which assesses a numerical scheme for shock instability in which a planar Mach 6 shock simply travels along a rectangular duct. For numerical solution, the duct is set up with a mesh of 20$\times$800 unit square cells. Now, the widthwise centerline is perturbed in the following manner:
\begin{equation*}
{y}_{i,{j}_{mid}} =
\begin{cases}
{y}_{i,{j}_{mid}} + 10^{-3} \ \text{for \emph{i} even}, \\
{y}_{i,{j}_{mid}} - 10^{-3} \ \text{for \emph{i} odd}
\end{cases}
\end{equation*}
With schemes like Godunov's exact Riemann solver and approximate Riemann solver of Roe, this perturbation promotes odd-even decoupling thereby destroying the planar shock structure. On the contrary, the shock captured (after a long time t=100) using the upwind scheme with the new DKS is stable to the perturbation, as shown in figure \ref{fig: Comparison_odd_even_decoupling}. In terms of accuracy, we note that the new scheme is less diffusive than URS.
\\
\subsubsection{Double-Mach reflection (DMR)}
\begin{figure}[h!]
\makebox[\textwidth][c]{%
\begin{subfigure}[b]{.5\textwidth}
\centering
\includegraphics[width=1.1\textwidth]{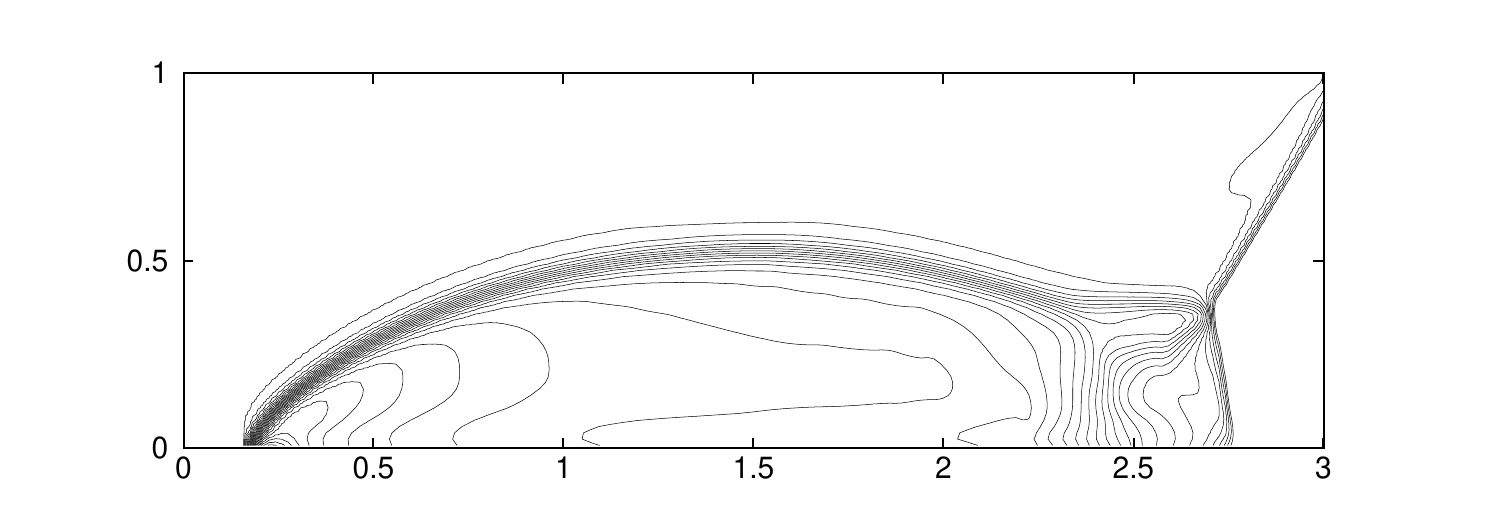}
\caption{URS}
\end{subfigure}%
\begin{subfigure}[b]{.5\textwidth}
\centering
\includegraphics[width=1.1\textwidth]{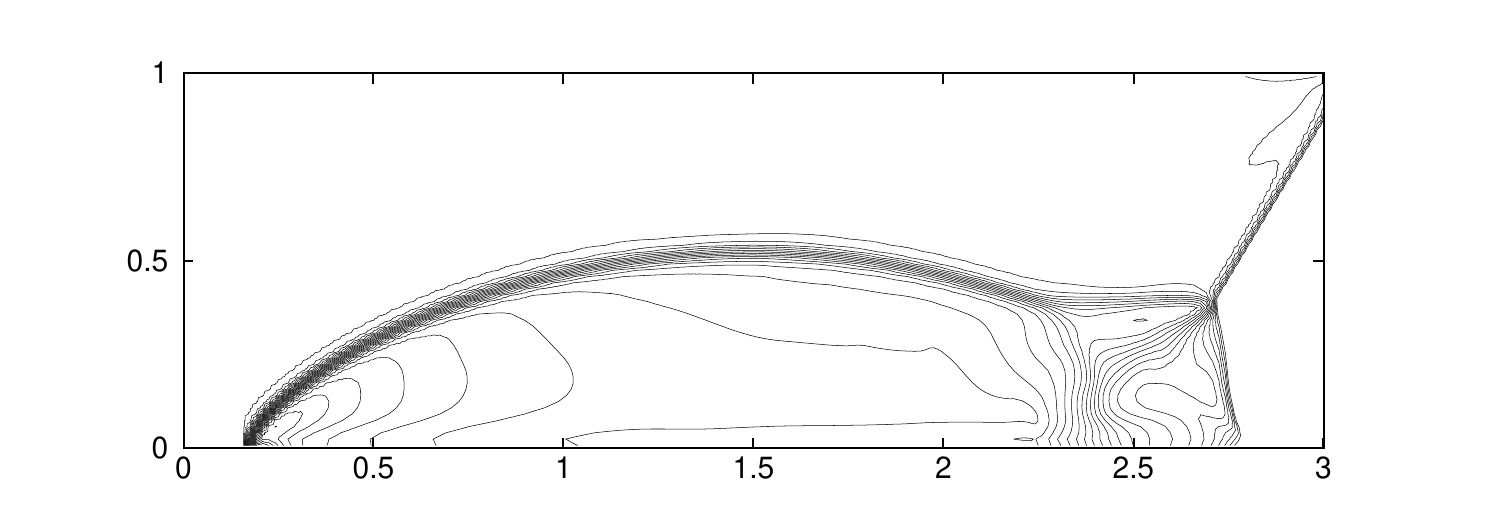}
\caption{Upwind scheme with new DKS}
\end{subfigure}
}
\caption{Comparison of second-order results - Density contours (5.0:0.517:20.0) at time=0.2 - for Double-mach reflection on a 240$\times$60 grid}
\label{fig:DMR_SO_240_60}
\end{figure}

In this unsteady test case~\cite{Woodward}, a Mach 10 shock is driven down a tube containing a wedge. At first the simple planar shock meets the walls of the tube at right angles, but on encountering the sloping surface of the wedge, a complicated shock reflection occurs resulting in the formation of reflected shocks, Mach stems, triple points and slip streams. This is one of the problems to test a numerical scheme against the shock-instability termed kinked Mach stem~\cite{Quirk}. Results for this unsteady test case (at time t=0.2) are presented in figure \ref{fig:DMR_SO_240_60}. The upwind scheme with the new DKS does not produce kinked Mach stem. Also, the various features of DMR test case are captured reasonably well. The new DKS based upwind scheme captures reflected shocks and Mach stems more crisply than URS.

\subsubsection{Shock diffraction}
\begin{figure}[h!]
\centering
\begin{subfigure}{.45\textwidth}
\centering
\includegraphics[trim = 4mm 4mm 10mm 10mm,clip,width=1.0\linewidth]{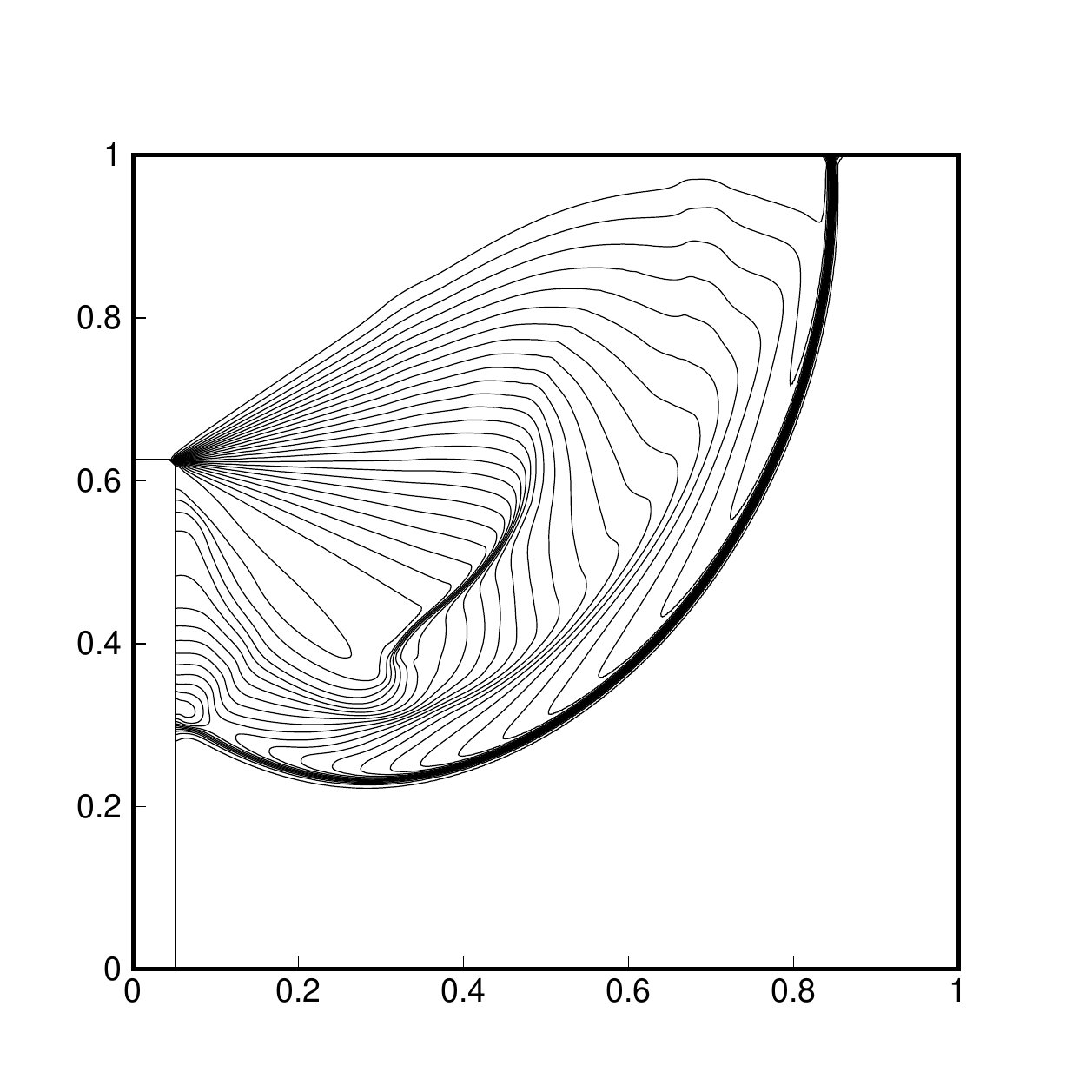}
  \caption{URS}
  \label{fig:URS_SO_2-D_Shock_Diffraction_400_400}
\end{subfigure}%
\begin{subfigure}{.45\textwidth}
\centering
\includegraphics[trim = 4mm 4mm 10mm 10mm,clip,width=1.0\linewidth]{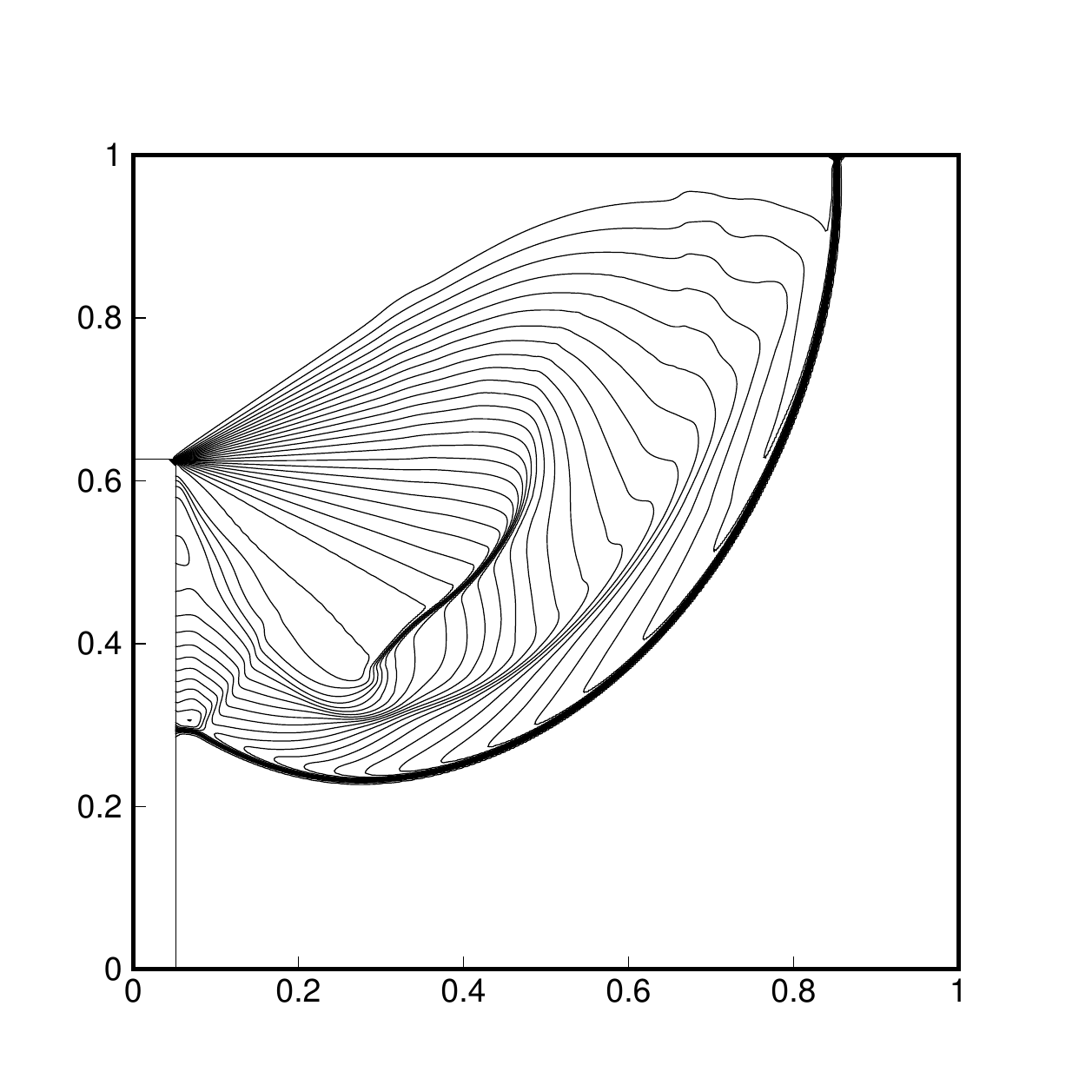}
  \caption{Upwind scheme with new DKS}
  \label{fig:Upwind_new_DKS_SO_2-D_Shock_Diffraction_400_400}
\end{subfigure}
\caption{Comparison of second-order results - Density contours (0.5:0.25:6.75) at time=0.1561 - for Shock diffraction test case on a 400$\times$400 grid}
\label{fig: Comparison_SO_Shock_diffr_TestCases}
\end{figure}

This is another test case~\cite{Huang} which assesses a numerical scheme for expansion shocks. This test case has complex flow features involving a planar shock wave moving with incident Mach number, a diffracted shock wave around the corner and a strong expansion wave. The strong shock wave accelerates the flow and interacts with post-shock fluid to further complicate the flow. Other distinct flow features are a slip stream and a contact surface. Godunov-type and Roe schemes are known to fail for this test case~\cite{Quirk} as they admit expansion shocks without a proper fix. Results for this unsteady test case (at time t=0.1561) are presented in figure \ref{fig: Comparison_SO_Shock_diffr_TestCases}. The upwind scheme with new DKS  does not produce any expansion shock. Also, the various features like the slip stream and contact surfaces are captured more crisply than URS.

\clearpage

\subsubsection{Hypersonic flow past a half-cylinder}
\begin{figure}[h!]
\makebox[\textwidth][c]{%
\begin{subfigure}[b]{.22\textwidth}
\centering
\includegraphics[width=0.8\textwidth]{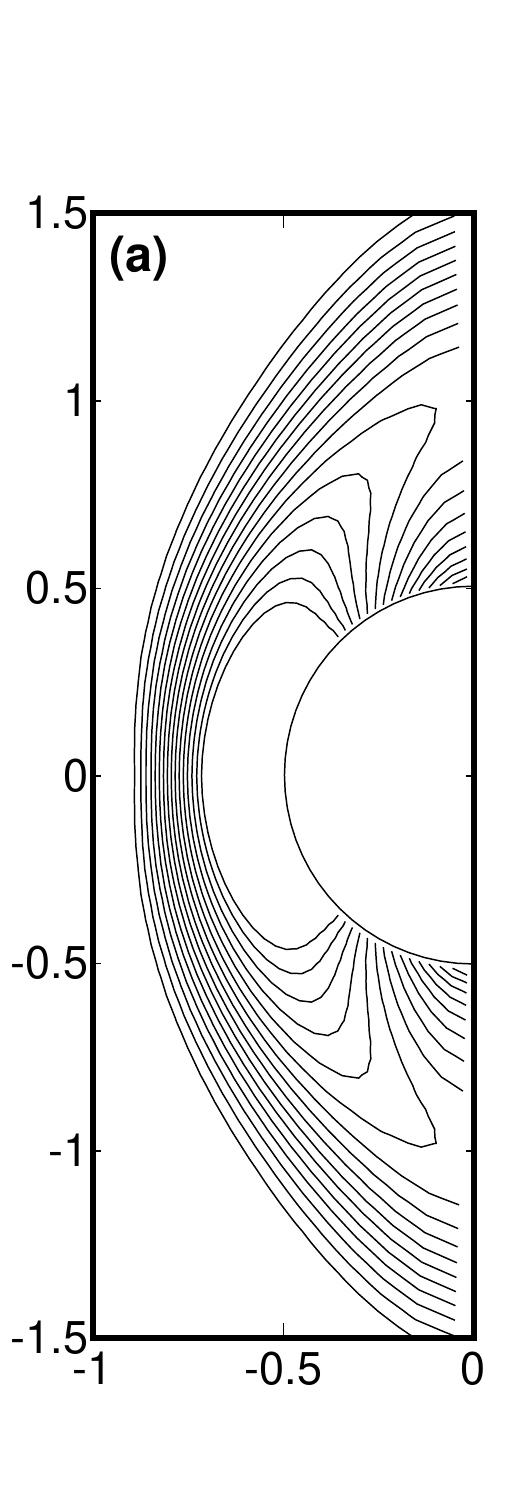}
\end{subfigure}%
\begin{subfigure}[b]{.22\textwidth}
\centering
\includegraphics[width=0.8\textwidth]{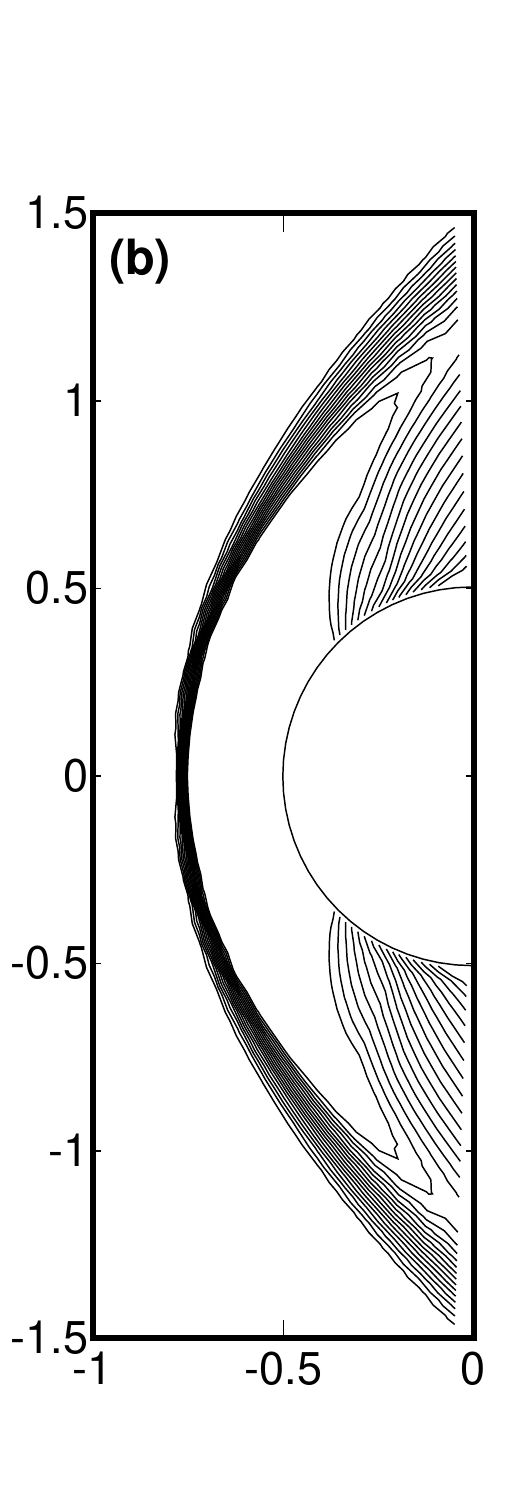}
\end{subfigure}
}\\
\makebox[\textwidth][c]{%
\begin{subfigure}[b]{.22\textwidth}
\centering
\includegraphics[width=0.8\textwidth]{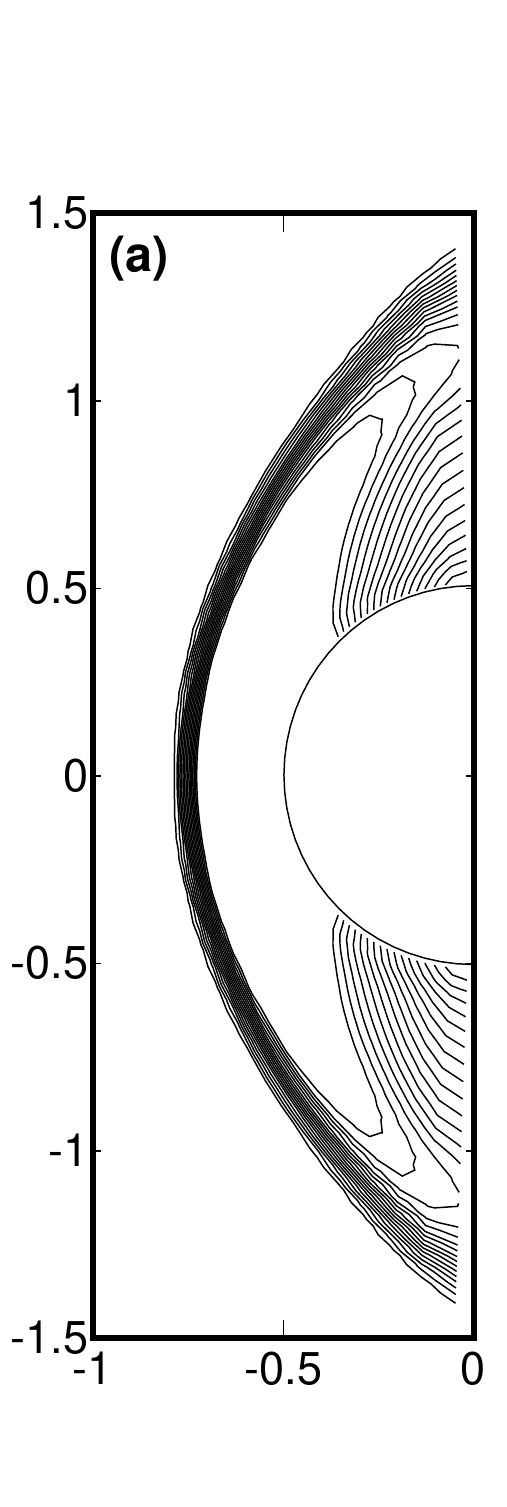}
\end{subfigure}%
\begin{subfigure}[b]{.22\textwidth}
\centering
\includegraphics[width=0.8\textwidth]{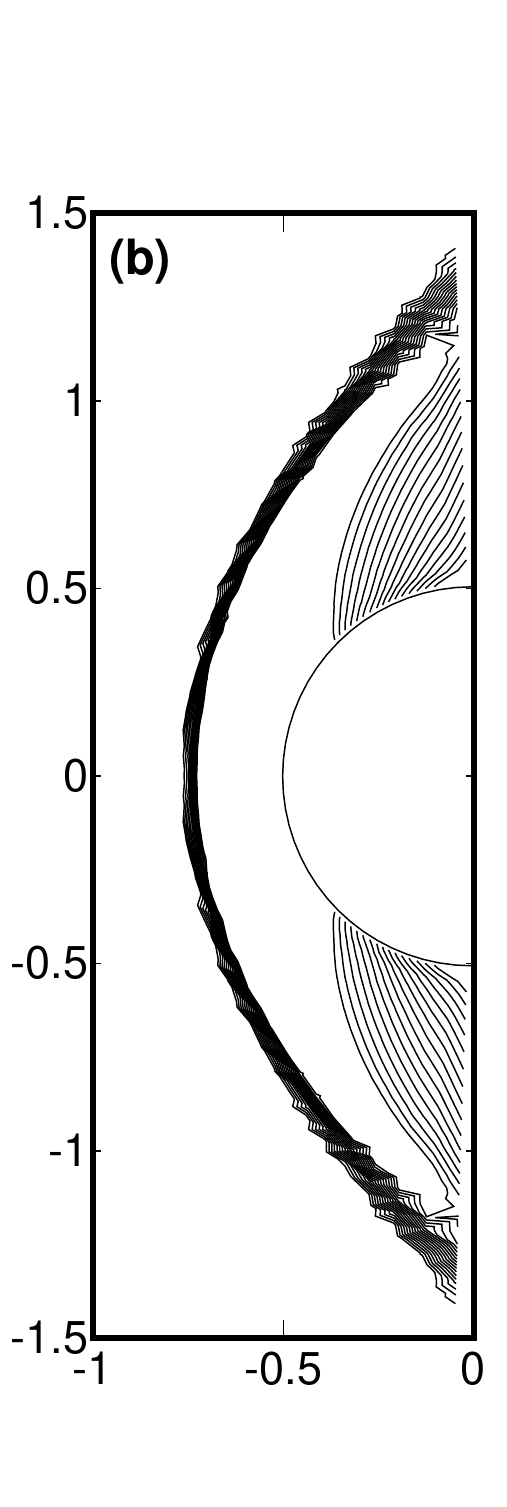}
\end{subfigure}
}
\caption{Comparison of first-order results (top row) and second-order results (bottom row) - Density contours (2.0:0.2:5.0) - for Half-cylinder on a 45$\times$45 grid using (a) URS (b) Upwind scheme with new DKS}
\label{fig: Compare_SO_half_cyl}
\end{figure}

This test case assesses a numerical scheme for the shock instability called carbuncle shock discussed by Quirk~\cite{Quirk} and Meng-Sing Liou~\cite{Liou}. Results for this test case are presented with first-order accuracy and second-order accuracy in figure \ref{fig: Compare_SO_half_cyl}. The upwind scheme with the new DKS does not produce carbuncle shocks. It also captures the bow shock more crisply than URS.

\clearpage

\subsubsection{Transonic flow past NACA0012 airfoil}
\begin{figure}[h!]
\makebox[\textwidth][c]{%
\begin{subfigure}[b]{.5\textwidth}
\centering
\includegraphics[width=1.0\textwidth]{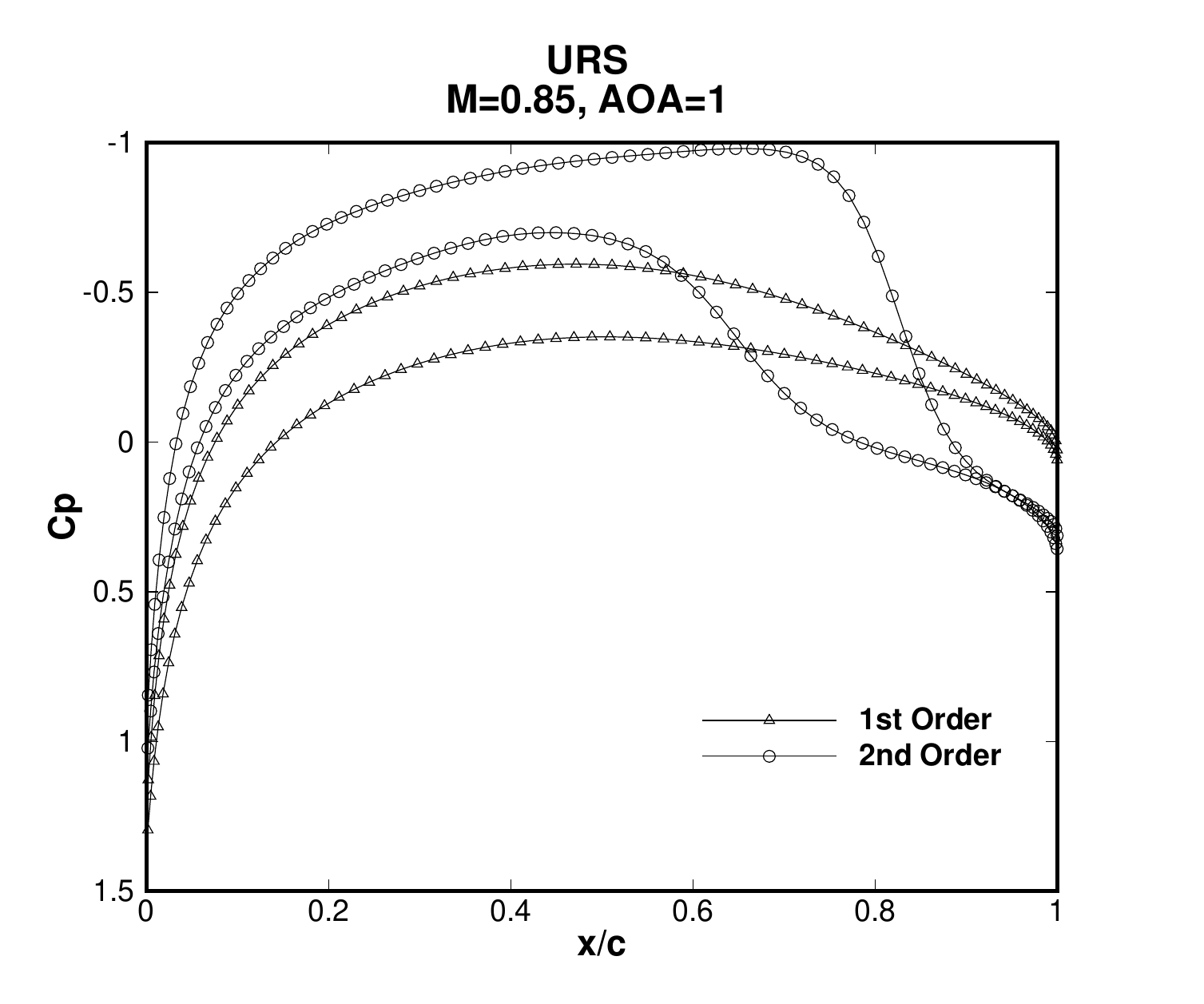}
\caption{URS}
\end{subfigure}%
\begin{subfigure}[b]{.5\textwidth}
\centering
\includegraphics[width=1.0\textwidth]{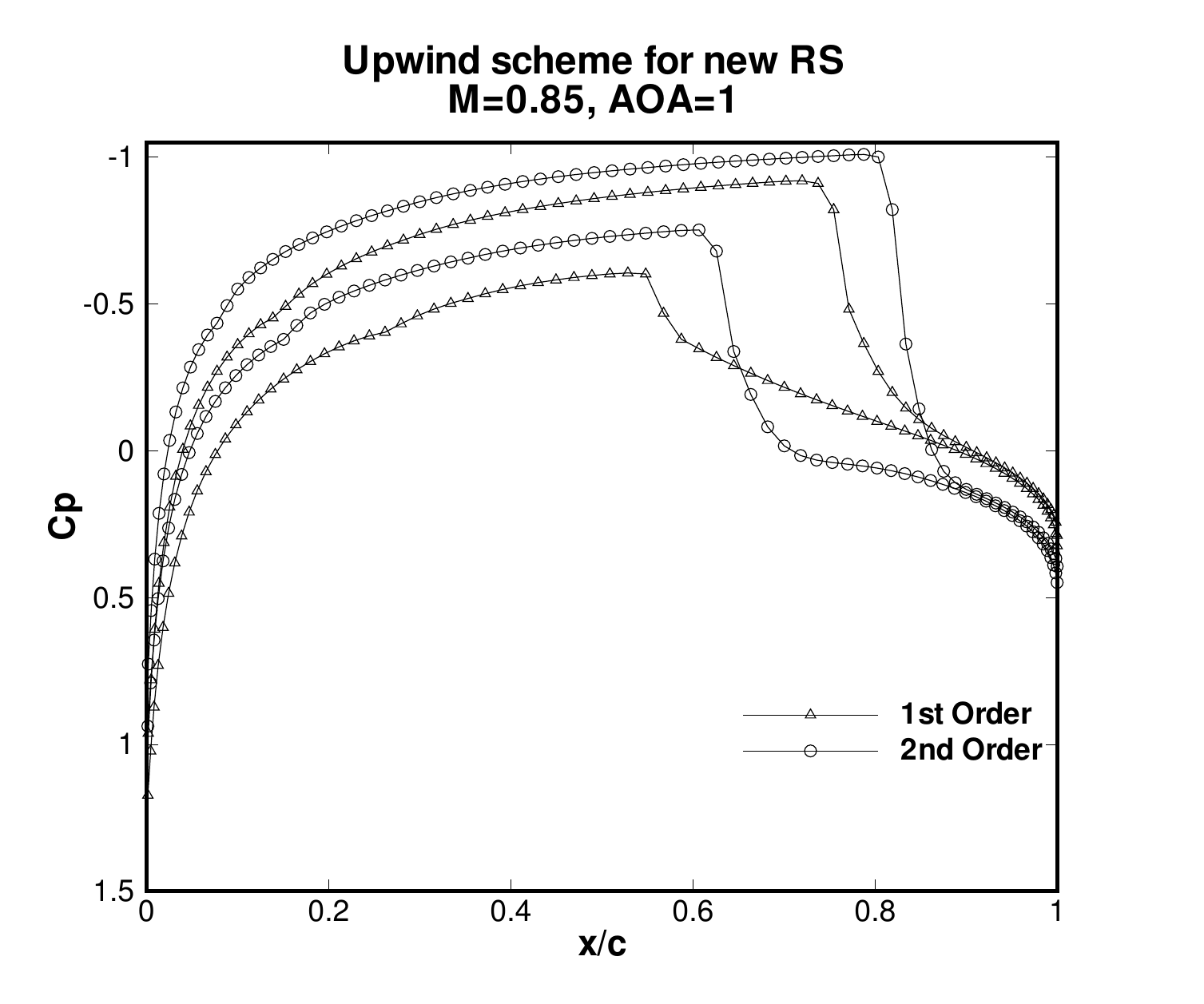}
\caption{Upwind scheme with new DKS}
\end{subfigure}
}
\caption{Comparison of Cp plots for transonic flow over airfoil, M=0.85, AOA=$1^\circ$}
\label{fig: Cp_transonic_airfoil}
\end{figure}

A benchmark case of transonic flow over NACA0012 airfoil~\cite{Viviand}, with an inflow Mach number 0.85 and angle of attack 1$^\circ$, is simulated using the present schemes.  C$_p$ plots of first-order and second-order accuracy are compared in figure \ref{fig: Cp_transonic_airfoil}. The better accuracy of the upwind scheme with new DKS compared to URS is evident from the plots of pressure contours and C$_p$ plots.    

\begin{figure}[h!]
\makebox[\textwidth][c]{%
\begin{subfigure}[b]{.5\textwidth}
\centering
\includegraphics[width=1.0\textwidth]{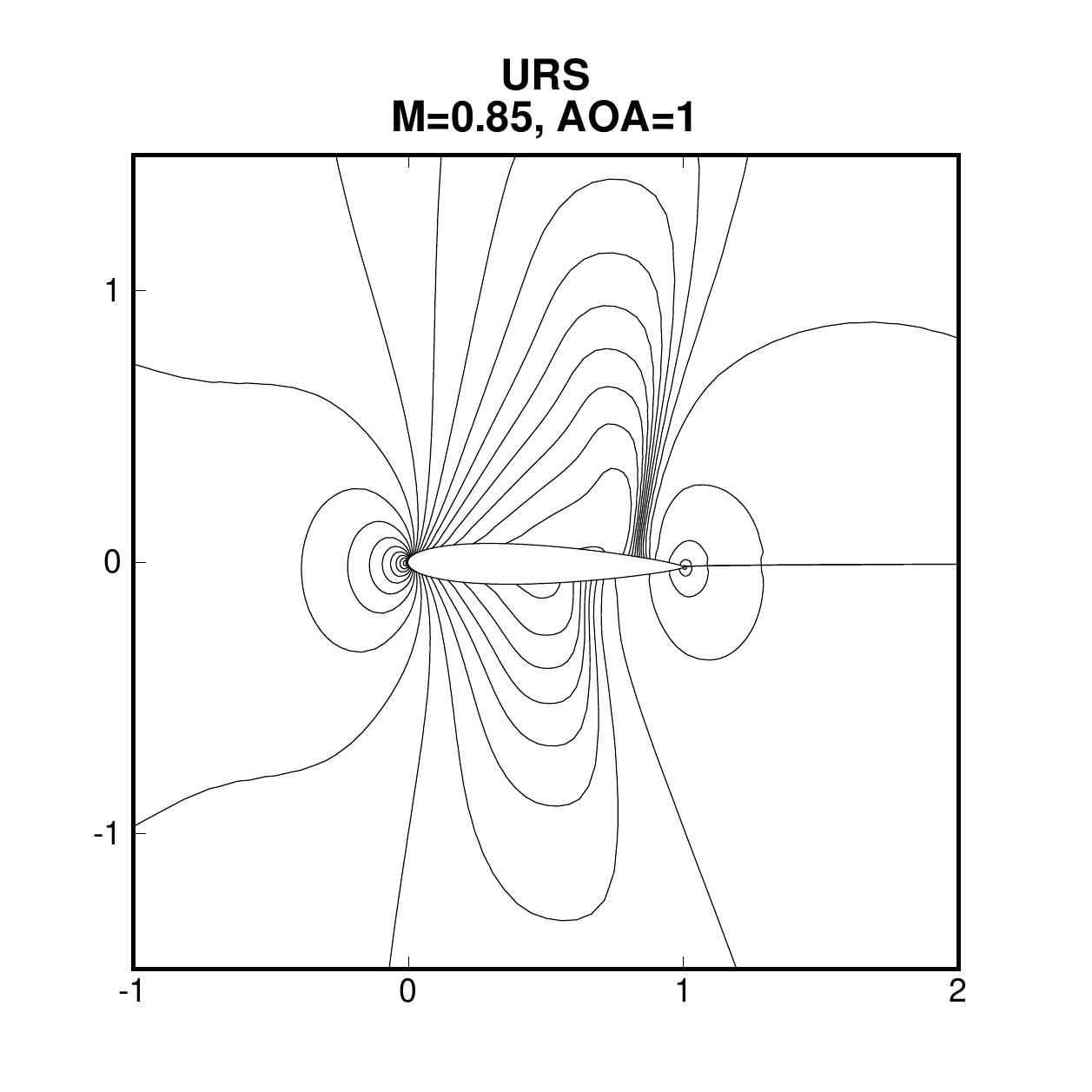}
\caption{URS}
\end{subfigure}%
\begin{subfigure}[b]{.5\textwidth}
\centering
\includegraphics[width=1.0\textwidth]{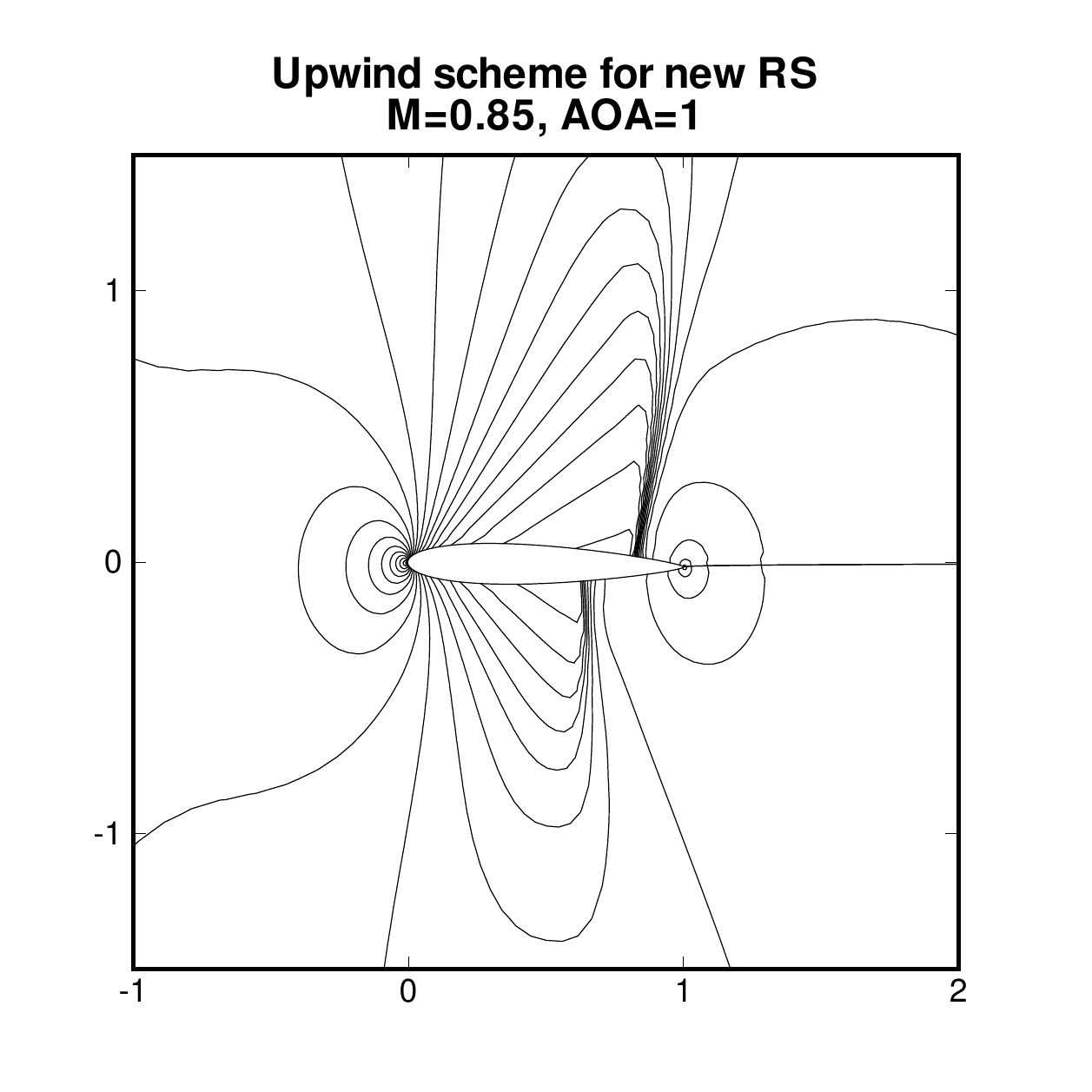}
\caption{Upwind scheme with new DKS}
\end{subfigure}
}
\caption{Comparison of (second order) pressure contours (0.405:0.05:1.805) for transonic flow over airfoil, M=0.85, AOA=1$^\circ$}
\label{fig: Compare_SO_transonic_airfoil}
\end{figure}

\subsubsection{Supersonic flow past NACA0012 airfoil}
\begin{figure}[h!]
\makebox[\textwidth][c]{%
\begin{subfigure}[b]{.5\textwidth}
\centering
\includegraphics[width=1.0\textwidth]{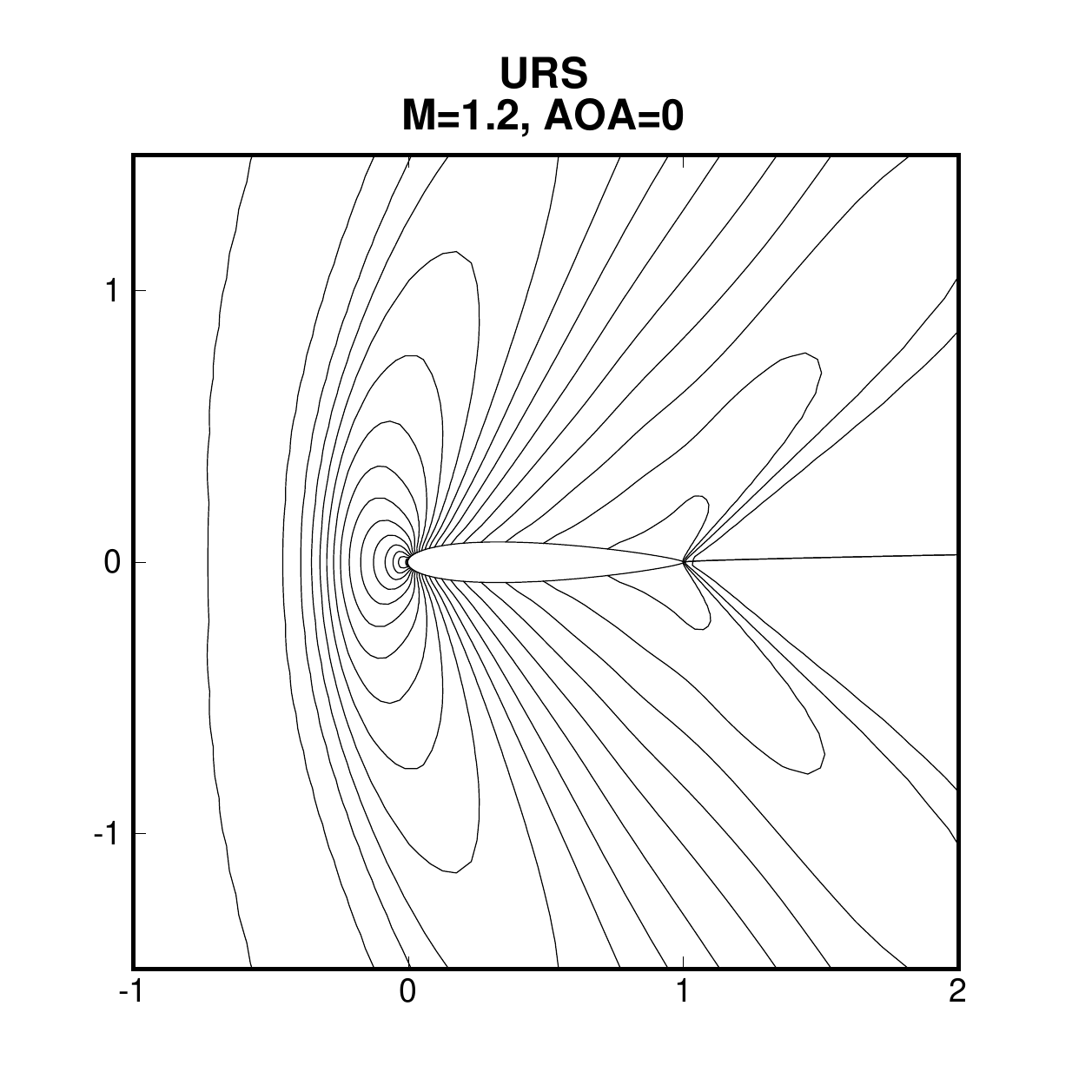}
\caption{URS}
\end{subfigure}%
\begin{subfigure}[b]{.5\textwidth}
\centering
\includegraphics[width=1.0\textwidth]{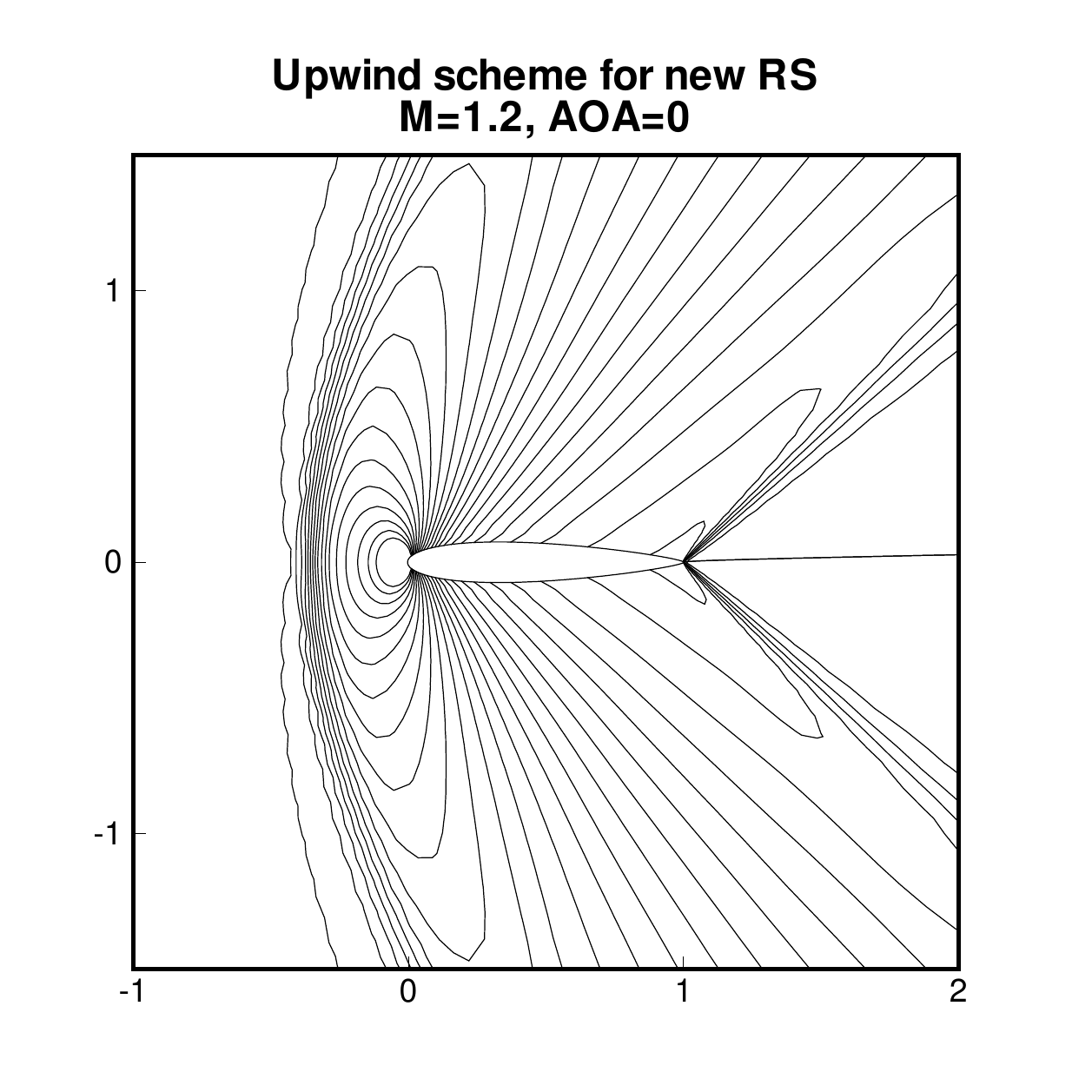}
\caption{Upwind scheme with new DKS}
\end{subfigure}
}
\caption{Comparison of (second order) pressure contours (0.405:0.05:1.805) for supersonic flow over airfoil, M=1.2, AOA=0}
\label{fig: Compare_SO_supersonic_airfoil}
\end{figure}

\begin{figure}[h!]
\makebox[\textwidth][c]{%
\begin{subfigure}[b]{.5\textwidth}
\centering
\includegraphics[width=1.0\textwidth]{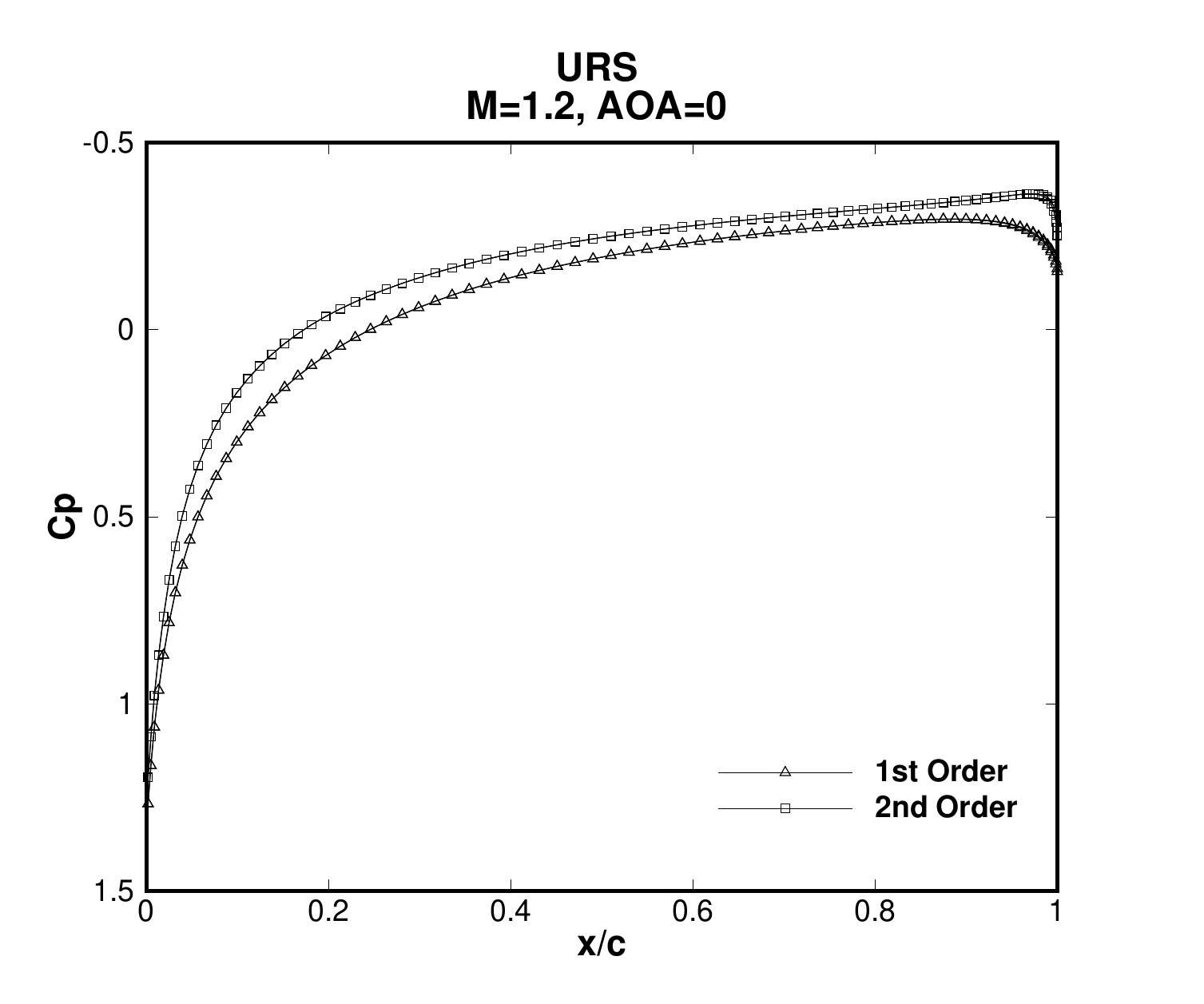}
\caption{URS}
\end{subfigure}%
\begin{subfigure}[b]{.5\textwidth}
\centering
\includegraphics[width=1.0\textwidth]{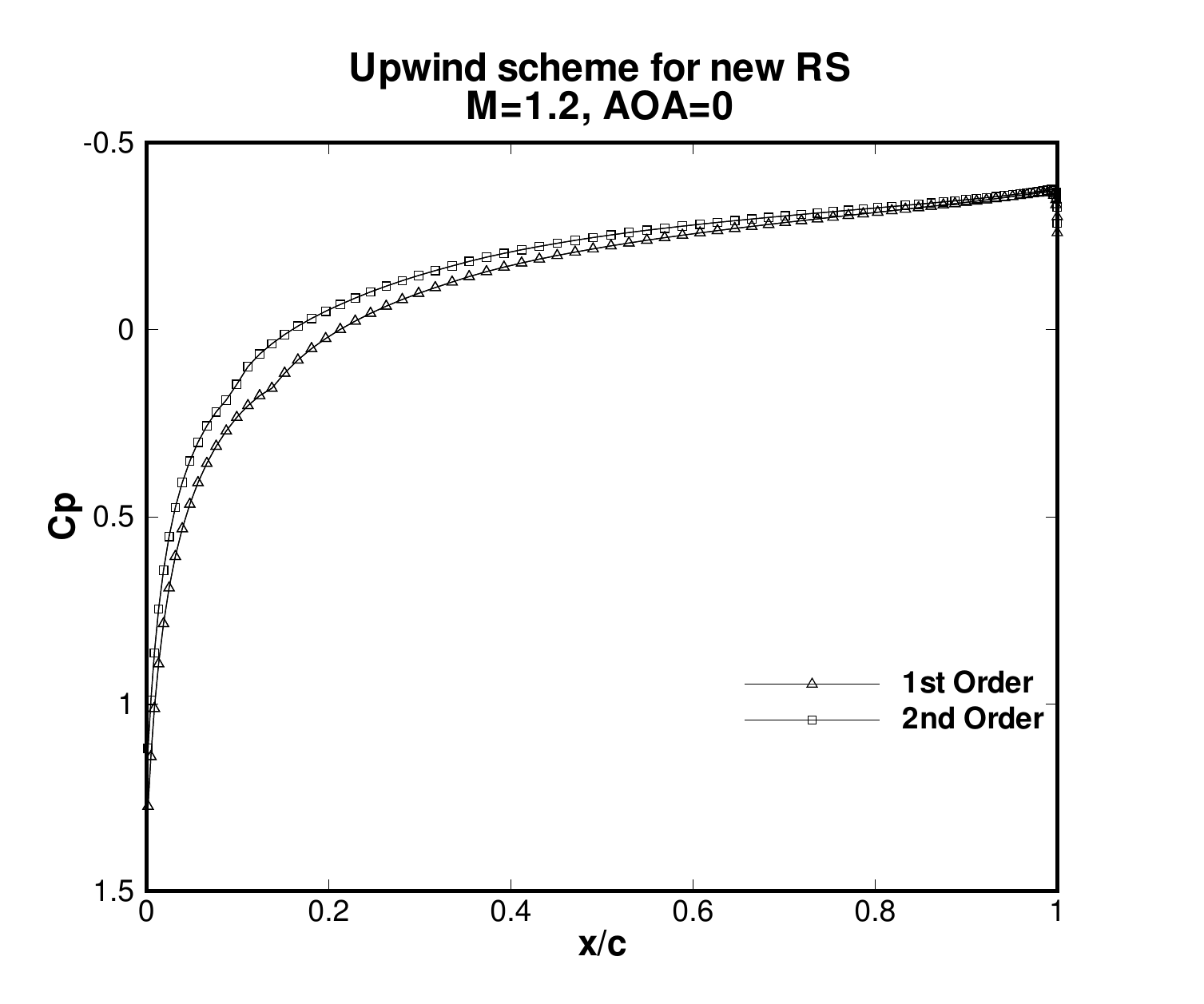}
\caption{Upwind scheme with new DKS}
\end{subfigure}
}
\caption{Comparison of Cp plots for supersonic flow over airfoil, M=1.2, AOA=0}
\label{fig: Cp_supersonic_airfoil}
\end{figure}

A benchmark case of supersonic flow over NACA0012 airfoil~\cite{Viviand}, with an inflow Mach number 1.2 and zero angle of attack, is simulated using the present schemes. Pressure contours are plotted in figure \ref{fig: Compare_SO_supersonic_airfoil}. C$_p$ plots are presented in figure \ref{fig: Cp_supersonic_airfoil}. C$_p$ values from the upwind scheme with new DKS are closer to those in the benchmark~\cite{Viviand}, compared to the same in URS.

\newpage

\subsubsection{Positivity-preservation test cases}
As discussed in section \ref{sec:positivity_pres}, to check positivity preservation of the upwind scheme for the new discrete kinetic system, we use the scheme on specifically designed test cases provided by Parent~\cite{B_Parent}. Computations with this new discrete kinetic scheme did not fail and negative values of density or pressure never developed.  The results are discussed below.
\begin{enumerate}
 \item {\bf Flow in a rectangular enclosure with a cut-out along the bottom wall}: The streamlines pertaining to the initial conditions are shown in figure \ref{fig:positivity_result_1_streamlines_ic}. After a certain time t = 0.00047, the flow starts to turn around upon hitting the surfaces of the cut-out. This situation can be seen in figure \ref{fig:positivity_result_1_streamlines_400_iter}. The corresponding pressure contours are also shown in figure \ref{fig:positivity_result_1_pressure_400_iter}. After a further time period, at t=0.00097, the flow bounces back from the boundary and meets flow from inside, from other directions. In some places, this forms slip lines. The streamlines and pressure contours for this situation are shown in figures \ref{fig:positivity_result_1_streamlines_2400_iter} and \ref{fig:positivity_result_1_pressure_2400_iter}.
 
\begin{figure}[h!]
\centering
\includegraphics[trim = 4mm 4mm 10mm 10mm,clip,width=0.7\linewidth]{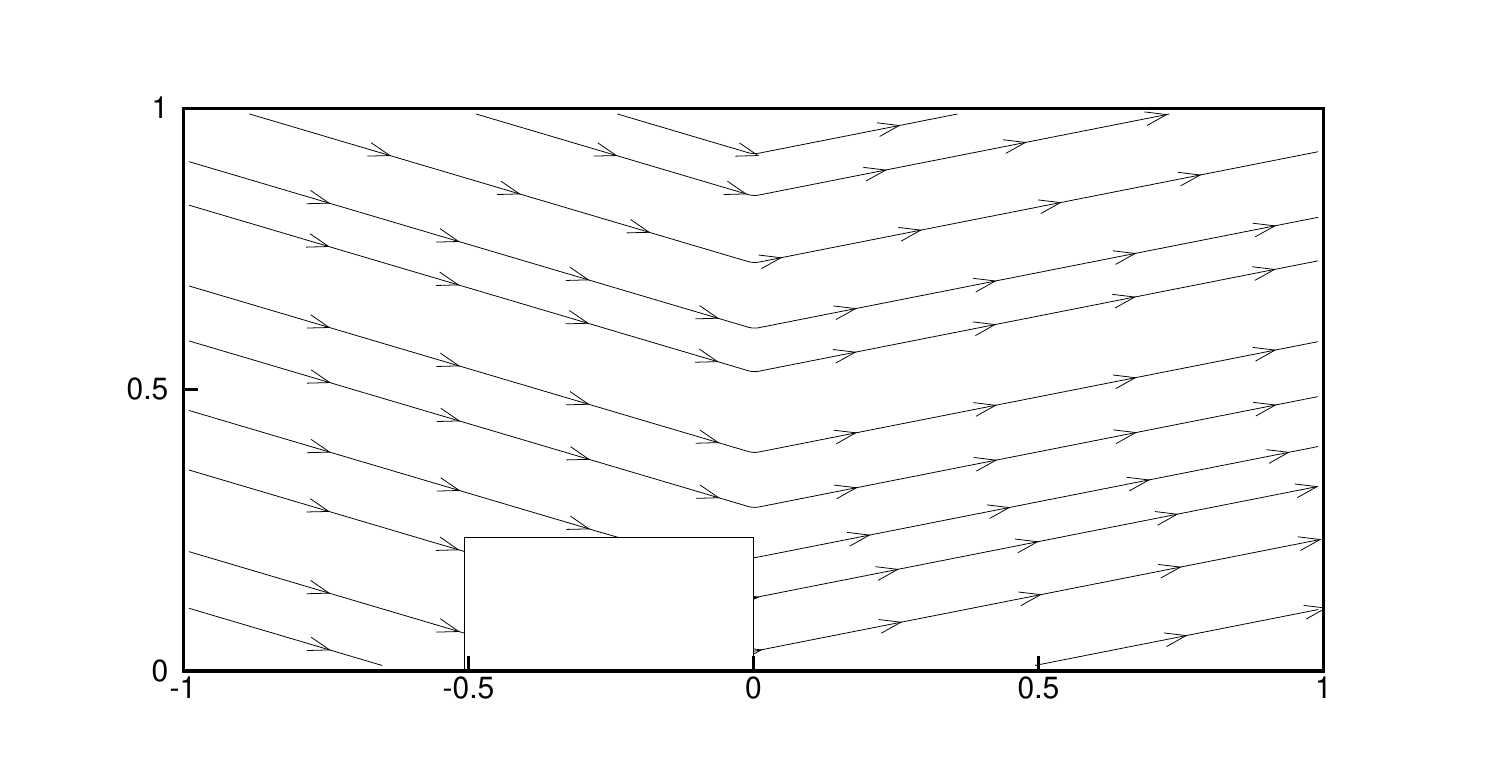}
\caption{Initial streamlines for test case \#8 of Parent~\cite{B_Parent}}
  \label{fig:positivity_result_1_streamlines_ic}
\end{figure}

\begin{figure}[h!]
\centering
\includegraphics[trim = 4mm 4mm 10mm 10mm,clip,width=0.7\linewidth]{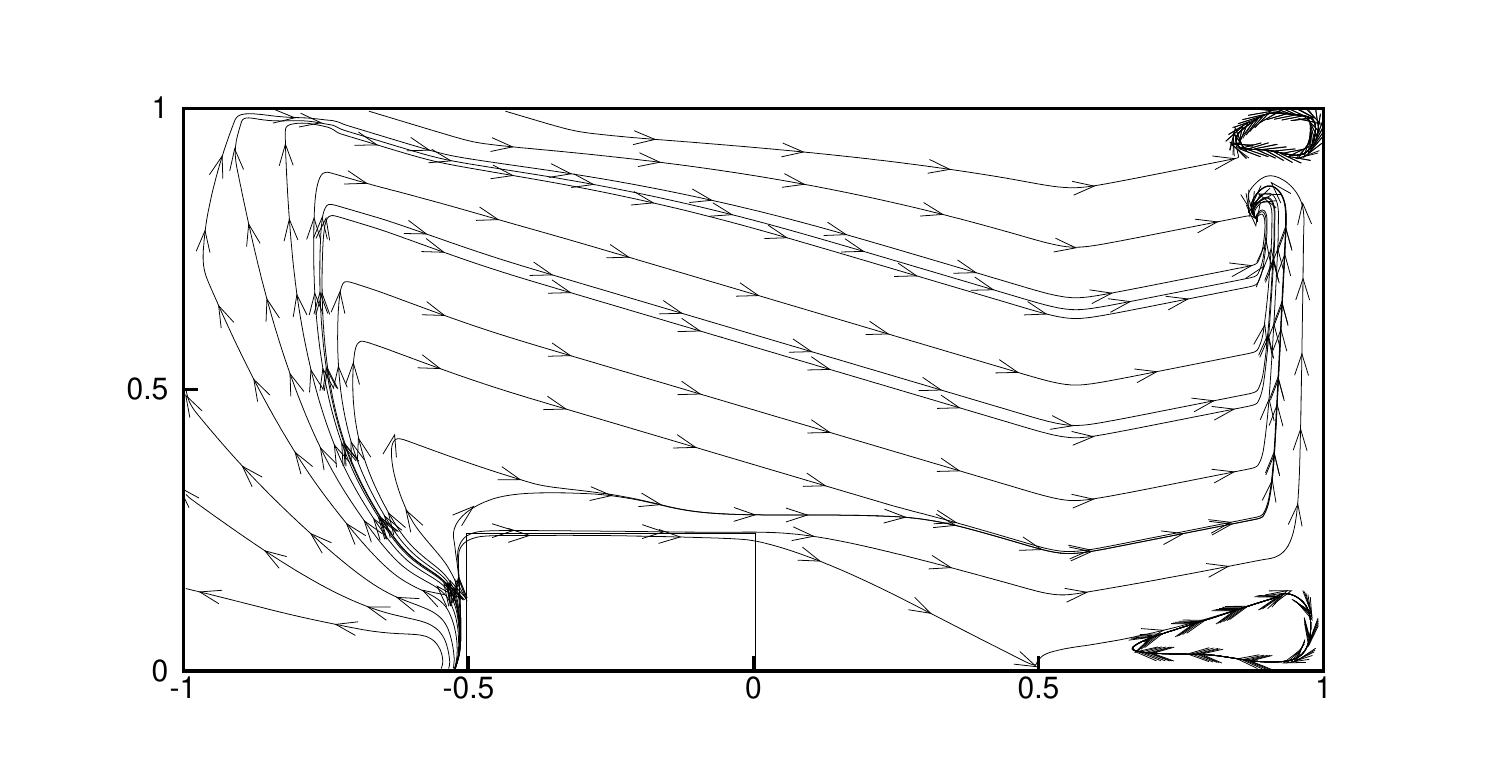}
\caption{Streamlines at t = 0.00047}
  \label{fig:positivity_result_1_streamlines_400_iter}
\end{figure}

\begin{figure}[h!]
\centering
\includegraphics[trim = 4mm 4mm 10mm 10mm,clip,width=0.7\linewidth]{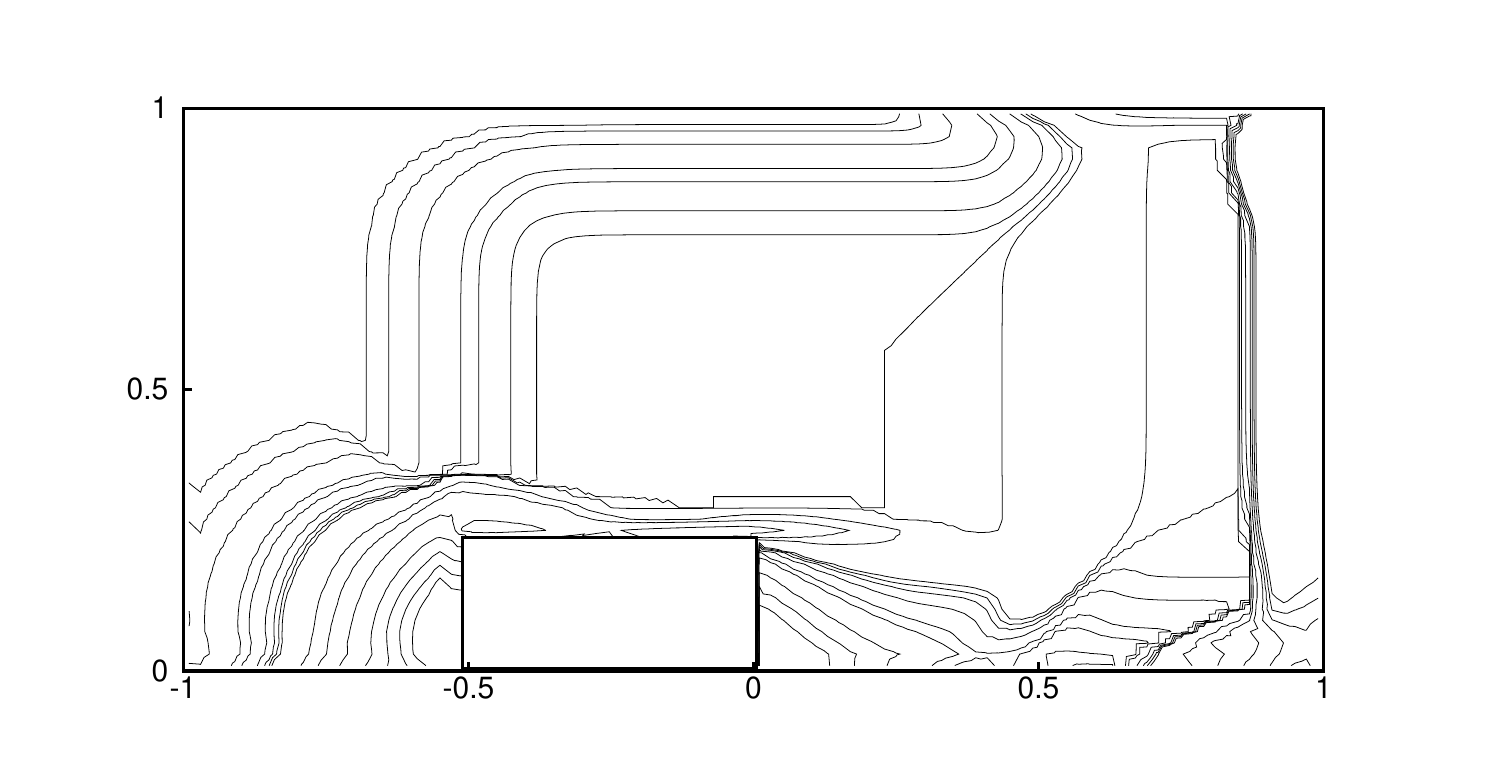}
\caption{Pressure contours at t = 0.00047}
  \label{fig:positivity_result_1_pressure_400_iter}
\end{figure}

\begin{figure}[h!]
\centering
\includegraphics[trim = 4mm 4mm 10mm 10mm,clip,width=0.7\linewidth]{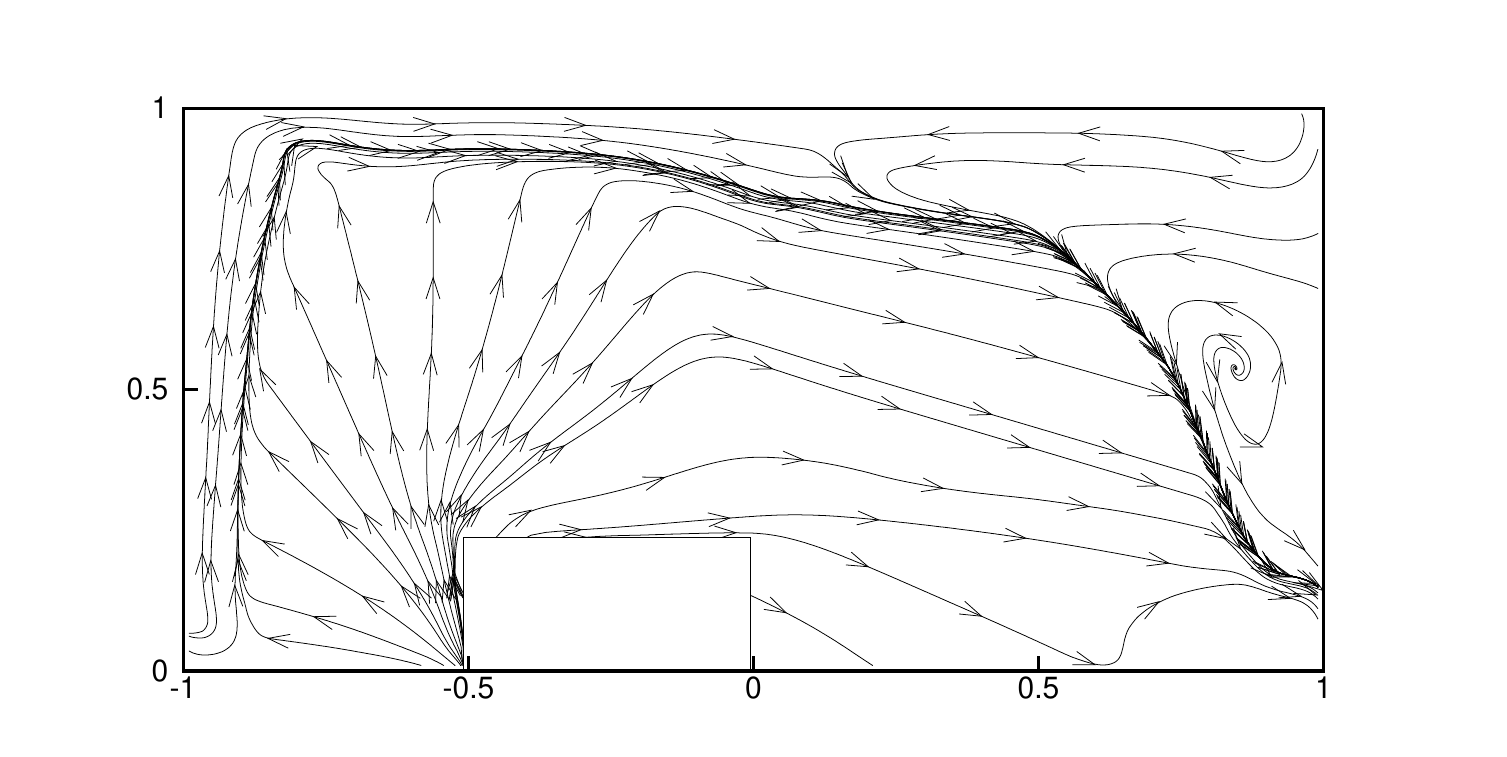}
\caption{Streamlines t = 0.00097}
  \label{fig:positivity_result_1_streamlines_2400_iter}
\end{figure}

\begin{figure}[h!]
\centering
\includegraphics[trim = 4mm 4mm 10mm 10mm,clip,width=0.7\linewidth]{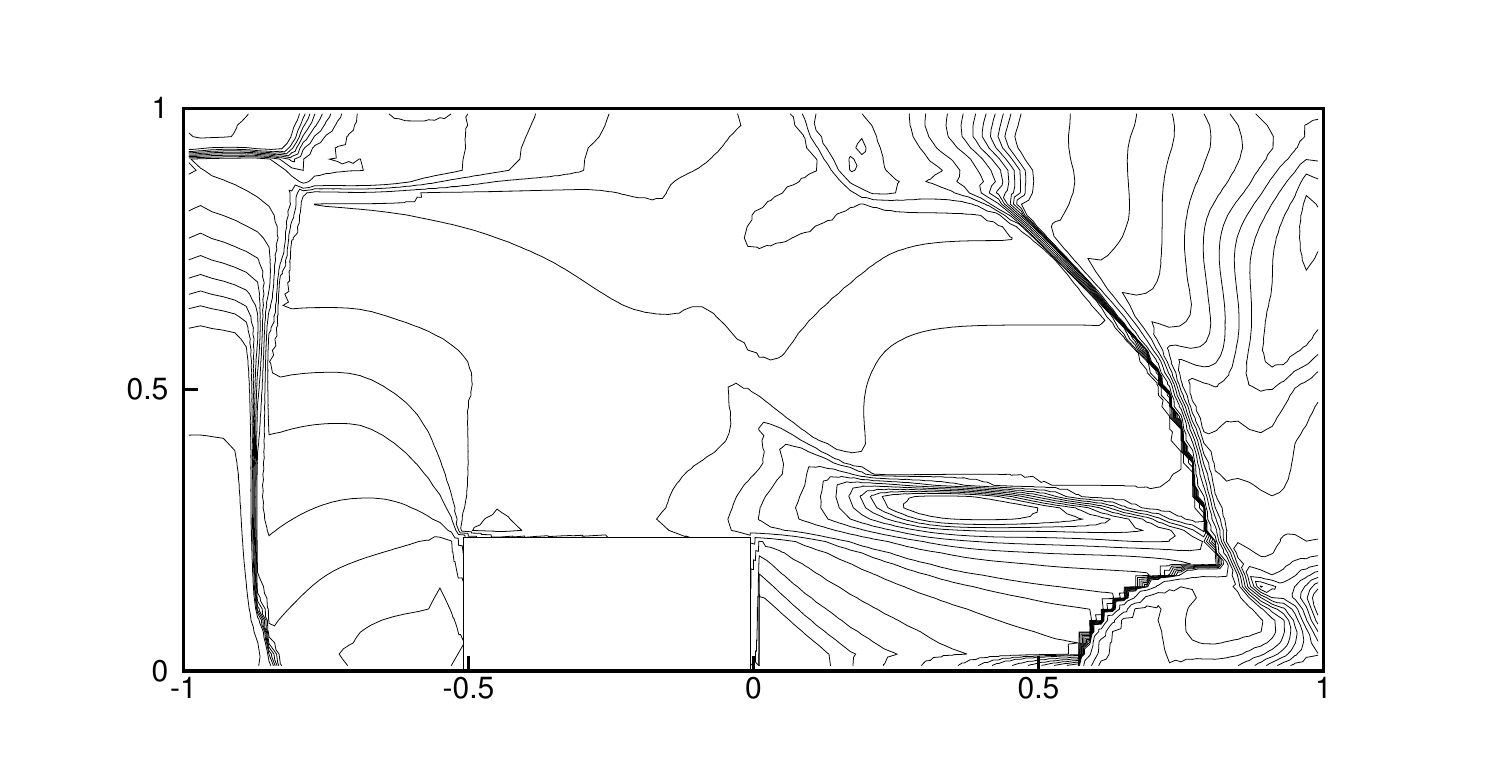}
\caption{Pressure contours at t = 0.00097}
  \label{fig:positivity_result_1_pressure_2400_iter}
\end{figure}

\newpage

\item {\bf Supersonic flow through a channel with a wavy wall at the bottom}: This test case is associated with generation of expansion waves along the bottom wall, their subsequent reflections at the top wall.  All the features are captured by the new discrete kinetic scheme as seen in the pressure contour plot in figure \ref{fig:positivity_result_3}.

\begin{figure}[h!]
\centering
\includegraphics[trim = 1mm 1mm 1mm 1mm,clip,width=0.95\linewidth]{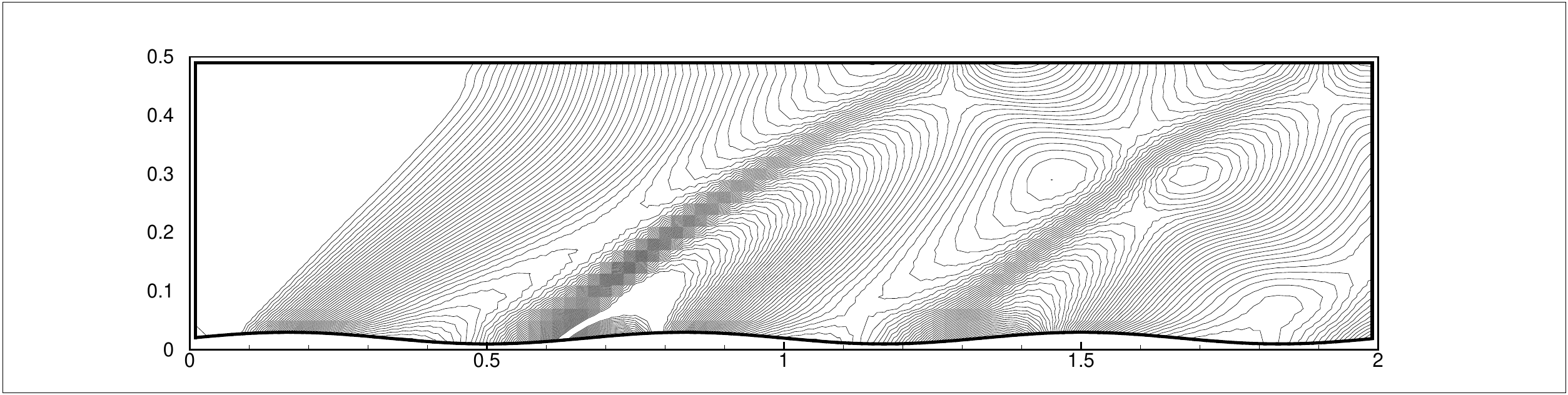}
\caption{Pressure contours for test case \#11 of Parent~\cite{B_Parent}}
  \label{fig:positivity_result_3}
\end{figure}

\end{enumerate}

\section{Summary}
Novel discrete kinetic systems for Euler equations with physically relevant discrete velocities are presented.  The kinetic theory based derivations of discrete velocities is introduced to match the eigenvalues of the corresponding flux Jacobian matrices for the Euler equations. The corresponding upwind schemes are quite accurate and robust as demonstrated for various benchmark 1-D and 2-D test cases.  





\end{document}